\documentclass[amsmath,amssymb,aps,pre,groupedaddress,superscriptaddress,twocolumn]{revtex4-2}
\usepackage[utf8]{inputenc}
\usepackage[english]{babel}
\usepackage[english]{varioref}
\usepackage{graphicx}
\usepackage{dcolumn}
\usepackage{bm}
\usepackage{amsmath}
\usepackage{amssymb}
\usepackage{xcolor}
\usepackage[font=small]{caption}
\usepackage{mathrsfs}
\usepackage{tabularx}
\usepackage[colorlinks=true,linkcolor=blue,urlcolor=blue,citecolor=blue,anchorcolor=blue]{hyperref}

\begin{document}

\title{Emerging Activity Temporal Hypergraph: a model for generating realistic time-varying hypergraphs
}

\author{Marco Mancastroppa}
\email{marco.mancastroppa@cpt.univ-mrs.fr}
\affiliation{Aix Marseille Univ, Université de Toulon, CNRS, CPT, Turing Center for Living Systems, 13009 Marseille, France}
\author{Giulia Cencetti}
\affiliation{Aix Marseille Univ, Université de Toulon, CNRS, CPT, Turing Center for Living Systems, 13009 Marseille, France}
\author{Alain Barrat}
\affiliation{Aix Marseille Univ, Université de Toulon, CNRS, CPT, Turing Center for Living Systems, 13009 Marseille, France}

\begin{abstract}
Time-varying group interactions constitute the building blocks of many complex systems.
The framework of temporal hypergraphs makes it possible to represent them by taking into
account the higher-order and temporal nature of the interactions. 
However, the corresponding datasets are often incomplete and/or limited in size and duration, 
and surrogate time-varying hypergraphs able to reproduce their statistical features
constitute interesting substitutions, especially to understand how dynamical processes unfold on group interactions.
Here, we present a new temporal hypergraph model, the Emerging Activity Temporal Hypergraph (EATH),
which can be fed by parameters measured in a dataset and 
create synthetic datasets with similar properties
. 
In the model, each node has an independent underlying activity dynamic and the overall system activity emerges from the nodes dynamics, with temporal group
interactions resulting from both the activity of the nodes and memory mechanisms. 
We first show that the EATH model can generate surrogate hypergraphs of several empirical datasets of face-to-face interactions, mimicking temporal and topological properties at the node and hyperedge level. We also showcase
the possibility to use the resulting synthetic data in simulations of higher-order contagion dynamics, 
comparing the outcome of such process on original and surrogate datasets. Finally, we
illustrate the flexibility of the model, which can generate synthetic hypergraphs with tunable
properties: as an example, we generate "hybrid" temporal hypergraphs, which mix properties of different empirical datasets. 
Our work opens several perspectives, from the generation of synthetic realistic hypergraphs
describing contexts where data collection is difficult 
to a deeper understanding of dynamical processes on temporal hypergraphs.
\end{abstract}

\maketitle

\section{Introduction}
Static network theory represents a powerful framework for the description of complex systems composed of interacting elements, providing crucial insights into the main features of many systems and into the understanding of dynamical processes unfolding on them \cite{newman_book,barrat2008dynamical}. However,  evidence has emerged that many complex systems present properties not captured by static pairwise networks: \textit{(i)} many real systems feature  group (higher-order) interactions \cite{battiston2020,bick2023,battiston2021physics}, and \textit{(ii)} such interactions evolve over time, with non-trivial activation, aggregation and disaggregation dynamics \cite{HOLME201297,masuda2016,Karsai2012,Cencetti2021,kirkley2024,Iacopini2024,Mancastroppa2024}. The higher-order and temporal nature of interactions requires a shift of representation from \textit{static pairwise networks} to \textit{temporal hypergraphs}, where nodes can interact through hyperedges (group interactions) of arbitrary size, each with its own activation times and durations. 
Taking into account these two levels of complexity allows to better characterize the structure 
and dynamics of complex systems and the properties of dynamical processes occurring on them. 
Indeed, explicitly representing group interactions by hyperedges provides new insights, through the definition of specific higher-order structure characterization tools 
and the inclusion of higher-order interaction mechanisms in dynamical processes \cite{battiston2020,bick2023,battiston2021physics,mancastroppa2023hyper,Contisciani2022,agostinelli2025higher,iacopini2019,cencetti23,Majhi2022}.
Moreover, the temporal dimension is needed to describe the system's evolution
\cite{HOLME201297,masuda2016,Mancastroppa2024}
and to take into account the impact on dynamical processes of features such as causality, non-simultaneity of interactions, and heterogeneities in the duration of interactions or between interactions \cite{Perra2012,Mancastroppa_2019,karsai2018bursty,rocha2013,moinet2016,loedy2024}. 

Taking into account both the temporal and higher-order nature of interactions is however challenging. Advances have recently been made in the characterization of temporal hypergraphs \cite{Mancastroppa2024,Cencetti2021,ceria2023temporal,Gallo2024,cencetti24_bea,Iacopini2024,digaetano2024} and in the study of dynamical processes on temporal hypergraphs \cite{neuhauser2021,chowdhary2021simplicial}. 
Models of temporal hypergraphs have been proposed, with varying levels of realism \cite{petri2018,Mancastroppa2024,Gallo2024,Iacopini2024,digaetano2024}. For instance, higher-order activity driven (HAD) models generalized the activity-driven networks \cite{Perra2012,Mancastroppa2024}.
Other models of temporal hypergraphs have been designed to reproduce specific properties observed in real systems, such as intra-order and cross-order correlation profiles \cite{Gallo2024} or the aggregation and disaggregation dynamics of groups \cite{Iacopini2024}.
Despite these important steps, 
we still lack systematic methods capable of generating surrogate hypergraphs, 
i.e., of measuring a set of features in a given empirical dataset and of producing
synthetic stochastic data mimicking its properties, as done for temporal networks \cite{cencetti2024generating,Presigny2021,Calmon2024}. 
Such methods are useful in particular when datasets are incomplete or limited in duration and size,
as they can generate realistic synthetic data on longer timescales or with larger populations
\cite{Genois2015_Nc,EAMES201572,cencetti2024generating,Presigny2021,Calmon2024}, 
making it possible to investigate dynamical processes \cite{cencetti2024generating,Calmon2024}, e.g. studying the importance of specific structures in a spreading process, or the impact of various containment measures. 

Here, we provide a contribution to this modelling endeavour by proposing a new model for temporal hypergraphs: the \textit{Emerging Activity Temporal Hypergraph (EATH)} model. This model takes as input an empirical system and extracts a certain set of features from it; then, to generate the dynamics of hyperedges, it relies on several theoretical mechanisms driving the creation of group interactions, previously shown to generate realistic features in temporal network models
\cite{zhao2011,vestergaard2014,cencetti2024generating}.
Specifically, here we consider a series of empirical hypergraphs describing time-resolved
face-to-face interactions between individuals in different contexts, and identify a set of common emergent
properties generally considered important for dynamical processes (Sec. \ref{sec:empirics}):
\textit{(i)} a heterogeneous, bursty and correlated dynamics both at the nodes and groups level \cite{Cencetti2021,karsai2018bursty,Karsai2012,Iacopini2024}; 
\textit{(ii)} a heterogeneous topological and temporal behaviour of single nodes and groups;
\textit{(iii)} non-trivial patterns of node participation in groups of different sizes. 

To produce synthetic data emulating such properties, we first assume that each node presents intrinsic features, with in particular an independent underlying dynamical modulation of social activity, transitioning between a low- and a high-activity phase \cite{poissonian_bursty_2008,Reis2020,Hiraoka2020,Karsai2012}. 
The system activity results from the superposition of these activity dynamics and from the mechanisms
determining the creation (and end) of hyperedges, which integrate nodes' activity 
\cite{Perra2012} with long- and short-term memory mechanisms \cite{vestergaard2014,zhao2011,stehle2010dynamical,laurent2015calls,Iacopini2024,kim2018,gelardi2021,lebail2023} (Sec. \ref{sec:model}). 
For several dataset, we show that these model ingredients are able to  
generate surrogate temporal hypergraphs replicating empirical temporal-topological properties
of the original data (Sec. \ref{sec:surrogate}). 
In particular, the surrogates produced by the EATH model capture the
hypergraph features shaping the higher-order contagion dynamics \cite{St-Onge2022}
on temporal hypergraphs (Sec. \ref{sec:process}). 

While its initial purpose is to produce surrogate hypergraphs of specific given datasets, 
the EATH model can also generate synthetic hypergraphs with tunable artificial properties,
describing a wide variety of potential behaviours. Such synthetic data 
can be used for instance to investigate the impact of specific temporal 
or structural properties on the outcome of dynamical processes \cite{HOLME201297,masuda2016,neuhauser2021,chowdhary2021simplicial}. Moreover, 
we illustrate how the EATH model can generate realistic temporal hypergraphs representing systems for which data are not available or cannot be collected easily (Sec. \ref{sec:hybrid}), for example \textit{"hybrid"} systems that mix properties of different contexts or populations \cite{cencetti2024generating,Genois2015_Nc,EAMES201572}.

\section{Empirical data}
\label{sec:empirics}

\subsection{Definitions and notations}

We consider a time-varying hypergraph $\mathscr{H}$ over the time interval $(0,\mathcal{T}]$ and we represent it as a sequence of $\mathcal{T}/\delta t$ static hypergraphs $\mathscr{H}=\{\mathcal{H}_t\}_{t=1}^{\mathcal{T}/\delta t}$, with time-resolution $\delta t$ 
{(the index $t$ goes from 1 to $\mathcal{T}/\delta t$, indicating each static snapshot starting from the first time-window at $t=1$)}. At each time $t$ the system is represented by the unweighted instantaneous hypergraph $\mathcal{H}_t=(\mathcal{V}_t,\mathcal{E}_t)$, with $N_t=|\mathcal{V}_t|$ active nodes participating in $E_t=|\mathcal{E}_t|$ hyperedges of arbitrary size $m \in [2,M]$
(we define a node as ``active'' if it is involved in at least one hyperedge of size $\geq 2$, i.e.,
when it is not isolated). The nodes and hyperedges within $\mathcal{H}_t$ are the ones that have been active at least once in the time interval $((t-1)\delta t, t \delta t]$ \cite{Mancastroppa2024,kirkley2024}. 

The weighted aggregated static hypergraph $\mathcal{H}=(\mathcal{V},\mathcal{E})$ is composed of all the $N=|\mathcal{V}|$ nodes and all the hyperedges that have been active at least once in $\mathscr{H}$, and the hyperedge weight $w_e$ is the number of snapshots in which the hyperedge $e$ has been active. We define the hyperedges size distribution as $\Psi(m)=\sum_{e \in \mathcal{E}| |e|=m} w_e/\sum_{e \in \mathcal{E}} w_e$: it thus takes into account the number of times that each hyperedge of size $m$ has been active in $\mathscr{H}$. 

We also consider the pairwise projection $\mathcal{G}=(\mathcal{V},\mathcal{L})$ of $\mathcal{H}$, obtained by projecting each hyperedge in $\mathcal{H}$ on the corresponding clique and by assigning to each (pairwise) link $l \in \mathcal{L}$ a weight $w_l = \sum_{e \in \mathcal{E} | l \subset e} w_e$. 

To each node $i$ we can finally assign its aggregated properties in the aggregated hypergraph and in its pairwise projection: namely, its hyperdegree $D(i)$, {i.e., the number of distinct hyperedges in which $i$ is involved in $\mathcal{H}$}, its hyper-strength $S(i)=\sum_{e \in \mathcal{E}|i \in e}w_e$, 
its degree $D_{proj}(i)$, {i.e., the number of distinct links in which $i$ is involved in $\mathcal{G}$}, and its strength $S_{proj}(i)=\sum_{l \in \mathcal{L}|i \in l} w_l$. 

\subsection{Empirical temporal hypergraphs}

\begin{figure*}[ht!]
\includegraphics[width=\textwidth]{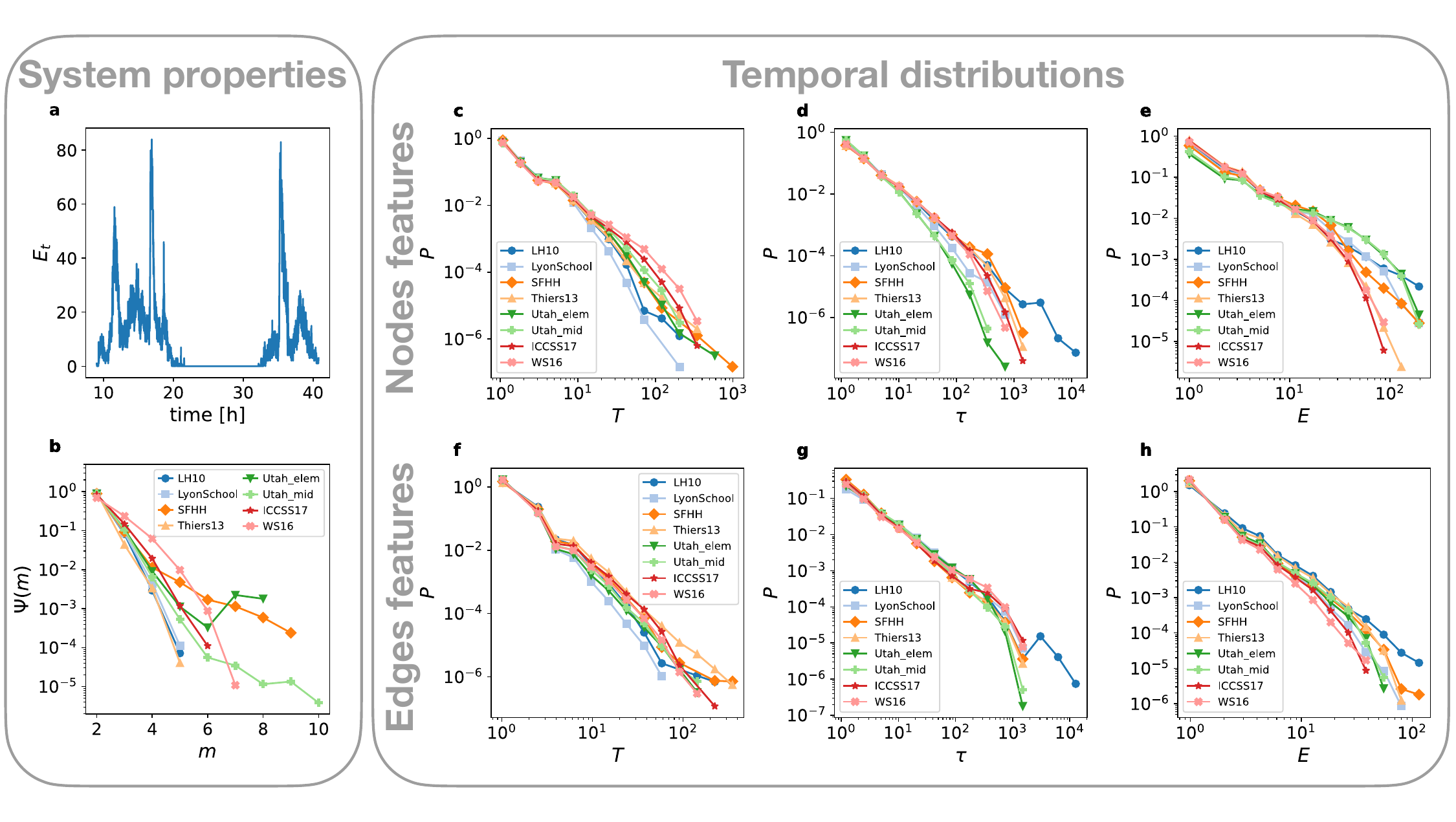}
\caption{\textbf{Properties of empirical temporal hypergraphs.} In panel \textbf{a} we report the evolution of the number of active hyperedges $E_t$ for the SFHH dataset, while in panel \textbf{b} we show the hyperedge size distribution $\Psi(m)$ for all the datasets. In panels \textbf{c}-\textbf{h}, the first and second rows show respectively the temporal properties of nodes and hyperedges in different datasets. \textbf{c},\textbf{f}: distribution of event durations $P(T)$;  \textbf{d},\textbf{g}: distribution of inter-event times $P(\tau)$;  \textbf{e},\textbf{h}: distribution of the number of events in a train of events $P(E)$, where a train is defined with $\Delta = 15 \delta t $ for all datasets, except for LH10 and SFHH where we consider $\Delta = 60 \delta t $, with $\delta t$ the hypergraph resolution.}
\label{fig:figure1}
\end{figure*}

Real time-varying networks and hypergraphs are characterized by non-trivial temporal and higher-order properties~\cite{karsai2018bursty,Cencetti2021,Iacopini2024}.
Here, we consider eight different temporal hypergraphs of face-to-face interactions between individuals, collected in various contexts (see Methods for their detailed description): schools in France and in the USA (``LyonSchool'', ``Thiers13'', ``Utah\_elem'', ``Utah\_mid''), hospital (``LH10''), and conferences (``SFHH'', ``ICCSS17'', ``WS16'').
For all these datasets the entire system’s activity undergoes important variations in time, as shown in Fig.~\ref{fig:figure1}\textbf{a}, where we report an example of the temporal evolution of the total number of interactions $E_t$ (see the {Sec. IA} of the Supplementary Material, SM, for the other datasets). 
Periods of high activity are followed by intervals of inactivity, and different orders of interactions can be more or less active at different times (see SM, {Sec. IA}). Different orders are also characterized by different frequencies of appearance: small interaction sizes are encountered more often, and large 
hyperedges are much less frequent, as shown by Fig.~\ref{fig:figure1}\textbf{b} where the hyperedges size distributions $\Psi(m)$ are reported for all the datasets.
These observations provide a first insight of the complexity of empirical hypergraphs and will constitute the basis on which we will start to build our generation model.

Figure~\ref{fig:figure1}\textbf{c-h} displays the distribution of six quantities characterizing the temporal activation of nodes and of hyperedges. 
Starting with the point of view of nodes, Fig. \ref{fig:figure1}\textbf{c} reports the distribution of node event durations $P(T)$ for all the nodes of the hypergraph. 
We recall that an event for node $i$ starts at time $t$ if $i$ goes from being isolated at $t-1$ to being involved in at least one hyperedge at $t$, and 
the event ends when $i$ becomes again isolated (when it is no longer involved in any hyperedge). 
Note that an event for node $i$ is considered to go on even if the hyperedges in which it is 
involved during the event change: the event 
describes a temporal interval between two times in which $i$ is isolated.
The duration of a node event hence corresponds to the number of consecutive time steps in which the node is active.
Figure \ref{fig:figure1}\textbf{c} shows that all the considered hypergraphs are characterized by heavy-tailed distributions of the node event durations.

Since each node usually experiences multiple events during the time span of the hypergraph, we also measure the number of time steps between two subsequent events, i.e. the inter-event times. 
The distribution of node inter-event times $P(\tau)$ is shown in Fig. \ref{fig:figure1}\textbf{d}, with again a heavy-tailed distribution for each dataset.
The heterogeneity of these distributions can be quantified through the burstiness parameter $B$ \cite{goh2008burstiness,karsai2018bursty}. Here we compare the burstiness of a distribution $P(x)$ of a quantity $x$ (event duration or inter-event time) with that of an exponential distribution with the same mean, $\langle x \rangle$, and lower cut-off, $x_m$: 
$\Delta B = B-B_{exp}(x_m,\langle x \rangle)$. $\Delta B>0$ and $\Delta B<0$ indicate then respectively a distribution more and less heterogeneous than the exponential baseline (see Methods).
In Table \ref{tab:table1} we report $\Delta B$ for the considered temporal distributions: 
all values are positive, indicating a heterogeneous dynamics.

\begin{figure*}[ht!]
\includegraphics[width=\textwidth]{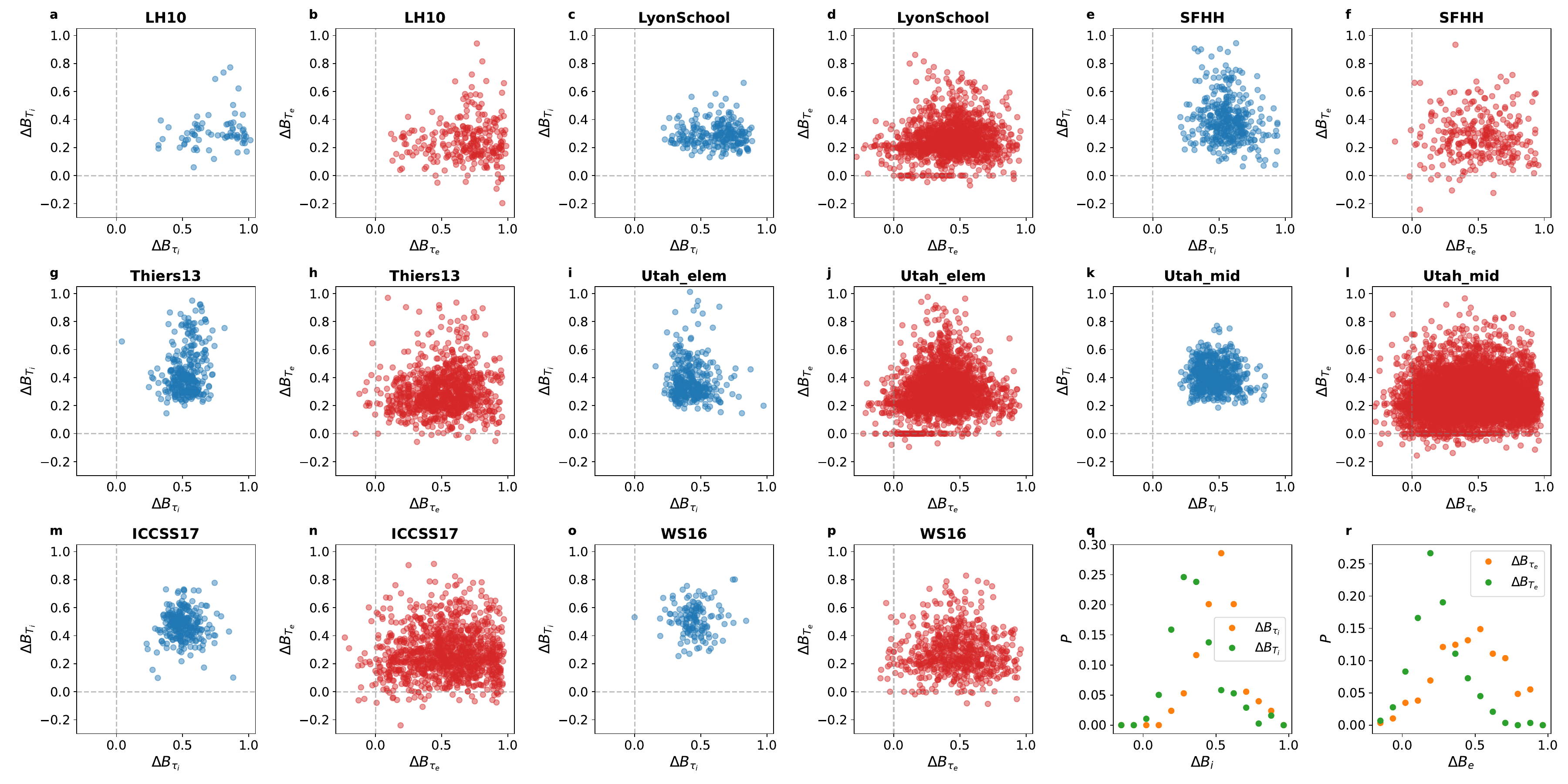}
\caption{\textbf{Nodes and hyperedges temporal heterogeneity.} For each node and hyperedge we compute the burstiness values $\Delta B_{\tau}$ and $\Delta B_{T}$ of their event and inter-event duration distributions. In each panel, each point corresponds to a node (\textbf{a},\textbf{c},\textbf{e},\textbf{g},\textbf{i},\textbf{k},\textbf{m},\textbf{o}) or a hyperedge (\textbf{b},\textbf{d},\textbf{f},\textbf{h},\textbf{j},\textbf{l},\textbf{n},\textbf{p}) and we show the scatterplot between $\Delta B_{\tau}$ and $\Delta B_{T}$. The gray dashed lines correspond to the reference $\Delta B=0$. Panels \textbf{q}, \textbf{r} show the distributions $P(\Delta B_{\tau})$ and $P(\Delta B_{T})$, respectively for nodes and hyperedges, for the SFHH dataset. We consider only the nodes and hyperedges with at least $10$ activation events.}
\label{fig:figure2}
\end{figure*}

Finally, we follow \cite{Karsai2012} and define a train of events as a series of consecutive events such that the interval between the end of an event and the beginning of the successive one is smaller than a parameter $\Delta$. We count the number $E$ of events in each such
train and report the resulting distribution $P(E)$ in Fig. \ref{fig:figure1}\textbf{e} 
for a specific value of $\Delta$ (see caption). The broad character of this distribution 
shows the presence of temporal correlations in the analyzed hypergraphs \cite{Karsai2012}. 

\begin{table}[t!]
    \begin{tabular}{c|c|c|c|c}
    \hline
    \hline
         & \multicolumn{2}{c}{Nodes} \vline & \multicolumn{2}{c}{Hyperedges} \\ \hline
         & $\Delta B_{T}$ & $\Delta B_{\tau}$ & $\Delta B_{T}$ & $\Delta B_{\tau}$ \\ \hline
         LH10 & 0.42 &0.81 & 0.56 & 0.60 \\
         LyonSchool & 0.33 & 0.66 & 0.49 & 0.41 \\
         SFHH & 0.73 &0.54 & 0.76 & 0.53 \\
         Thiers13 & 0.63 & 0.55 & 0.69 & 0.50 \\
         Utah\_elem & 0.55 & 0.50 & 0.64 & 0.37 \\
         Utah\_mid & 0.44 & 0.54 & 0.62 & 0.50 \\
         ICCSS17 & 0.52 & 0.51 & 0.59 & 0.44 \\
         WS16 & 0.52 & 0.48 & 0.62 & 0.36 \\
    \hline
    \hline
    \end{tabular}
    \caption{\textbf{Burstiness of temporal distribuitions - Data.} For each dataset we report the burstiness $\Delta B_T$ and $\Delta B_{\tau}$ of the duration distribution, $P(T)$, and of the inter-event times distribution, $P(\tau)$, for both nodes and hyperedges.} 
    \label{tab:table1}
\end{table}

The second row of Fig. \ref{fig:figure1} reports the same quantities relative to the hyperedges dynamics. Note that an event from $t_1$ to $t_2$ 
for a hyperedge simply corresponds to the existence of that hyperedge
in all snapshots $\mathcal{E}_t$ for $t \in [t_1,t_2]$, with the hyperedge being instead absent
from both $\mathcal{E}_{t_1-1}$ and  $\mathcal{E}_{t_2+1}$.
We report the distribution of hyperedge events durations, $P(T)$ (Fig. \ref{fig:figure1}\textbf{f}),  of hyperedge inter-event times, $P(\tau)$ (Fig. \ref{fig:figure1}\textbf{g}), and the distribution of the number of events in trains of hyperedge events, $P(E)$ (Fig. \ref{fig:figure1}\textbf{h}).
Again, we observe long tails, bursty dynamics (see also Table \ref{tab:table1}) and temporal correlations.

The distributions shown in Fig. \ref{fig:figure1} are built considering all nodes or all 
hyperedges. These broad distributions could however result from a superposition of either narrow or broad distributions of durations at the single node or hyperedge level, and the dynamical properties
of different nodes and hyperedges could potentially be either homogeneous or heterogeneous.
We thus investigate the dynamical properties of individual nodes and hyperedges by computing the burstiness of the distribution of event and inter-event durations of each node and hyperedge. Fig.~\ref{fig:figure2} reports for each dataset a scatterplot of the resulting burstiness values ($\Delta B_T$ vs. $\Delta B_{\tau}$): these plots highlight the complex
temporal behavior of nodes (resp. groups) 
{by illustrating simultaneously: (i) the heterogeneity in the nodes (resp. hyperedges) behaviour; (ii) the correlation (or absence thereof) 
between the two burstiness measures.} Figure 
\ref{fig:figure2}\textbf{q}, \textbf{r} report moreover the distributions of the node and 
hyperedge burstiness values for the SFHH conference dataset~\cite{sociopatterns,genois2018,ISELLA2011166,genois2018} (see the SM, {Sec. IC}, for other datasets).
Interestingly, these distributions span a broad interval of values (almost all positive): this reveals a clear heterogeneity of nodes and hyperedges temporal behaviors, some exhibiting a very bursty activity while other have statistics of event or inter-event durations closer to a Poissonian dynamics. Moreover, the scatterplots do not exhibit any correlation between the burstiness value for the event durations and the one for the inter-event durations{: this suggests that the systems are characterized by the presence of nodes and hyperedges 
with very different temporal behaviour, ranging from very regular statistics in the durations and inter-event times (with $\Delta B_T \lesssim 0$ and $\Delta B_\tau \lesssim 0$) to very irregular dynamics (with both $\Delta B_T$ and $\Delta B_\tau$ taking large values), 
passing through all intermediate combinations of behaviors.}

\begin{figure*}[ht!]
\includegraphics[width=\textwidth]{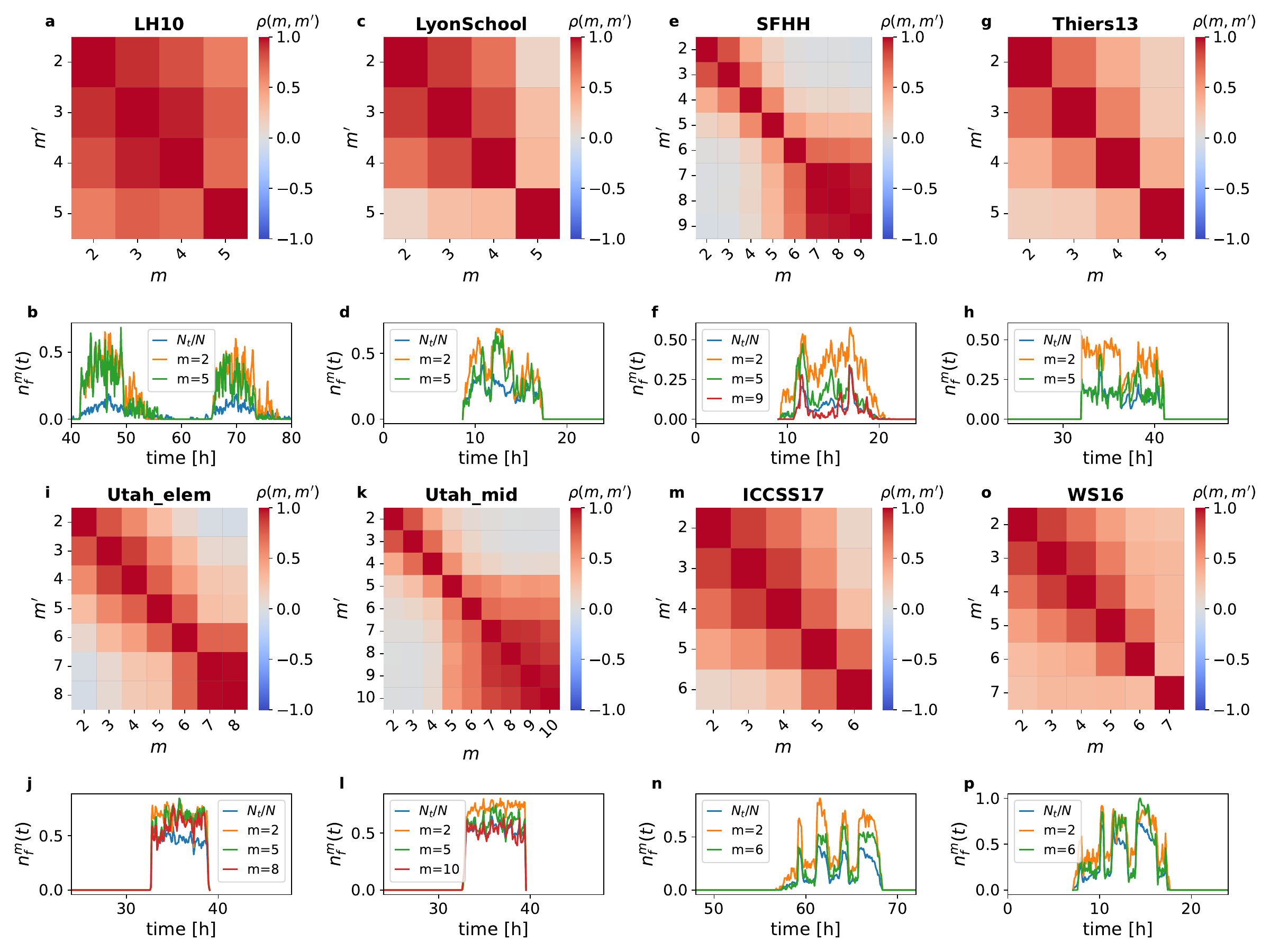}
\caption{\textbf{Participation of nodes at different interaction orders.} For each dataset, at each order $m$ we rank the nodes based on the time they spend interacting at size $m$, and 
we show the Pearson's correlation coefficient $\rho(m,m')$ between the rankings obtained at order $m$ and $m'$ (panels \textbf{a},\textbf{c},\textbf{e},\textbf{g},\textbf{i},\textbf{k},\textbf{m},\textbf{o}). We also compute the fraction $n_f^m(t)$ of nodes active at time $t$ (in hyperedges of any size) among the nodes occupying the top $f N$ positions of the nodes ranking at order $m$,  and show the temporal evolution of $n_f^m(t)$ for different $m$ (see legend), fixing $f=0.1$ (panels \textbf{b},\textbf{d},\textbf{f},\textbf{h},\textbf{j},\textbf{l},\textbf{n},\textbf{p}); we also plot the total fraction $N_t/N$ of active nodes in the population at time $t$ in each panel.}
\label{fig:figure3}
\end{figure*}

While the statistics measured above do not distinguish between hyperedges of different sizes,
Figure~\ref{fig:figure1}\textbf{b} shows that not all sizes are equally probable in
empirical hypergraphs. As Fig.~\ref{fig:figure1}\textbf{b} displays a global
distribution, we also investigate whether different nodes display different behaviors in terms of
the sizes of the hyperedges in which they participate. To this aim, we consider the total 
time spent by each node $i$ in hyperedges of order $m$, ${T}_i^m=\sum_{e \in \mathcal{F}(i) | |e|=m} w_e$, with $\mathcal{F}(i)$ the set of hyperedges in which $i$ is involved in $\mathcal{H}$.
We then rank the nodes according to this measure for each order $m$
and we compare the rankings at different orders: Fig. ~\ref{fig:figure3} gives the Pearson's correlation coefficient $\rho(m,m')$ between the rankings at orders $m$ and $m'$.
In general, adjacent orders tend to have similar node rankings, but in some cases the matrices are organized in blocks, with very low correlation values between orders belonging to different blocks. 
For instance, in the SFHH dataset the node rankings at orders between 2 and 4 are very different from those at orders between 6 and 9.
Thus, the nodes that are the most important for large interactions are not the most present in small interactions, and vice-versa, revealing a further heterogeneity among nodes, which can be divided into categories depending on how they allocate their social activity in groups of different sizes. 
Note that these differences in behaviour emerge from comparing the rankings at different group sizes, since the nodes allocate differently their activity at the various orders; however,
for all nodes the largest $T_i^m$ are obtained for small sizes $m$ (see SM, {Sec. ID}). 

Finally, Fig.~\ref{fig:figure3} also presents temporal plots of the activity of the most
important nodes at different orders $m$, i.e., those occupying the top $f N$ positions of the nodes ranking at order $m$. Specifically, each curve gives the fraction $n_f^m(t)$ of the $f N$ nodes most
present in hyperedges of size $m$ that are active in hyperedges of any size at each time $t$. 
The activation dynamics is quite correlated for all orders (and correlated with the global activity timeline). The fractions values however differ, and in some datasets the activation dynamics is different depending on the considered order: nodes of different classes tend to activate differently over time, suggesting that they participate in different system activities. For example, in SFHH the nodes more exposed to larger groups tend to activate only during the phases of high activity (e.g. conference coffee breaks), while nodes more important in low orders are often active even outside such phases; in Thiers13 (highschool), the
active fraction of nodes more important in larger groups can become larger than the one of nodes
important in small groups during the school breaks.

The above analysis confirms and expands previous results on the dynamical properties of
empirical hypergraphs \cite{zhao2011,Cencetti2021,Iacopini2024}. In particular, while the bursty dynamics of nodes activity has been considered in several models of temporal network generation~\cite{stehle2010dynamical,rocha2013,vestergaard2014}, only a few models also take into account that of network edges~\cite{Reis2020,Hiraoka2020} and, up to our knowledge, none of the existing hypergraph generation models reproduces the bursty dynamics of hyperedges.
Moreover, the heterogeneity of single hyperedges and nodes in terms both of temporal behaviours
and of relative importance in the hyperedges of different sizes was not addressed in previous 
analyses. All these results highlight important properties that a realistic hypergraph generation model should reproduce.

Let us notice that the empirical characteristics have been observed in face-to-face contact datasets and may be different in other types of temporal datasets of interactions. Even if guided
by these data, the model that we propose in the next Section is in fact highly versatile, aiming at reproducing in a surrogate the features measured in the original dataset that one wants to mimic. 

\begin{figure*}[ht!]
\includegraphics[width=\textwidth]{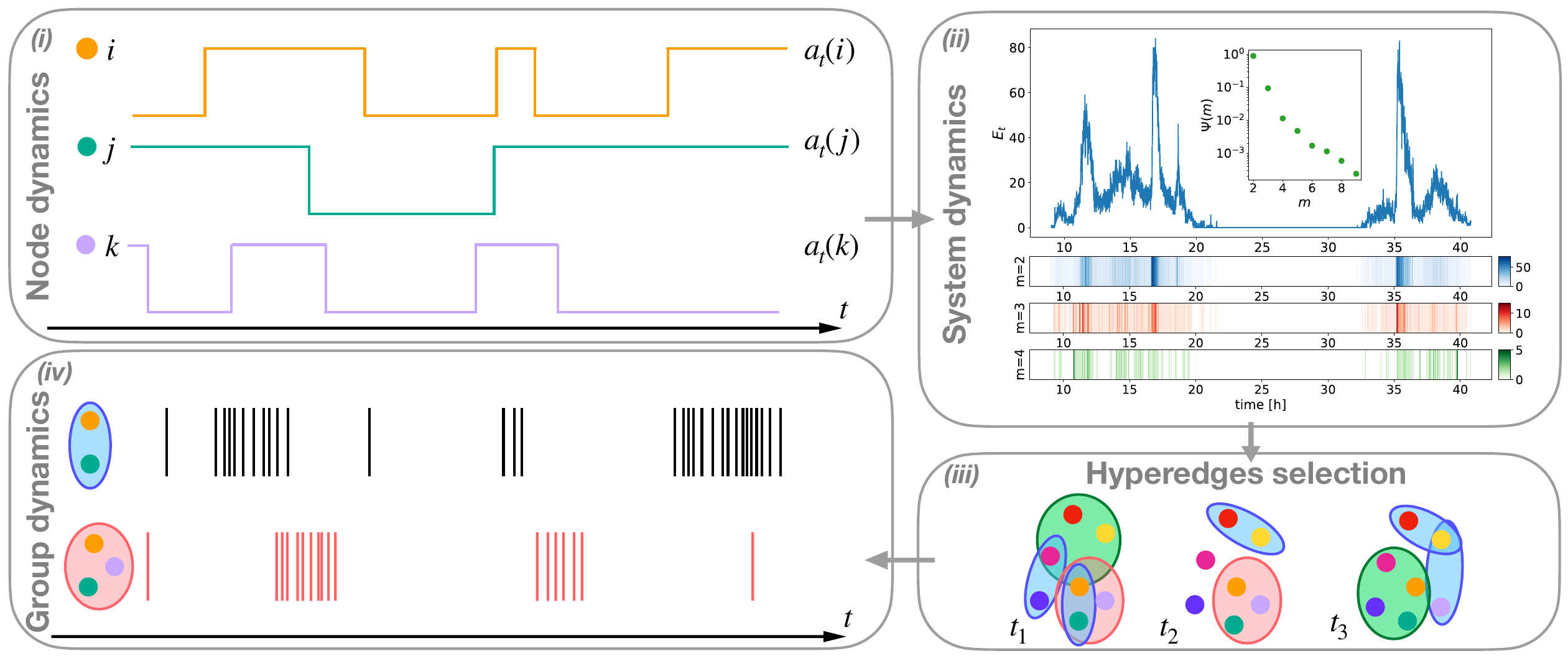}
\caption{\textbf{Sketch of the Emerging Activity Temporal Hypergraph (EATH) model.} The EATH model consists in: \textit{(i)} generation of the activity dynamics of individual nodes, which transition between low- and high-activity states; \textit{(ii)} the overall activity of the system emerges from the activity dynamics of individual nodes and from the size distribution $\Psi(m)$; \textit{(iii)} the nodes that are part of active interactions are selected with memory and activity mechanisms, hence producing the \textit{(iv)} groups dynamics. The panel \textit{System dynamics} shows the evolution of the number of active hyperedges $E_t$, also divided in sizes for $m \in [2,3,4]$, and the hyperedge size distribution $\Psi(m)$ (inset) for the SFHH dataset.
}
\label{fig:figure4}
\end{figure*}

\section{Emerging Activity Temporal Hypergraph (EATH) model}
\label{sec:model}
We propose a model to generate a synthetic time-varying hypergraph $\mathscr{H}=\{\mathcal{H}_t\}_{t=1}^{\mathcal{T}/\delta t}$ over the time interval $(0,\mathcal{T}]$, with time-resolution $\delta t$, with hyperedges of size $m \in [2,M]$ and with a population of $N$ nodes. 

Each node in the system is characterized by three parameters, which set the nodes behaviour in terms of both temporal and topological features: \textit{(i)} the \textit{persistence activity} $a_T(i)$, describes a node tendency to be active over time, i.e. the larger it is the more time $i$ spends interacting; 
\textit{(ii)} the \textit{instantaneous activity} $a_h(i)$, describes the node propensity to participate in different hyperedges simultaneously when active, i.e. the larger it is the more numerous are the groups $i$ participates in when active (i.e., the larger the hyperdegree of $i$
in the snapshot hypergraph); 
\textit{(iii)} the \textit{order propensity} $\varphi_i(m)$, accounts for the node relative
inclination to participate in hyperedges of size $m$, with $\sum_{m=2}^M \varphi_i(m)=1$. 
The two activities, $a_T(i)$ and $a_h(i)$, provide complementary information: the persistence activity describes the temporal behaviour of the node, affecting the time spent interacting, while the instantaneous activity describes the topological behaviour, affecting the node structural centrality
at the snapshot level. 
This allows to describe a wide variety of social behaviors empirically observed  \cite{Mancastroppa2024}, from nodes often active in many interactions to nodes rarely active in a few interactions, and also intermediate behaviors of nodes very active, but in a few groups simultaneously, or on the contrary of nodes rarely active but participating in many simultaneous
hyperedges when active. 

The hypergraph generation involves two independent steps (see Fig. \ref{fig:figure4}). The first step (Sec. \ref{sez:phase1}) consists in building for each node its activity dynamics, which establishes how they become more or less socially active over time based on their specific features, and in building the overall activity of the system, which results from the activity of single agents (see Fig. \ref{fig:figure4}). The fact that the system activity emerges from the nodes activity dynamics is a central point of our model, hence we dub the model as the \textit{Emerging Activity Temporal Hypergraph (EATH)} model. The second step (Sec. \ref{sez:phase2}) consists in selecting the nodes that participate in the active hyperedges: this selection is performed 
taking into account the nodes activity and their order propensity, as well as memory mechanisms 
\cite{vestergaard2014,zhao2011,stehle2010dynamical,laurent2015calls,Iacopini2024,kim2018,gelardi2021,lebail2023} (see Fig. \ref{fig:figure4}).

These steps make it possible to generate heterogeneous and correlated dynamics at the nodes and groups level, with variability in single node and group behaviours. 
In particular, each node activity dynamics is driven both by internal agent features and by an external modulation due to the system schedule: their combined effects create a bursty dynamics with temporal correlations, and result in heterogeneous agents behaviour \cite{Hiraoka2020,Reis2020}. This underlying node dynamics induces temporal correlations and heterogeneity in the groups dynamics, since hyperedges are activated with higher probability when all their nodes (or a majority) have higher activity \cite{Hiraoka2020,Reis2020}.

\subsection{Nodes and system activity dynamics}
\label{sez:phase1}
The core idea of the model consists in assuming that each node presents an independent dynamic switching between a low-activity phase, in which the node tends not to interact, and a high-activity phase, in which it has a higher propensity for interactions \cite{Hiraoka2020,Reis2020,Karsai2012,poissonian_bursty_2008}. This dynamics for each node $i$ is encoded in the activity $a_t(i)$ (see Fig. \ref{fig:figure4}):
\begin{equation}
    a_t(i)=\begin{cases}
        a_h(i) & \text{if } i \in h(t)  \\
        \gamma a_h(i) & \text{if } i \in l(t)
        \end{cases},
\end{equation}
where $\gamma \in [0,1]$, $h(t)$ and $l(t)$ are respectively the set of nodes in the high- and low-activity phase at time $t$. Hereafter we will fix $\gamma = 10^{-3}$ (see SM, {Sec. III}, for other values).

Each node features an independent dynamic in discrete time, with probability $r_{l \to h}(i,t)$ of transitioning from the low- to the high-activity phase in each time-step, and probability $r_{h \to l}(i,t)$ for the opposite transition. The transition probabilities can a priori depend on time and on the node, 
to take into account both the heterogeneity in the agents' propensity to be active and 
a potential external system-dependent modulation affecting the activity of all the nodes, 
which depends on the context and its specific schedule 
(e.g. the alternation of working days/weekends or working hours/nights). 
The transition probabilities are:
\begin{equation}
    r_{l \to h}(i,t)=\Lambda_t \varrho_l \frac{a_T(i)}{\langle a_T \rangle},
    \label{eq:rate1}
\end{equation}
where the average $\langle a_T \rangle$ refers to the average of $a_T(j)$ over all nodes $j$, and
\begin{equation}
    r_{h \to l}(i,t)=(1-\Lambda_t) \varrho_h ,
    \label{eq:rate2}
\end{equation}
where $\Lambda_t \in [0,1]$ is the external modulation, which impacts all nodes transition probabilities in the same way; the parameters $\varrho_l$ and $\varrho_h$ set the time-scales of the evolution. {Indeed, the average time nodes spend in the low-activity state is $\Delta_l = \left\langle \frac{1}{r_{l \to h}(i,t)} \right\rangle $, and the average time they spend in the high-activity state is $\Delta_h = \left\langle \frac{1}{r_{h \to l}(i,t)} \right\rangle$ (averages over both nodes and time). Once $\Delta_l$ and $\Delta_h$ are extracted from the empirical 
data (or tuned synthetically to desired values), the equations (\ref{eq:rate1}) and (\ref{eq:rate2}) 
allow us to obtain the values of the 
coefficients $\varrho_l$ and $\varrho_h$ needed to 
reproduce the desired time-scales (see Methods for more details).} Note that nodes with higher persistence activity $a_T(i)$ spend more time in the high-activity state as they have a larger ratio $r_{l \to h}(i,t)/r_{h \to l}(i,t)$.

The timelines of the nodes activity dynamics are created 
assuming all nodes to be in the low-activity phase at $t=-\infty$, and
letting the activity dynamics evolve independently for each node using the transition probabilities of
Eqs. \eqref{eq:rate1},\eqref{eq:rate2}, with $\Lambda_t=\Lambda_0 \forall t \in [-\infty,0]$. 
We run the dynamics until a stationary state is reached, obtaining a configuration of nodes
in high- or low-activity, that we use as initial condition for $t=0$. The dynamics for 
$t \geq 0$ is then generated for each node.

The overall system activity, i.e. the evolution of the number of group interactions, results 
from the combined  dynamics of the nodes' activity, so that times with more nodes in high-activity status correspond to snapshots with a higher number of hyperedges. 
We impose that the number of hyperedges active in each snapshot $t$ is:
\begin{equation}
    E_t={\xi} \frac{A_t}{\langle A_t \rangle},
\end{equation}
where $A_t$ is the cumulative contribution of all nodes' activities, 
$A_t= \sum_{i \in \mathcal{V}} a_t(i)$, 
and the parameter ${\xi}$ determines the average number of hyperedges generated 
(as $\langle E_t \rangle = {\xi}$). The sizes of the $E_t$ active hyperedges at time $t$
are distributed according to $\Psi(m)$ as described below.

\subsection{Creation of the active hyperedges at each snapshot}
\label{sez:phase2}

The second step of the temporal hypergraph generation consists in selecting the nodes involved in each active hyperedge in each snapshot $t$ (see Fig. \ref{fig:figure4}). For each hyperedge, the node selection is driven by: 
\textit{(i)} the single nodes features, i.e. their propensity to be active at time $t$ ($a_t(i)$) and to belong to specific group sizes ($\varphi_i(m)$); 
\textit{(ii)} a reinforcement mechanism, based on both short- and long-term memory mechanisms \cite{vestergaard2014,zhao2011,stehle2010dynamical,Iacopini2024,kim2018,gelardi2021,lebail2023}.

The short-term memory considers a \textit{"long-gets-longer"} mechanism, that favors the continuation of long-lasting interactions, possibly with an aggregation or disaggregation mechanism \cite{Iacopini2024,laurent2015calls,vestergaard2014,zhao2011}. The long-term memory favors interactions among nodes who have already met in the past and hence feature a stronger social connection encoded
in an underlying \textit{"social bond network"} \cite{kim2018,gelardi2021,lebail2023}. 
Note that, for simplicity, we encode this mechanism in a pairwise ``memory'', and assume that the generation of a group can be driven by the binary-memory among all the couples of nodes in the group.
In practice, the long-term memory is implemented through a memory matrix $\omega_0$, where the element $\omega_0(i,j)$ encodes the strength of the connection $(i,j)$, assumed to represent a previous long-term dynamics of interactions between $i$ and $j$: we assume that this long-term memory matrix is frozen and not updated during the generation of the temporal hypergraph, i.e. the time scales of this
memory are much larger than those of the temporal hypergraph considered (see Methods for
the generation of $\omega_0$). 
We then consider the normalized memory matrix $\Tilde{\omega}_0$, in which  $\Tilde{\omega}_0(i,j)=\omega_0(i,j)/\sum_{j \in \mathcal{V}} \omega_0(i,j)$ estimates what fraction of $i$'s memory is shared with $j$. Note that $\Tilde{\omega}_0$ is not symmetric, reflecting the fact that nodes have different activity and can distribute their interactions differently among the other nodes \cite{gelardi2021,lebail2023}. 

To generate the snapshot $\mathcal{H}_t$ at time $t$, we
repeat the following steps until $E_t$ active hyperedges have been created. 
First, we draw a size $m$ from $\Psi(m)$. We then generate a hyperedge $e$ of size $m$
according to one of two possible mechanisms, each taking into account activities, order
propensities and long-term memory:
\begin{itemize}
    \item with probability $p$, $e$ is generated from scratch randomly: this models the generation of new groups not related to interactions active at the previous time $t-1$. The nodes forming $e$ are selected with probability proportional to their activity $a_t(i)$, to their order propensity $\varphi_i(m)$ and to the memory $\Tilde{\omega}_0$ within the group (see Methods for details).
    
    \item with probability $(1-p)$, a short-term memory mechanism is instead used, namely, 
    the hyperedge is generated as the continuation of a hyperedge $e'$ that was active at $t-1$, and had size $m'$ equal to $m-1$, $m$ or $m+1$. $e$ will then be either equal to $e'$ (if $m'=m$), 
    or consist in a group $e'$ gaining (if $m'=m-1$) or losing (if $m'=m+1$) a member. This allows
    to model aggregation and disaggregation mechanisms. Among all the possible processes
    $e' \to e$ (corresponding each to a hyperedge of $\mathcal{E}_{t-1}$
    with size $\in \{m-1,m,m+1\}$), the actual one is chosen with probability proportional to the activities $a_t(i)$ of $e$'s nodes, to their order propensities $\varphi_i(m)$, to the long-term memory $\Tilde{\omega}_0$ within $e$, and to the duration of the hyperedge $e'$. See the Methods for more details. Note that here we consider that only one node can leave/join an existing group at each time step, however more complex splitting and merging mechanisms could potentially be considered \cite{Iacopini2024}.
\end{itemize}

\subsection{EATH model parameters}
The parameters of the EATH model 
can be extracted from data, to generate surrogate copies of empirical systems, or can be arbitrarily tuned. These parameters can be divided into two categories: the \textit{system parameters}, which describe the system and the context in which the interactions take place; the \textit{population parameters}, which describe the behaviours of individual agents. 

The system parameters include: the average number of interactions $\langle E_t \rangle$ and their size distribution $\Psi(m)$, defining the connectivity and the type of group activity; the modulation $\Lambda_t$ and the time scales $\mathcal{T}, \Delta_h, \Delta_l$, which describe temporal patterns due to specific schedules; the memory matrix heterogeneity $\langle \omega_0^2 \rangle/\langle \omega_0 \rangle^2$ and the probability $p$ that determine the balance between the generation of new hyperedges and the reinforcement of already activated groups.
The population properties are: the number of nodes $N$, which defines the size of the population; the distribution of activities $a_h(i)$, $a_T(i)$, and of the order propensity $\varphi_i(m)$, which describe the agents' heterogeneity. In the next Section we describe how to extract these parameters
from a dataset in order to generate a surrogate temporal hypergraph of that dataset.

\section{EATH surrogate hypergraphs}
\label{sec:surrogate}
\begin{figure*}[ht!]
\includegraphics[width=\textwidth]{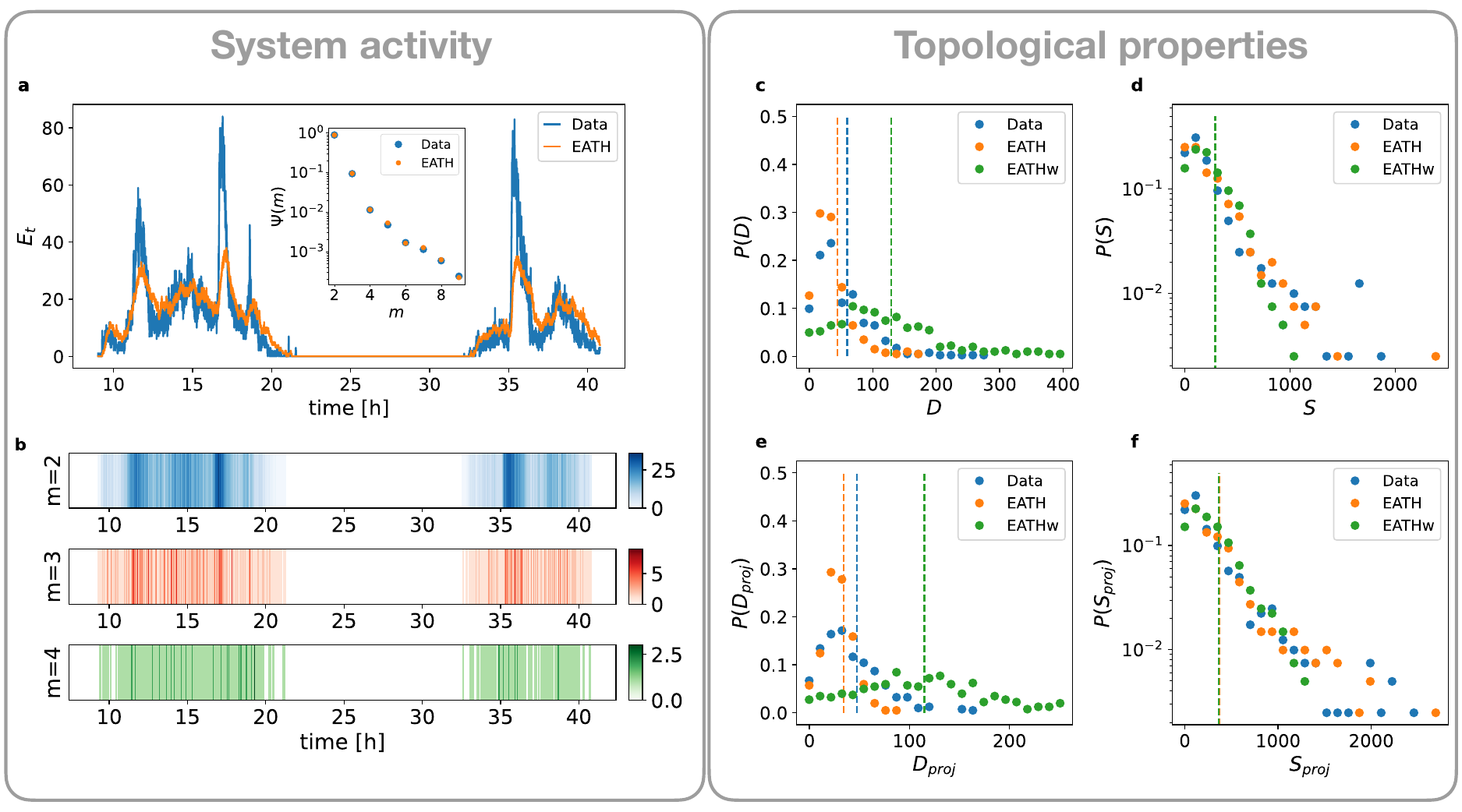}
\caption{\textbf{Generated system activity and topology.} \textbf{a}: evolution of the number of active hyperedges $E_t$ and distributions of the hyperedges' sizes $\Psi(m)$ (inset), for the empirical dataset and for the EATH model. \textbf{b}: evolution of the number of active hyperedges of size $m \in [2,3,4]$ for the EATH model. \textbf{c}, \textbf{d}: respectively, distribution $P(D)$ of the total hyperdegree $D$ and distribution $P(S)$ of the total hyperstrength $S$, for the aggregated hypergraphs of the empirical dataset and of the surrogate hypergraphs obtained in the model with (EATH) and without memory (EATHw). \textbf{e}, \textbf{f}: same as \textbf{c},\textbf{d} for the degree and the strength in the aggregated projected pairwise graph. In all panels we consider the SFHH dataset, with model parameters extracted from the empirical hypergraph as described in the main text. In panels \textbf{c}-\textbf{f} the dashed vertical lines indicate the average values of the corresponding distributions.}
\label{fig:figure5}
\end{figure*}

\begin{figure*}[ht!]
\includegraphics[width=0.75\textwidth]{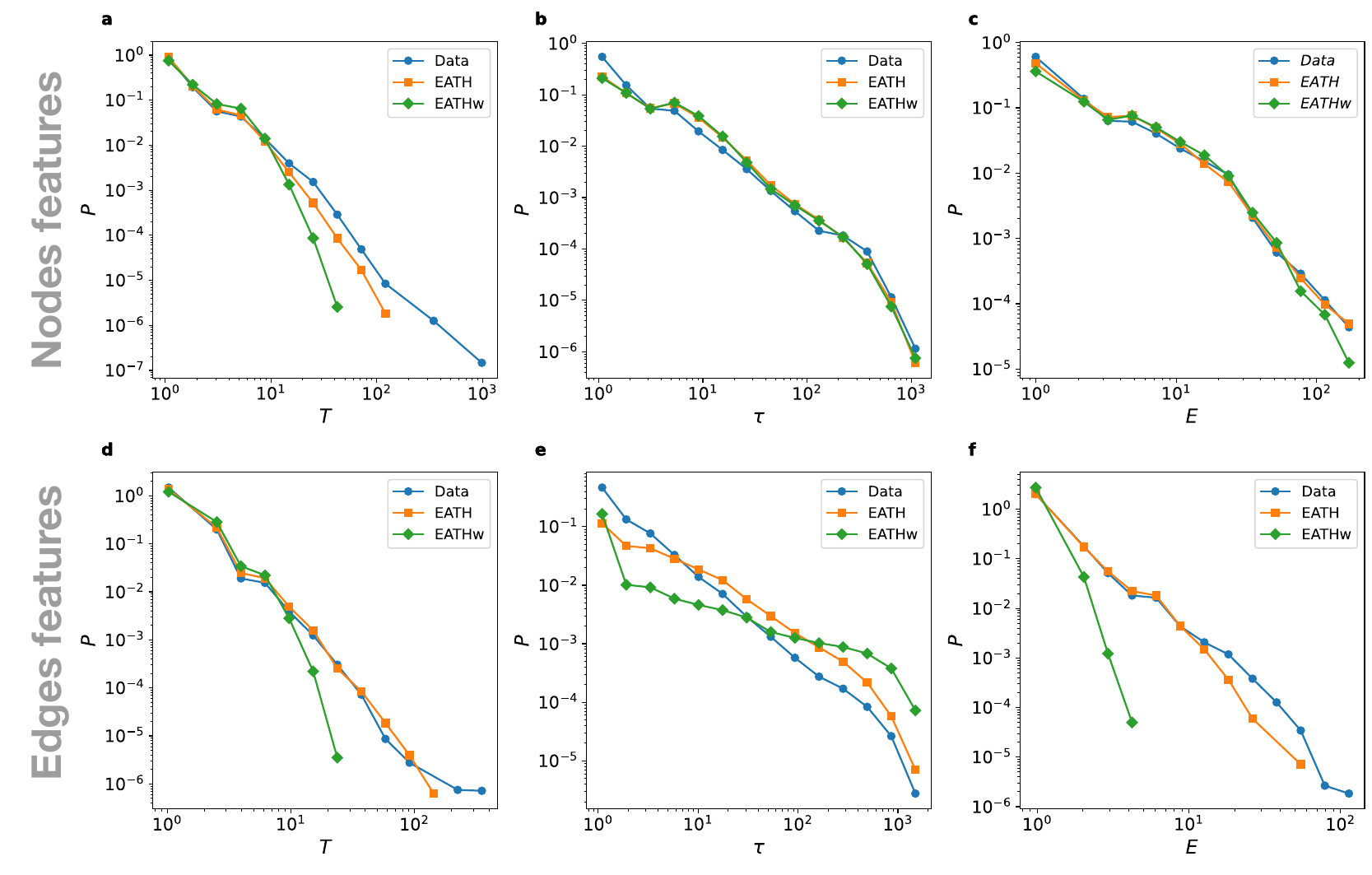}
\caption{\textbf{Generated temporal statistics.} The first and second rows show respectively the statistics of temporal properties for nodes and hyperedges. \textbf{a},\textbf{d}: distribution of event durations $P(T)$; \textbf{b},\textbf{e}: distribution of inter-event times $P(\tau)$; \textbf{c},\textbf{f}: distribution of the number of events in a train of events $P(E)$, with $\Delta = 60 \delta t $. In all panels we consider the SFHH dataset and we compare it with the hypergraphs generated by the model with (EATH) and without memory (EATHw), 
with the parameters extracted from the empirical hypergraph.} 
\label{fig:figure6}
\end{figure*}

\subsection{From data to model parameters}

The EATH model's main goal is to generate surrogate temporal hypergraphs emulating empirical systems.
To this aim, given an empirical dataset, we first need to extract the model parameters from the
data in order to be able to tune the generation process. 

First, several parameters are simple quantities that can be straightforwardly extracted from the datasets, such as the population size $N$, the duration $\mathcal{T}$ and
the average number of interactions $\langle E_t \rangle$. 

The population parameters are extracted for each node as follows:
\begin{itemize}
    \item the persistence activity $a_T(i)$ is estimated as the fraction of the total interaction time in which $i$ was active (counting the multiplicity if $i$ is active on several hyperedges simultaneously): 
    \begin{equation}
        a_T(i)=\sum_{e \in \mathcal{E}| i \in e} w_e/\sum_{e \in \mathcal{E}} w_e;
    \label{eq:a_T}
    \end{equation}
    \item the instantaneous activity $a_h(i)$ is measured as the average number of simultaneous interactions  node $i$ participates in: 
    \begin{equation}
        a_h(i)=\sum_{e \in \mathcal{E}| i \in e} w_e/{T}_i,
    \label{eq:a_h}
    \end{equation}
    where ${T}_i$ is the total time in which $i$ was active; 
    \item the order propensity $\varphi_i(m)$ is determined as the fraction of time the node spent interacting in groups of size $m$, with respect to all the orders: 
    \begin{equation}
        \varphi_i(m)=\sum_{e \in \mathcal{F}(i) | |e|=m} w_e/\sum_{e \in \mathcal{F}(i)} w_e ,
    \label{eq:phi}
    \end{equation}
    where $\mathcal{F}(i)$ is the set of hyperedges in which $i$ is involved.
\end{itemize}

The system parameters are extracted globally:
\begin{itemize}
    \item the hyperedge size distribution is by definition $\Psi(m)=\sum_{e \in \mathcal{E}| |e|=m} w_e/\sum_{e \in \mathcal{E}} w_e$; 
    \item the modulation is set as $\Lambda_t=N_t/N$, i.e. the fraction of active nodes at time $t$; 
    \item the probability of generating a new hyperedge $p$ is determined by counting at each time-step $t$ the number of hyperedges $O_t$ that already existed at time $t-1$ or that have lost/gained one node with respect to time $t-1$: 
    \begin{equation}
        p=\sum_{t=1}^{\mathcal{T}/\delta t}(E_t-O_t)/\sum_{t=1}^{\mathcal{T}/\delta t} E_t.
    \end{equation}
\end{itemize}

To obtain the long-term memory matrix $\omega_0$, we first run the model starting from an initial condition of equal memory between all nodes ($\omega_{0,ic} (i,j) = 1 - \delta_{i,j}$), increasing 
dynamically $\omega_{0,ic} (i,j)$ by one each time $i$ and $j$ interact. When the 
heterogeneity $\langle \omega_{0,ic}^2 \rangle/\langle \omega_{0,ic} \rangle^2$ reaches
a desired level (see Methods for details), we stop the simulation and use the resulting 
$\omega_{0,ic}$ as the frozen memory $\omega_0$ to generate the model. 

Finally, the average times that nodes spend in the high- or low- activity phases, $\Delta_h$ and $\Delta_l$, are extracted from the data assuming that a node is in the high-activity phase when involved in a sequence of events, while it is in the low-activity phase otherwise (see Methods for details).

In the following subsections, we show the results of the generation of surrogate
temporal hypergraphs when using the SFHH dataset as original data \cite{sociopatterns,genois2018,ISELLA2011166,genois2018}: 
this dataset describes face-to-face interactions occurring during a conference with 403 participants, collected using wearable proximity sensors by the SocioPatterns collaboration 
over a time span of two days, with a temporal resolution of 20 seconds (see Methods for details).
In the SM, we show analogous results for all the other datasets. Moreover,
in addition to hypergraphs generated with the EATH model described above, we also
consider a baseline version in which no memory effect is taken into account in the
generation process (EATHw version): this will allow us to test the importance of the memory 
mechanisms in reproducing empirical properties.

\subsection{Higher-order temporal and topological properties}

Figure \ref{fig:figure5}\textbf{a},\textbf{b} first shows that 
the model reproduces well the overall system dynamics, both in terms of its modulation and the activation dynamics of the different sizes over time, with larger groups appearing more frequently in high activity phases. In particular, Fig. \ref{fig:figure5}\textbf{b} should be compared with Fig. \ref{fig:figure4} where the same dynamics is shown for the original dataset.

Moreover, the EATH model captures the empirical structure both for the higher-order and pairwise topology at the aggregated level. 
The generated surrogate hypergraph replicates indeed well the empirical distributions of the hyperdegree and hyperstrength in the aggregated hypergraph $\mathcal{H}$, $P(D)$ and $P(S)$, 
both qualitatively and quantitatively (Fig. \ref{fig:figure5}\textbf{c},\textbf{d}). This also holds when considering the corresponding distributions in the projected graph $\mathcal{G}$ (Fig. \ref{fig:figure5}\textbf{e},\textbf{f}) and the hyperdegree distribution for specific group sizes (see SM, {Sec. IB}). The EATH model also reproduces the evolution of the distributions of the interactions weights and of their heterogeneity (see SM, {Sec. IB}), obtained considering the hypergraph aggregated from time $0$ to time $t$. 
We note that the topology generated without the memory effects (EATHw) is very different: the 
average hyperdegree and projected degree are much higher, the distributions span a much broader interval, and the (hyper)degree reach in particular very large values; some nodes participate
in a very large number of groups, having a large neighborhood, and only few groups and links 
are recurring (see Fig. \ref{fig:figure5} and SM, {Sec. IB}). The empirical higher-order structure is thus
reproduced by the combination of the memory mechanisms and the heterogeneity in node activity,
as also noted in \cite{Mancastroppa2024}. 
Their combined effect also allows to reproduce modularity values of the pairwise projection of the aggregated hypergraph close to the ones observed in the data, while the EATHw model
leads always to very small modularity values (see SM, {Sec. IB}).

\begin{figure*}[ht!]
\includegraphics[width=\textwidth]{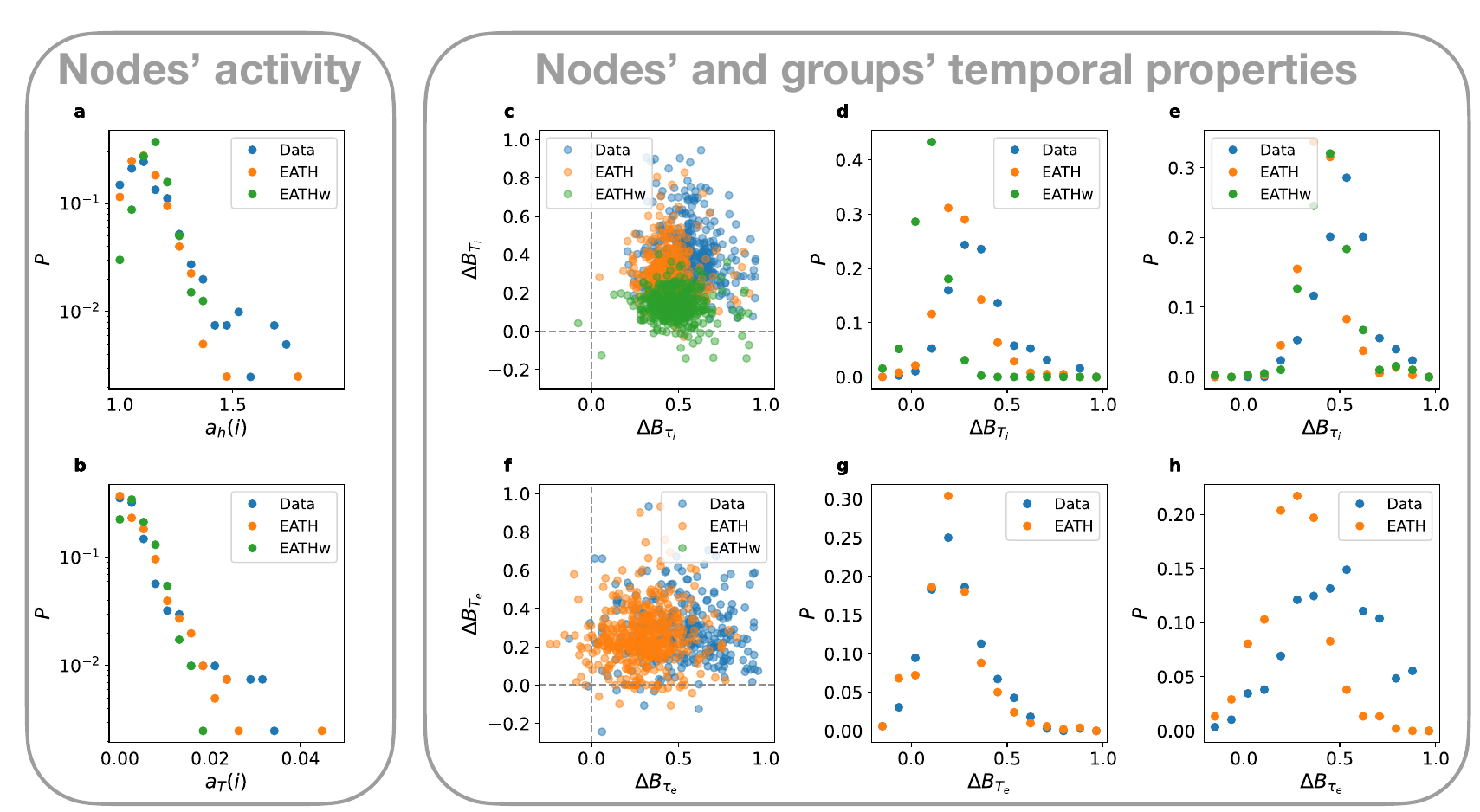}
\caption{\textbf{Generated nodes and hyperedges heterogeneity.} Panels \textbf{a} and \textbf{b} show respectively the distributions of the instantaneous activity, $P(a_h(i))$, and of the persistence activity, $P(a_T(i))$, extracted from the empirical dataset (which are provided as input to the generation) and measured in the generated hypergraphs. 
In panel \textbf{c} (resp. \textbf{f}) each point corresponds to a node (resp. hyperedge) and we show the scatterplot 
between the burstiness of their inter-event time distributions, $\Delta B_{\tau}$, and of their event duration distributions, $\Delta B_{T}$. 
The gray dashed lines are the reference $\Delta B=0$. Panels \textbf{d}, \textbf{e} (resp. \textbf{g}, \textbf{h}) show the distributions $P(\Delta B_{T})$ and $P(\Delta B_{\tau})$ for nodes (resp. hyperedges): we consider only nodes and hyperedges with at least $10$ events. 
Note that in panels \textbf{g} and \textbf{h}
there are no results for the EATHw case, since no group activated enough times to define its burstiness parameter. 
In all panels we consider the SFHH dataset and we compare it with hypergraphs generated using the model with (EATH) and without memory (EATHw), with the parameters extracted from the empirical hypergraph. }
\label{fig:figure7}
\end{figure*}

Figure \ref{fig:figure6} focuses on the distributions of the quantities characterizing
the dynamics of nodes and groups, at the global level: the
dynamics generated for groups and nodes is heterogeneous, bursty and correlated, as empirically observed (see also the SM, {Sec. IC}, for the burstiness $\Delta B$ of the generated distributions). 
We note that,
at the node level, the distributions of the number of events in trains of events and the inter-event times are well replicated even with the memory-less model EATHw. Indeed, the interaction dynamics of nodes is largely governed by their activity dynamics, which is built in the first step
of the model (without reference to memory mechanisms). However, 
the EATHw generation process produces temporal dynamics that do not reproduce the empirical
distributions of event durations for nodes nor any of the distributions characterizing the
dynamics of groups (see Fig. \ref{fig:figure6}). 
In this case indeed, the active groups are selected at each time-step only 
according to the activity of the nodes, thus it is not possible to obtain long durations for group
interactions nor temporal correlations. Memory mechanisms are thus responsible
for the generation of heterogeneous group durations, and allow also to reactivate the same groups
at short intervals, creating trains of events \cite{vestergaard2014,laurent2015calls}.

\begin{figure*}[ht!]
\includegraphics[width=\textwidth]{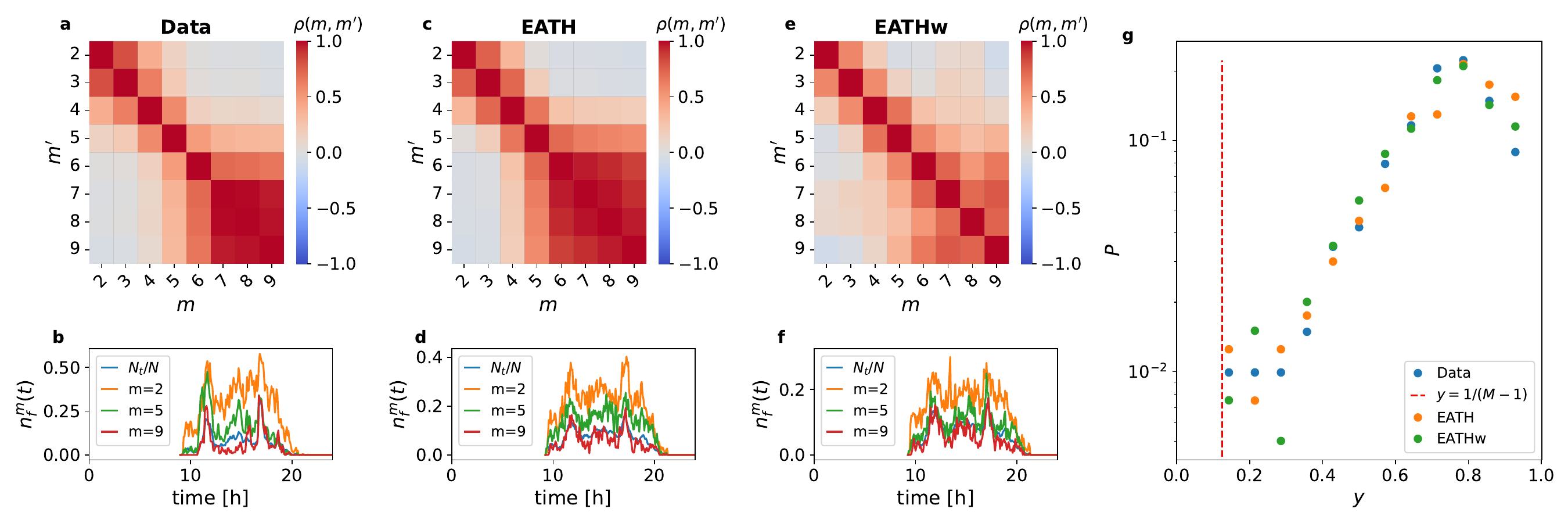}
\caption{\textbf{Generated nodes participation to different group sizes.} 
We consider the empirical SFHH hypergraph (Data) and the hypergraphs generated by the model with (EATH) and without memory (EATHw), with parameters extracted from the empirical hypergraph. 
In each case, we rank the nodes at each order $m$ according to the time spent interacting at that order.
We then show the Pearson's correlation coefficient $\rho(m,m')$ between the rankings at order $m$ and $m'$ (\textbf{a},\textbf{c},\textbf{e}). 
We show the evolution of the fraction $n_f^m(t)$ of active nodes among the ones occupying the top $f N$ position of the ranking at order $m$, for different $m$ (see legend), and $f=0.1$, focusing on the same time window (\textbf{b},\textbf{d},\textbf{f}). We also plot the total fraction $N_t/N$ of active nodes in the population. Panel \textbf{g} shows the distributions of the node participation ratios $P(y)$ (the red dashed line represents $y=1/(M-1)$).} 
\label{fig:figure8}
\end{figure*}

Let us now focus on the individual node and hyperedge level. 
First, we measure the actual persistence and instantaneous activities, $\Tilde{a}_T(i)$ and $\Tilde{a}_h(i)$, in the generated hypergraphs (as defined in Eqs. \eqref{eq:a_T},\eqref{eq:a_h}), and we compare them with the empirical ones, $a_T(i)$ and $a_h(i)$, provided as input to the generation: both models reproduce the distributions and values, effectively replicating their heterogeneous behaviour (see Fig. \ref{fig:figure7}\textbf{a},\textbf{b}), hence validating the chosen model mechanisms. 

As shown in Fig. \ref{fig:figure7}\textbf{c}-\textbf{h}, nodes and groups present heterogeneous temporal behaviours in the 
surrogate temporal hypergraph generated by the EATH model, as was the case in the empirical data:
the burstiness of the distributions obtained measuring the event and inter-event durations
of individual nodes and hyperedges span a large interval, showing that the hypergraph elements can have either regular or irregular dynamics. As in the empirical data, no correlations are observed 
between the burstiness values for event and inter-event durations, and the model reproduces quite well the empirical distributions of burstiness $\Delta B_{T}$ and $\Delta B_{\tau}$ for both nodes and groups. For the EATHw model, the nodes have much narrower distributions of event durations, and 
each hyperedge can be reactivated only a few times, hence lacking enough statistics to characterize
its temporal distributions (see Fig. \ref{fig:figure7}\textbf{c}-\textbf{h}).

The EATH model also generates a realistic pattern of nodes interactions at different orders, reconstructing the empirical heterogeneity with which nodes participate 
in groups of different sizes (see Fig. \ref{fig:figure8}). First, we compute the actual node propensity of each node $i$,
$\varphi_i(m)$, in each generated hypergraph (as defined in Eq. \eqref{eq:phi}). We then compute for each node the participation
ratio $y(i)=\sum_{m=2}^M \varphi_i(m)^2$, which quantifies how the activity of a node is distributed across the different interaction orders (as $\sum_{m=2}^M \varphi_i(m)=1$;
we recall that $y(i) \in [1/(M-1),1]$: if $y(i) \sim 1$ the node interacts at only one specific size; if $y(i) \sim 1/(M-1)$ the node acts equally across all orders).
Figure \ref{fig:figure8}\textbf{g} shows that, in both models as well as in the empirical dataset,
most nodes activate preferentially on specific orders (large values of $y(i)$); however, the system is actually characterized by a wide variability, with some nodes interacting uniformly across all orders.
Both EATH and EATHw models yield surrogate hypergraphs that
replicate the distribution of the participation ratio and 
therefore the variability in the behaviour of the nodes of the empirical data.
Figure \ref{fig:figure8} also shows that the 
correlations between the ranking of the importance of nodes for group
interactions of different sizes are well reproduced by the EATH model. The model also 
captures the link between how nodes allocate their activity at different orders and the temporal pattern of activation of nodes. The model without memory (EATHw) partially reproduces these patterns (Fig. \ref{fig:figure8}\textbf{e}), still replicating the block structure of the
matrix of correlation values, but with lower levels of correlation. 

Finally, we highlight a limitation of the model. 
Not surprisingly given the creation rules described above, the surrogate temporal hypergraphs generated do not reproduce the dynamics of group aggregation and disaggregation (see SM, {Sec. IE}): indeed the model's group selection mechanisms only consider processes in which a group can lose or gain 
a single node. However, the empirical dynamics is more complex, with merging and splitting
of groups of different sizes, as described in \cite{Iacopini2024}. To reproduce such dynamics, additional mechanisms should be introduced in the model, with rules
making it possible for a group at time $t-1$ to lose or gain several nodes at once.


\subsection{Higher-order SIR dynamics}
\label{sec:process}

\begin{figure*}[ht!]
\includegraphics[width=\textwidth]{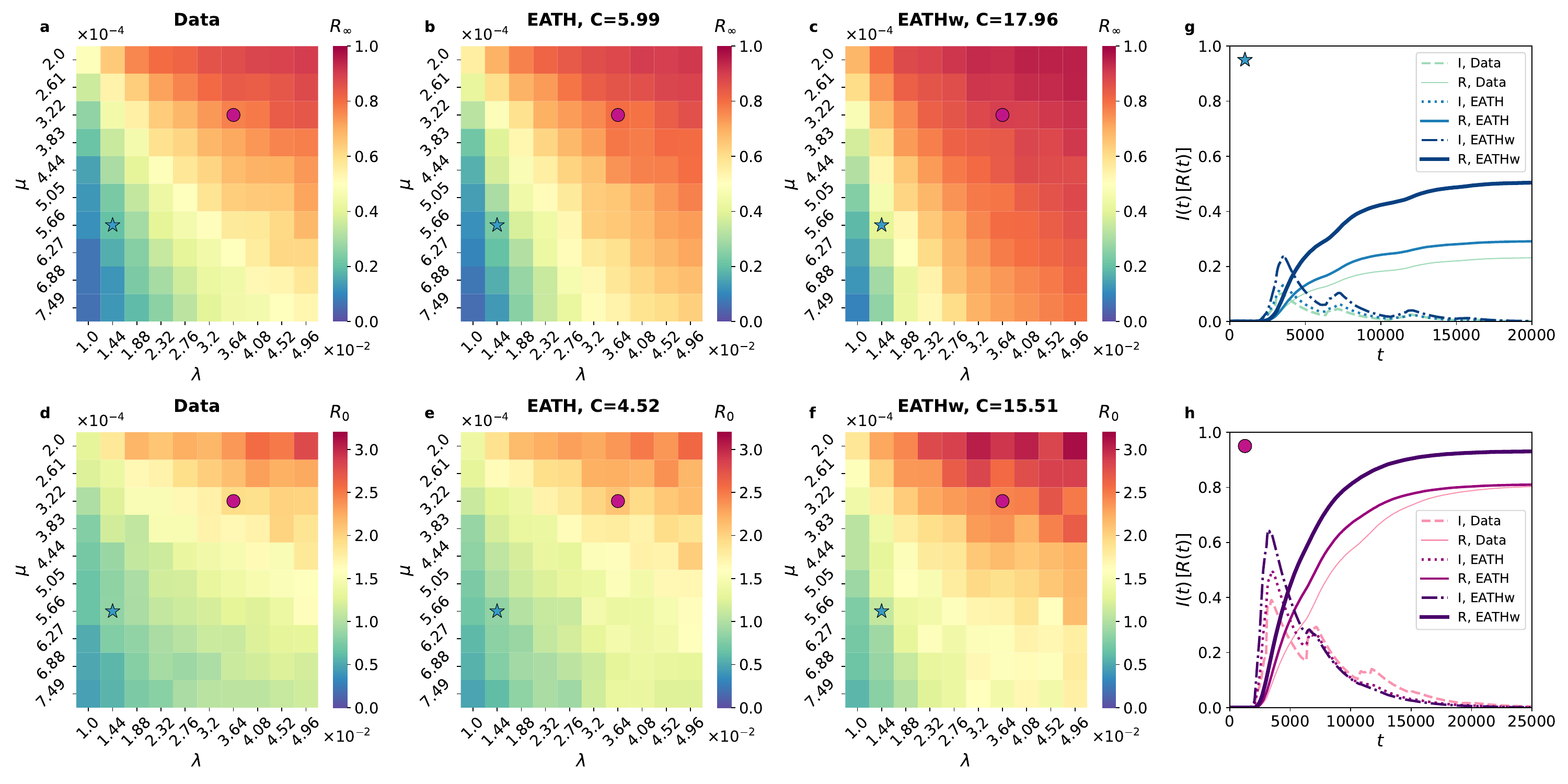}
\caption{\textbf{Higher-order SIR process.} Panels \textbf{a}-\textbf{c} and \textbf{d}-\textbf{f} show respectively the epidemic final size $R_{\infty}$ and the basic reproduction number $R_0$ as a function of the epidemiological parameters $(\lambda,\mu)$. The results are obtained by averaging over 400 numerical simulations, fixing $\nu=4$, and considering the spreading over the SFHH empirical hypergraph (Data) and over the corresponding temporal hypergraphs generated using the model with (EATH) and without memory (EATHw). $C$ indicates the Canberra distance \cite{canberradist} between the matrices obtained by simulating the process on the data and on the corresponding synthetic hypergraphs. In panels \textbf{g},\textbf{h} we show the fraction of infected $I(t)$ and recovered $R(t)$ nodes as a function of time, when a random initial seed is infected at $t_0=0$, and averaging the curves over 200 realizations: panel \textbf{g}, $(\lambda,\mu)=(1.44 \, 10^{-2},5.66 \, 10^{-4})$ (see the blue star in panels \textbf{a}-\textbf{f}); panel \textbf{h}, $(\lambda,\mu)=(3.64 \, 10^{-2},3.22 \, 10^{-4})$ (see the purple circle in panels \textbf{a}-\textbf{f}).} 
\label{fig:figure9}
\end{figure*}

The goal of the EATH model is to produce surrogate temporal hypergraphs that can be used to simulate
dynamical processes on hypergraphs. We therefore consider here one such process, namely a prototypical
spreading process including higher-order mechanisms, and compare its dynamics and outcome when
simulated on either the 
empirical or the surrogate hypergraphs.
Specifically, 
we consider the higher-order SIR (susceptible-infected-removed) epidemic process based on a non-linear infection rate \cite{St-Onge2022}, which assumes a higher probability of contagion in case of simultaneous exposure to multiple sources of infection \cite{St-Onge2022,iacopini2019}. Each node of the hypergraph can be susceptible, $S$, infected, $I$, or recovered, $R$. The dynamics evolves in discrete time, with a time-step $\delta t$ equal to the resolution of the hypergraph $\mathscr{H}=\{(\mathcal{V}_t,\mathcal{E}_t)\}_{t=1}^{\mathcal{T}/\delta t}$. At each time-step $t$, a susceptible node $j \in \mathcal{V}_t$ that is part of a hyperedge $e \in \mathcal{E}_t$ with $i_{e,t}$ infected nodes, gets infected with probability $(1-e^{-\lambda i_{e,t}^\nu})$ within the group $e$, where $\lambda \in [0,1]$ and $\nu \geq 1$. Hence, overall a susceptible node $j$ at time $t$ is infected with probability: 
\begin{equation}
    p_j= 1- \prod\limits_{e \in \mathcal{F}_t(j)} e^{-\lambda i_{e,t}^\nu},
\end{equation}
where $\mathcal{F}_t(j)$ is the set of hyperedges in which $j$ is involved at time $t$ and $i_{e,t}$ is the number of infected nodes in the hyperedge $e$ at time $t$. Infected nodes recover spontaneously with probability $\mu$ in each time-step. 

We numerically simulate this process for $\nu=4$ and various values of $\lambda$ and $\mu$,
and compute (see Methods): \textit{(i)} the epidemic final size $R_{\infty}$, i.e. the fraction of the population reached by the spread; \textit{(ii)} the basic reproduction number $R_0$, i.e. the average number of secondary infections generated by a single infected node in a fully susceptible population; \textit{(iii)} the temporal dynamics of the spread, i.e. the evolution of the fraction of infected $I(t)$ and recovered $R(t)$ agents. 

Figure \ref{fig:figure9} shows the results of the simulations performed on both empirical and surrogate hypergraphs for the SFHH dataset (see SM, {Sec. IF}, for other datasets). The SIR dynamics on the EATH model replicates well the empirical one for all the parameters configurations, both in terms of the total impact of the epidemic $R_{\infty}$ and of the initial spreading $R_0$ (see Fig. \ref{fig:figure9}\textbf{a}-\textbf{f}). Furthermore, the epidemic evolution on the EATH model is similar to the empirical one, replicating the timing and incidence levels of the epidemic peaks, despite
a slight overestimation of the peak heights
(see Fig. \ref{fig:figure9}\textbf{g},\textbf{h}). Simulations on the hypergraphs generated without memory mechanisms (EATHw) yield a much faster dynamics in the early stages, with a much higher first
peak and overall a substantial overestimation of the impact of the spread.

\section{EATH artificial and hybrid hypergraphs}
\label{sec:hybrid}

\begin{figure*}
\includegraphics[width=0.9\textwidth]{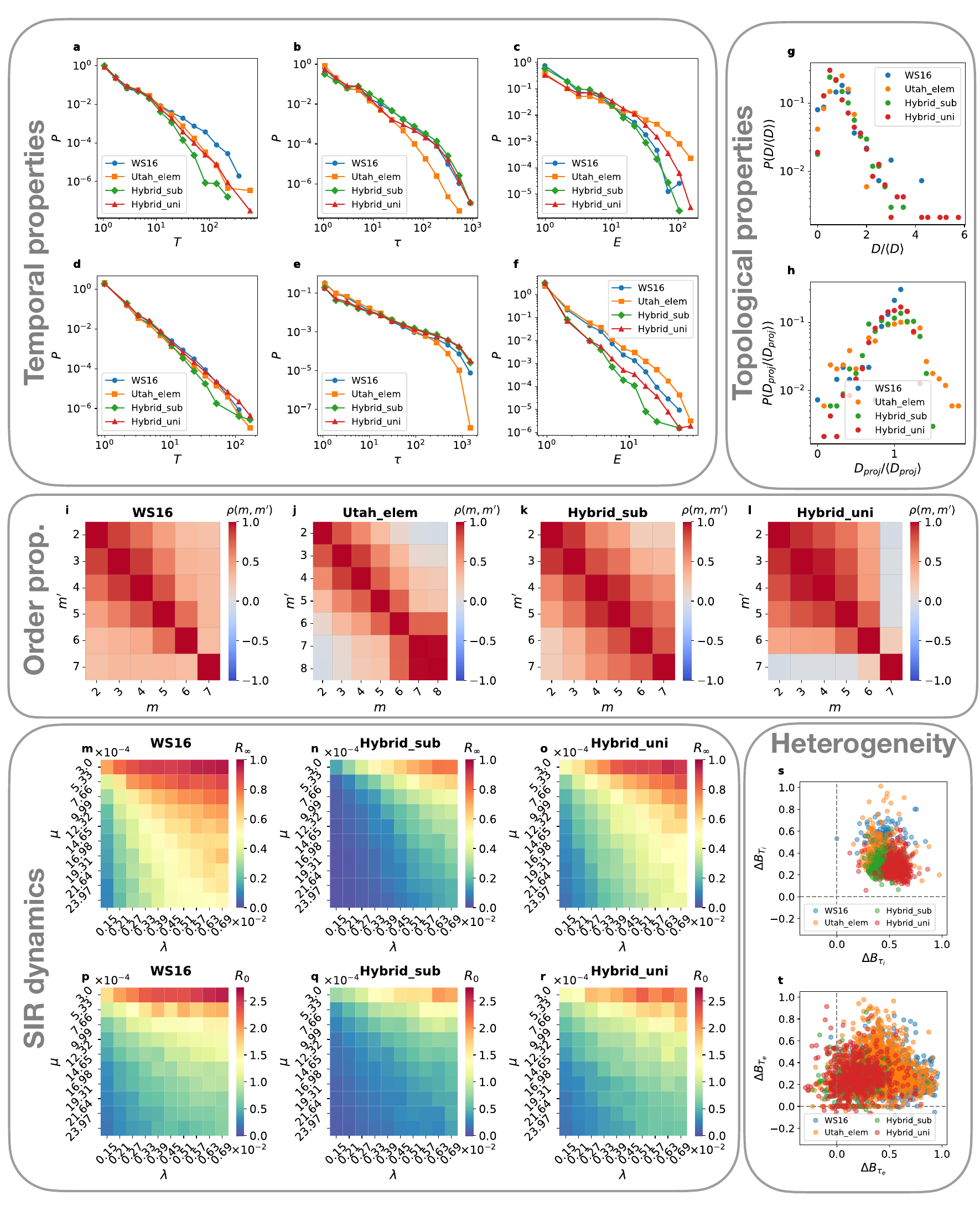}
\caption{\textbf{Hybrid hypergraphs - mixing WS16 (system) and Utah\_elem (population).} We consider two hybrid hypergraphs (\texttt{Hybrid\_sub} and \texttt{Hybrid\_uni}) generated with the EATH model considering the properties of the datasets WS16 (conference context; $d_1$ - system properties) and  Utah\_elem (school context; $d_2$ - population properties). \textbf{a}-\textbf{f}: distributions of inter-event times, durations and numbers of events in a train of events, for nodes (\textbf{a}-\textbf{c}) and hyperedges (\textbf{d}-\textbf{f}); \textbf{g},\textbf{h}: 
normalized hyperdegree distribution in the aggregated hypergraph, $P(D/\langle D \rangle)$, and 
normalized degree distribution in the projected aggregated graph, $P(D_{proj}/\langle D_{proj} \rangle)$; 
\textbf{i}-\textbf{l}: Pearson's correlation coefficient $\rho(m,m')$ between the node rankings at size $m$ and $m'$, obtained considering the time spent by nodes interacting at each order; 
\textbf{m}-\textbf{r}: epidemic final-size $R_{\infty}$ and basic reproduction number $R_0$ as a function of the SIR epidemiological parameters $(\lambda, \mu)$, obtained in the same conditions of Fig. \ref{fig:figure9}; 
\textbf{s}-\textbf{t}: scatterplots of the burstiness of inter-event time distributions $\Delta B_{\tau}$ and that of duration distributions $\Delta B_{T}$ for single nodes and hyperedges, considering only the nodes and hyperedges with at least 10 events.} 
\label{fig:figure10}
\end{figure*}

The interest and use of the EATH model actually go beyond the generation of surrogate temporal hypergraphs. The model parameters and mechanisms governing the different generative phases can be  tuned in order to generate hypergraphs with various properties. 
This opens the possibility to build temporal hypergraphs with arbitrary features, possibly
even different from the empirical ones, and to test their impact, e.g. on dynamical processes. 
At the context/system level, we can for instance build a hypergraph: 
\textit{(i)} with a different temporal activity dynamics, imposing an arbitrary modulation $\Lambda_t$; 
\textit{(ii)} with groups of different sizes, by changing the distribution $\Psi(m)$; 
\textit{(iii)} with different temporal and interaction scales, e.g. a longer time interval $\mathcal{T}$ or a higher connectivity $\langle E_t \rangle$. 
At the population level, we can moreover: 
\textit{(i)} consider a larger population $N$; 
\textit{(ii)} consider arbitrary distributions of the activities, $a_h(i)$ and $a_T(i)$; 
\textit{(iii)} set different order propensities $\varphi_i(m)$. 
In the SM, {Sec. III}, we generate such cases and explore how these modifications affect the temporal, topological and dynamical features of the system.

Thanks to the EATH model flexibility, it is also possible to generate \textit{hybrid hypergraphs},
which mix properties of different empirical systems. This might in particular be useful to study hypothetical systems whose properties combine the features of different empirical datasets, to understand if such systems still possess realistic statistical features, 
and to investigate the temporal, topological and dynamical features of systems for 
which there are no available data or data collection is not possible.
To this aim, we consider the generation of a hypergraph by mixing the properties of two datasets, $d_1$ and $d_2$, respectively with $N_1$ and $N_2$ nodes, $\langle E_{t,1} \rangle$ and $\langle E_{t,2} \rangle$ connectivity level, $M_1$ and $M_2$ maximum group size. In particular, we consider two possible hybrid generation procedures:

\textit{(i)} \textit{Hybrid substitution} (\texttt{Hybrid\_sub}). The hypergraph is generated with the system parameters of $d_1$ and the population parameters of $d_2$. This case models a population of a specific environment (e.g. students of a school), which preserves its features but interacts in a different context (e.g. a conference), with different temporal patterns, activity levels and sizes of interactions. The connection between the system and population properties requires modifying the order propensity $\varphi_i(m)$ according to the maximum interactions sizes: if $M_2 \geq M_1$,  $\varphi_i(m)$ is kept as in $d_2$; if $M_2<M_1$ we first fix $\varphi_i(m)=\varphi_i(M_2) \forall m \in [M_2+1,M_1]$, so that nodes preserve their propensity for interactions at large sizes and we then normalize the resulting $\varphi_i(m)$ on the interval $[2,M_1]$. We keep the level of interactions $\langle E_t \rangle = \langle E_{t,1}\rangle$, even if $N_2 \neq N_1$, to focus on the pure effect of a population that interacts in a different environment, possibly characterized by a different interaction level;

\textit{(ii)} \textit{Hybrid union} (\texttt{Hybrid\_uni}). The hypergraph is generated with the system parameters of $d_1$ and a population that is the union of the populations of $d_1$ and $d_2$. This case models two different populations that interact in the same environment, preserving their features. The connection between the system and population properties requires setting $\langle E_t\rangle$: this can be done in several ways, depending on the goal in generating such a hybrid model. Among the various possibilities, we consider here $\langle E_t\rangle= (N_1+N_2) \langle E_{t,1} \rangle /N_1$, tuning the level of interactions $\langle E_t \rangle$ to the actual size of the population in the hybrid hypergraph: this allows us to simply investigate the effect of modifying the population of a system by adding new individuals that behave differently.

In Fig. \ref{fig:figure10} we show the properties of hybrid hypergraphs generated by mixing a conference (WS16, $d_1$) and a school dataset (Utah\_elem, $d_2$). The generated hypergraphs differ from the empirical ones, but they still present realistic statistical properties for all the measures considered: a heterogeneous and bursty dynamics with temporal correlations both at the node and group level (Fig. \ref{fig:figure10} \textbf{a}-\textbf{f}), a realistic shape of the hyperdegree and degree distributions in the aggregated hypergraph and in its pairwise projection (Fig. \ref{fig:figure10} \textbf{g},\textbf{h}), heterogeneity in the behaviours of the nodes both in the propensity to interact at different orders (Fig. \ref{fig:figure10} \textbf{i}-\textbf{l}) and temporally (Fig. \ref{fig:figure10} \textbf{s},\textbf{t}). 
The outcome of a SIR dynamics presents as well a realistic pattern even if some differences are observed with respect to the original dataset $d_1$ (Fig. \ref{fig:figure10} \textbf{m}-\textbf{r}): in \texttt{Hybrid\_sub} the epidemic is less effective, since the average number of interactions per node is lower; in \texttt{Hybrid\_uni} the dynamics is more similar, but still differs due to the presence of $d_2$ nodes with different behaviours. In the SM, {Sec. II}, we also consider other hybrid hypergraphs.

\section{Discussion}
In this article, we have proposed the Emerging Activity Temporal Hypergraph (EATH), a 
generative model for surrogate temporal hypergraphs. 
In the model, each node presents an independent underlying dynamics, switching between a low-activity  and a high-activity phases \cite{poissonian_bursty_2008,Reis2020,Hiraoka2020,Karsai2012}, driven by internal agent features and by external system factors. The system global activity emerges 
from the nodes activity dynamics, and the groups activation is based on activity and memory mechanisms \cite{vestergaard2014,zhao2011,stehle2010dynamical,Iacopini2024,kim2018,gelardi2021,lebail2023}, 
in order to generate a heterogeneous and correlated dynamics at the nodes and groups level.

The EATH model can generate surrogate temporal hypergraphs emulating empirical systems and their different levels of heterogeneity. It captures and reproduces a wide set of empirical temporal and topological properties. Moreover, the EATH model is able to replicate the higher-order properties of the system that shape higher-order infection dynamics, generating thus
surrogate temporal hypergraphs useful to investigate contagion processes. {We note that 
we have here leveraged the fact that 
the model parameters for surrogates generation can be easily extracted from the empirical data to be emulated.
It would in principle be possible to retrieve them using methods based on maximum likelihood. 
This represents a promising direction for future research.}

The EATH modelling capacities go beyond the generation of surrogate hypergraphs: the model mechanisms and parameters are flexible, allowing to build hypergraphs with arbitrary desired temporal-topological features, describing a wide variety of behaviours.  We illustrate this ability in the SM by generating various artificial hypegraphs, for example by synthetically setting the distribution of hyperedge sizes, the external modulation of activities and the distribution of node activity, hence probing the structural and dynamical role played by these specific features. 
Moreover, by exploiting this flexibility, we generated hybrid hypergraphs that mix properties from different datasets, while still producing realistic temporal hypergraphs. This allows exploring hypothetical systems, for which no empirical data exist or can be collected.

Our work opens several future perspectives.
First, the EATH model provides tools to gain insights in the understanding of higher-order dynamical processes on temporal hypergraphs. Indeed, it can be used to generate surrogate temporal hypergraphs
of arbitrary temporal length, and with different population sizes with respect to the original
data considered, as well as hypergraphs describing hypothetical but realistic systems, or instead
temporal hypergraphs with features specifically tuned to explore the role of a given property
on the processes of interest. 
Moreover, while our analysis mainly focused on face-to-face social interaction systems, systems of different nature could be considered, for example scientific collaborations, online interactions and online communications \cite{Mancastroppa2024,mancastroppa2023hyper}. These systems might present different empirical properties, however the modularity and independence of the EATH generation steps provide enough flexibility to generate surrogates also of these systems, when the model parameters are appropriately extracted from the corresponding empirical datasets.
Finally, 
the model can also be extended by adding further levels of complexity, acting on some of its limitations: the model does not explicitly account for a community structure, which could be introduced into the memory mechanisms \cite{Contisciani2022}; the model does not either reproduce the dynamics of aggregation and disaggregation of groups, to fully reproduce it one could introduce additional mechanisms \cite{Iacopini2024}, e.g. by 
considering the gain or loss of several nodes for a hyperedge in a single time step; 
finally, the model could realistically introduce correlations in the size of the groups activated at consecutive times. All these examples represent promising directions for introducing further realism into the model.

\section{Methods}
\subsection{Datasets and pre-processing}
We consider several publicly available datasets of face-to-face interactions collected by different collaborations. The datasets consist in pairwise interactions collected with a resolution of 
20 seconds through RFID proximity sensors. We consider six datasets collected by the SocioPatterns collaboration \cite{sociopatterns,genois2018,ISELLA2011166} describing interactions: in a hospital (LH10 \cite{Vanhems2013}, with $N=75$ and $\mathcal{T}=96$ hours), in an elementary school (LyonSchool \cite{Stehle2011}, with $N=242$ and $\mathcal{T}=2$ days), in a high-school (Thiers13 \cite{Mastrandrea2015}, with $N=327$ and $\mathcal{T}=5$ days), in three conferences (SFHH \cite{genois2018}, with $N=403$ and $\mathcal{T}=2$ days; ICCSS17 \cite{Genois2023}, with $N=274$ and $\mathcal{T}=4$ days; WS16 \cite{Genois2023}, with $N=138$ and $\mathcal{T}=2$ days); and two datasets collected by the Contacts among Utah’s School-age Population (CUSP) project \cite{toth2015}, describing interactions in an elementary school (Utah\_elem \cite{toth2015}, with $N=339$ and $\mathcal{T}=2$ days) and in a middle school (Utah\_mid \cite{toth2015}, with $N=591$ and $\mathcal{T}=2$ days). These datasets describe social interactions in different environments, mediated by different mechanisms and activities.

We perform a pre-processing procedure to build a temporal hypergraph in the snapshot representation with resolution $\delta t$ \cite{Mancastroppa2024}: \textit{(i)} we aggregate the data over time windows of $\delta t$; \textit{(ii)} we identify the cliques within each window, i.e. the fully connected clusters; \textit{(iii)} in each time window $t$ we identify the maximal cliques, i.e. not completely contained in other larger cliques, and we promote them to hyperedges. Here we consider $\delta t$ as the original data resolution, i.e. $\delta t=20$ seconds.

\subsection{Burstiness parameter}
The burstiness parameter $B$ estimates temporal heterogeneity of a system, for example in the distribution of inter-event times or event durations \cite{karsai2018bursty,burstinessKim2016}. We consider the burstiness $B$ of an empirical distribution $P(t)$, with mean value $\langle t \rangle$ and standard deviation $\sigma$, obtained observing $n$ events:
\begin{equation}
    B= \frac{\sqrt{n+1}r-\sqrt{n-1}}{(\sqrt{n+1}-2)r+\sqrt{n-1}},
\end{equation}
where $r=\sigma/\langle t \rangle$. Note that this definition takes into account a correction for the number of observed events $n$, so that $B=-1$ for the perfectly regular case and $B=0$ in case of an exponential distribution \cite{karsai2018bursty,burstinessKim2016}. Here we consider $\Delta B = B-B_{exp}(t_m,\langle t \rangle)$, comparing the empirical burstiness $B$ with that of an exponential distribution $P(t)= \theta(t-t_m) \frac{1}{\langle t \rangle - t_m}e^{- \frac{t-t_m}{\langle t \rangle - t_m}}$ with the same mean value $\langle t \rangle$ and lower cut-off $t_m$ (where $\theta(x)$ is the Heaviside function). In particular:
\begin{equation}
    B_{exp}(t_m,\langle t \rangle)=-\frac{1}{2 \langle t \rangle/t_m-1}.
\end{equation}

\subsection{EATH groups selection mechanisms}
The second step of the EATH generation process consists in selecting the groups to be activated in each snapshot. If a size $m$ is extracted from $\Psi(m)$ at time $t$, we create a hyperedge of size $m$, that we call $e$:
\begin{itemize}
    \item with probability $p$, the hyperedge is generated from scratch randomly. In this case, the hyperedge $e$ is built progressively:
    \begin{enumerate}
        \item The first node $i$ is chosen with probability $q_i = a_t(i) \varphi_i(m)/\sum_{l \in \mathcal{V}} a_t(l) \varphi_l(m)$. The chosen $i$ is added to $e$;
        \item The second node $j$ is selected with probability $z_j = \alpha a_t(j) \varphi_j(m) \Tilde{\omega}_0(i,j) \Tilde{\omega}_0(j,i)$, thus taking into account the asymmetric long-term memory between $i$ and $j$, with $\alpha = 1/ \sum_{l \in \mathcal{V}} a_t(l) \varphi_l(m) \Tilde{\omega}_0(i,l) \Tilde{\omega}_0(l,i)$. Then the chosen $j$ is added to $e$;
        \item  at the $n$-th step, the node $k$ is chosen with probability $x_k =\beta a_t(k) \varphi_k(m) \prod\limits_{s \in e} \Tilde{\omega}_0(s,k) \Tilde{\omega}_0(k,s) $, thus accounting for the long-term memory between $k$ and all the nodes in the incomplete hyperedge $e=\{i,j,...\}$ of size $n-1$, with $\beta=1/\sum_{l \in \mathcal{V}} a_t(l) \varphi_l(m) \prod\limits_{s \in e} \Tilde{\omega}_0(s,l) \Tilde{\omega}_0(l,s)$. Then
        $k$ is added to $e$. 
    \end{enumerate}  
    The process is iterated until the desired size $|e|=m$ is reached. Note that if at any step there are no nodes available to be selected with these rules, we select them without the order propensity (i.e. only based on their activity and long-term memory).
    
    \item with probability $(1-p)$, the hyperedge is generated starting from a hyperedge already active at time $t-1$. A hyperedge $e'$ that exists at time $t-1$ can be the source of the new hyperedge $e$ at time $t$ only if its size $|e'| \in [m,m \pm 1]$, i.e. if $e' \in \mathcal{E}_{t-1}^{m,m \pm 1}$. In particular, the hyperedge generation $e' \to e$ can occur with three different continuation mechanisms, depending on the size of $e'$:
    \begin{enumerate}
        \item if $|e'|=m$, the hyperedge can be continued, i.e. reactivated, and in that case $e=e'$;
        \item if $|e'|=m+1$, the hyperedge can lose a node: the node $i$ to be removed is selected with probability $q_i = \alpha (\max\limits_{j \in e'}(a_t(j))-a_t(i))(1-\varphi_i(m))(1-\prod\limits_{j \neq i \in e'} \Tilde{\omega}_0(i,j) \Tilde{\omega}_0(j,i))$, where $\alpha$ is fixed so that $\sum_{j \in e'} q_j=1$. In this case $e=e' \setminus \{i\}$. 
        \item if $|e'|=m-1$, the hyperedge can be joined by a node: the node $i$ to be added is selected with probability $z_i = \beta a_t(i) \varphi_i(m) \prod\limits_{j \in e'} \Tilde{\omega}_0(i,j) \Tilde{\omega}_0(j,i)$, where $\beta$ is fixed so that $\sum_{j \in \mathcal{V}} z_j=1$. In this case $e=e' \cup \{i\}$. 
    \end{enumerate}
    According to these rules, each hyperedge $e' \in \mathcal{E}_{t-1}^{m,m \pm 1}$ is assigned with a possible continuation process $e' \to e$, eventually with the loss/gain of a node. Among all the possible hyperedge prolongation processes $e' \to e$, the effective one is selected with probability $h_{e' \to e} = \xi \frac{\sqrt{\Delta_{e'}}}{\langle \sqrt{\Delta_{e'}} \rangle_{e'}} \prod\limits_{i \in e} a_t(i) \varphi_i(m) \left[\prod\limits_{j \neq i \in e} \Tilde{\omega}_0(i,j) \Tilde{\omega}_0(j,i) \right]$, where $\Delta_{e'}$ is the duration of the hyperedge $e'$ and $\xi$ is fixed so that $\sum_{e' \in \mathcal{E}_{t-1}^{m,m \pm 1}} h_{e' \to e}=1$. Note that if no group can be generated from an already existing hyperedge (e.g. because of size constraints or already existing groups), the hyperedge is generated randomly as described in the previous point.
\end{itemize}

\subsection{Extracting EATH parameters from data}

As discussed in the main text, the model parameters can be extracted from the datasets to generate surrogate hypergraphs. Here we detail how to extract the memory and time-scales parameters. 

The long-term memory matrix $\omega_0$ is generated in a transient time through the model dynamics, so that it is compatible with the properties of the nodes, e.g. their activities. 
We fix an initial weight matrix $w$, where $w(i,j)=(1-\delta_{i,j})$ (with $\delta$ the Kronecker delta) so that initially all the pairs of nodes have equal memory; 
then we generate the temporal hypergraph with its dynamics by using the weight matrix $w$ as the memory matrix and by updating it at each time-step:  
the element $w(i,j)$ is simply increased by one every time $i,j$ meet in a hyperedge of any size. 
The evolution continues until the heterogeneity of the weight matrix reaches $\langle w^2 \rangle/\langle w \rangle^2 = g \langle w_l^2 \rangle/\langle w_l \rangle^2$, where $w_l$ is the weight matrix of the weighted projected aggregated network $\mathcal{G}$ of the empirical dataset. 
Then the frozen memory matrix of the model is set to $\omega_0=w$. The parameter $g$ is tuned for each dataset to reproduce the evolution of the weight distribution and its heterogeneity, by setting it inversely proportional to $\eta E$, where $\eta=|\mathcal{L}|/\binom{N}{2}$ is the density in the aggregated projected graph $\mathcal{G}=(\mathcal{V},\mathcal{L})$, and $E=\sum_{t=1}^{\mathcal{T}/\delta t} E_t$ is the total number of interactions activated. 
We have indeed observed that this approach reproduces well the empirical evolution of the weights (see SM, {Sec. IA,B}, for more details).

Finally, for each node we define an active sequence of events as a set of consecutive events whose time-distance is lower than $\overline{\Delta}$ and whose sum of the durations is larger than $\overline{\Delta}$: 
the node is then considered as in the high-activity phase when it is involved in such an active sequence of events, otherwise it is in the low-activity phase. Then we estimate the average time that nodes spend in the high-, $\Delta_h$, and low-activity phases, $\Delta_l$. $\overline{\Delta}$ depends on the dataset, and is set inversely proportional to $\eta E$ (see SM, {Sec. IA}, for more details). {Once obtained $\Delta_h$ and $\Delta_l$, the coefficients $\varrho_h$ and $\varrho_l$ which set the nodes transition rates (see Eqs. \eqref{eq:rate1},\eqref{eq:rate2}) can be derived. In particular $\varrho_h$ and $\varrho_l$ can be obtained, by inverting Eqs. \eqref{eq:rate1},\eqref{eq:rate2} and considering the definitions of $\Delta_l = \langle \frac{1}{r_{l \to h}(i,t)} \rangle $ and of $\Delta_h = \langle \frac{1}{r_{h \to l}(i,t)} \rangle$, as}:
$$
\varrho_l = \frac{1}{\Delta_l} \left\langle \frac{1}{\Lambda_t} \right\rangle 
\left\langle \frac{1}{a_T} \right\rangle
\langle a_T \rangle \ , 
\varrho_h = \frac{1}{\Delta_h} \left\langle \frac{1}{1 - \Lambda_t} \right\rangle ,
$$
where the average is either over nodes (for $a_T$ and $1/a_T$) or over time
(for $1/\Lambda_t$ and $1/(1-\Lambda_t)$).

\subsection{Higher-order SIR process simulations}
We simulate the higher-order SIR process on a temporal hypergraph $\mathscr{H}=\{\mathcal{H}_t\}_{t=1}^{\mathcal{T}/\delta t}$. Initially, we consider a fully susceptible population with only one infected node chosen uniformly at random in the entire population. The epidemic starts at time $t_0$ and is simulated until there are no more infected: if the epidemic 
is not ended when the network reaches its last snapshot, we repeat the network from the initial snapshot, taking into account periodicity due to nights or weekends.

For each of the parameters' configurations and temporal hypergraphs, we run simulations to estimate the epidemic final size $R_{\infty}$ and the basic reproduction number $R_0$. The initial simulation time $t_0$ is chosen uniformly at random in the entire time-span $[0,\mathcal{T}]$. $R_{\infty}$ is obtained by considering only the simulations that produce at least one infection, while $R_0$ considering also the simulations in which no transmission occurs \cite{cencetti2024generating}. The simulations to observe the temporal dynamics of the epidemic, $I(t)$ and $R(t)$, are instead run using as initial time $t_0=0$. 

In the higher-order SIR process, the contagion of a susceptible node results from 
the combined effect of different simultaneous groups and of different infected nodes within each group.
The evaluation of $R_0$ requires identifying the sole contribution of the infection seed $I_0$ to the contagion of nodes infected during its infectious period \cite{cencetti23,cencetti24_complex,cencetti2024generating}. Among the nodes $j$ that get infected during the infectious period of $I_0$, i.e. $j \in \mathcal{S}(I_0)$, $I_0$ contributes to the infection of those who participated in at least one group with it at the time of their infection $t_j$. The independent contribution to their infection given by each hyperedge $e$ is $(1-e^{-\lambda i_{e,t_j}^\nu})$, and this can be divided equally among its $i_{e,t_j}$ infected nodes. Therefore:
\begin{equation}
    R_0= \sum\limits_{j \in \mathcal{S}(I_0)} r_0^j,
\end{equation}
and the contribution of each single node $j \in \mathcal{S}(I_0)$ is: 
\begin{equation}
    r_0^j= 
    \begin{cases}
        \frac{\sum\limits_{e \in \mathcal{I}_{t_j}(j,I_0)} (1-e^{-\lambda i_{e,t_j}^\nu})/i_{e,t_j}}{\sum\limits_{e \in \mathcal{F}_{t_j}(j)} (1-e^{-\lambda i_{e,t_j}^\nu})}  & \text{if } |\mathcal{I}_{t_j}(j,I_0)|>0  \\
        0  & \text{if } |\mathcal{I}_{t_j}(j,I_0)|=0  \\
    \end{cases},
\end{equation}
where $\mathcal{F}_{t_j}(j)$ is the set of hyperedges in which $j$ is involved at the time $t_j$ of its infection, $\mathcal{I}_{t_j}(j,I_0)=\mathcal{F}_{t_j}(j) \cap \mathcal{F}_{t_j}(I_0)$ is the set of hyperedges active at $t_j$ that contain both $j$ and $I_0$.

\section*{Data availability}
The data that support the findings of this study are publicly available. The SocioPatterns data sets are available at \url{http://www.sociopatterns.org/}, and at \url{https://search.gesis.org/research_data/SDN-10.7802-2351?doi=10.7802/2351};  the Contacts among Utah’s School-age Population data set at \url{https://royalsocietypublishing.org/doi/suppl/10.1098/rsif.2015.0279}.

\begin{acknowledgments}
M.M. and A.B. acknowledge support from the Agence Nationale de la Recherche (ANR) project DATAREDUX (ANR-19-CE46-0008). G.C. acknowledges the support of the European Union’s Horizon research and innovation program under the Marie Sklodowska-Curie grant agreement No 101103026. The funders had no role in study design, data collection and analysis, decision to publish, or preparation of the manuscript.
\end{acknowledgments}

\section*{Competing interests}
The authors declare no competing interests.


\end{document}


\title{Supplementary Material for "Emerging Activity Temporal Hypergraph: a model for generating realistic time-varying hypergraphs"
}

\author{Marco Mancastroppa}
\affiliation{Aix-Marseille Univ, Universit\'e de Toulon, CNRS, CPT, Turing Center for Living Systems, 13009 Marseille, France}
\author{Giulia Cencetti}
\affiliation{Aix-Marseille Univ, Universit\'e de Toulon, CNRS, CPT, Turing Center for Living Systems, 13009 Marseille, France}
\author{Alain Barrat}
\affiliation{Aix-Marseille Univ, Universit\'e de Toulon, CNRS, CPT, Turing Center for Living Systems, 13009 Marseille, France}

\maketitle



In this Supplementary Material we present additional results regarding the EATH model generation. In Section \ref{sez:section1} we provide results on the generation of surrogate temporal hypergraphs using the EATH model for several datasets, with also additional results on the dynamical and temporal-topological characterization of the datasets and their corresponding surrogates.
In Section \ref{sez:section2}, we present the generation of hybrid hypergraphs, obtained by combining properties of different datasets. In Section \ref{sez:section3} we present results of the EATH generation when some model parameters are fixed synthetically to generate specific features, investigating their impact on the topological, temporal and dynamical properties of the hypergraph.

\section{EATH surrogates generation}
\label{sez:section1}
Here we present results of the generation of surrogate temporal hypergraphs using the EATH model, comparing the empirical hypergraphs with the corresponding synthetic ones. In Section \ref{sez:section1a} we show the overall activity of the system, also divided into different orders, and how some empirical parameters are extracted from data; in Sections \ref{sez:section1b}, \ref{sez:section1c} we show respectively the topological and temporal properties of the empirical datasets and of the generated hypergraphs; in Section \ref{sez:section1d} we show the patterns of nodes participation in interactions of different sizes; in Section \ref{sez:section1e} we show the dynamics of aggregation and disaggregation of groups \cite{Iacopini2024}; finally, in Section \ref{sez:section1f} we show results of numerical simulations of the higher-order SIR dynamics \cite{St-Onge2022} on the empirical and surrogate hypergraphs.

\newpage
\subsection{System activity and parameters}
\label{sez:section1a}
\begin{figure*}[ht!]
\includegraphics[width=\textwidth]{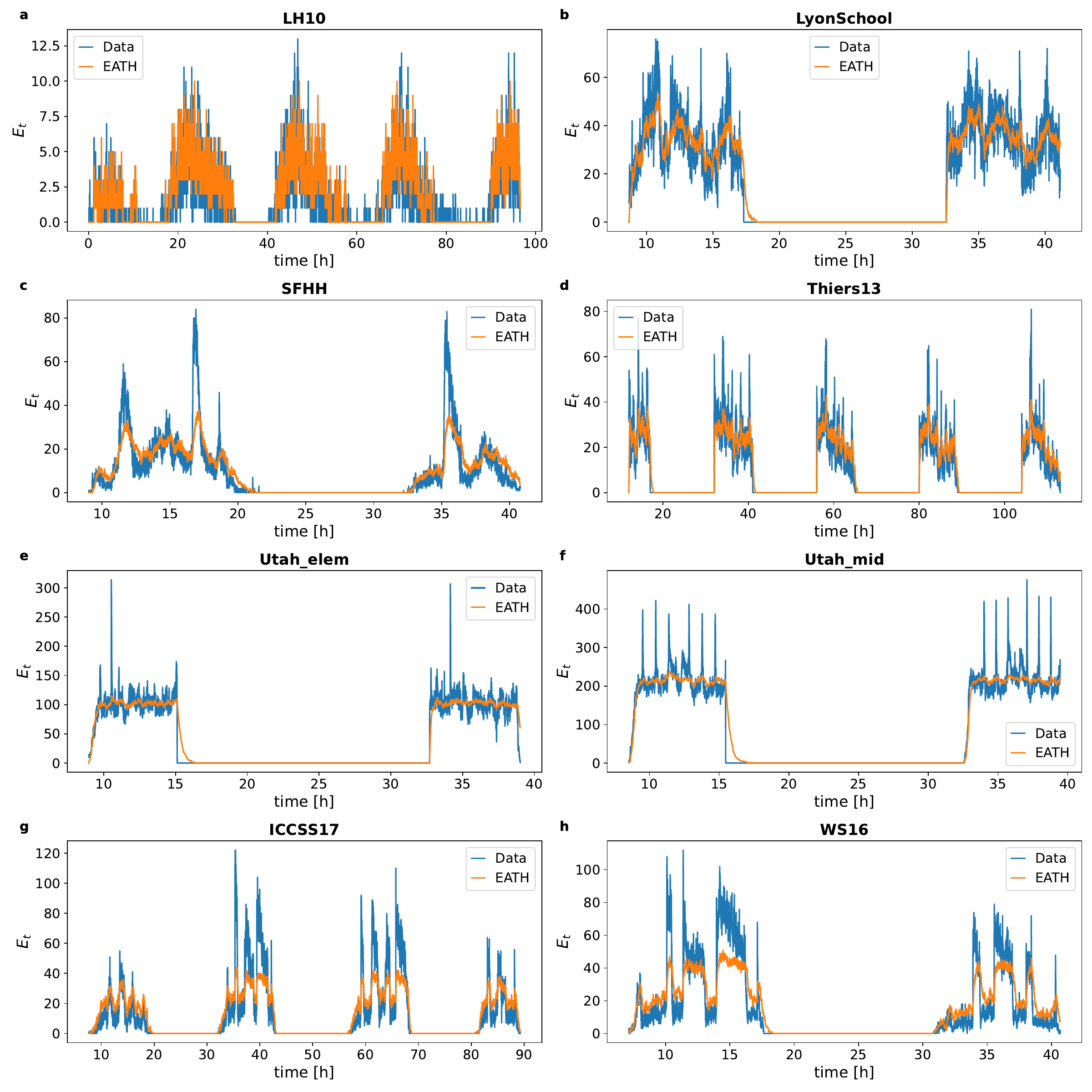}
\caption{\textbf{Overall system activity - I.} In each panel we show the number of active hyperedges $E_t$ (of any size) as a function of time for the empirical hypergraphs and for the EATH model. In each panel we consider a different dataset (see title).}
\label{fig:figure1}
\end{figure*}

\begin{figure*}[ht!]
\includegraphics[width=\textwidth]{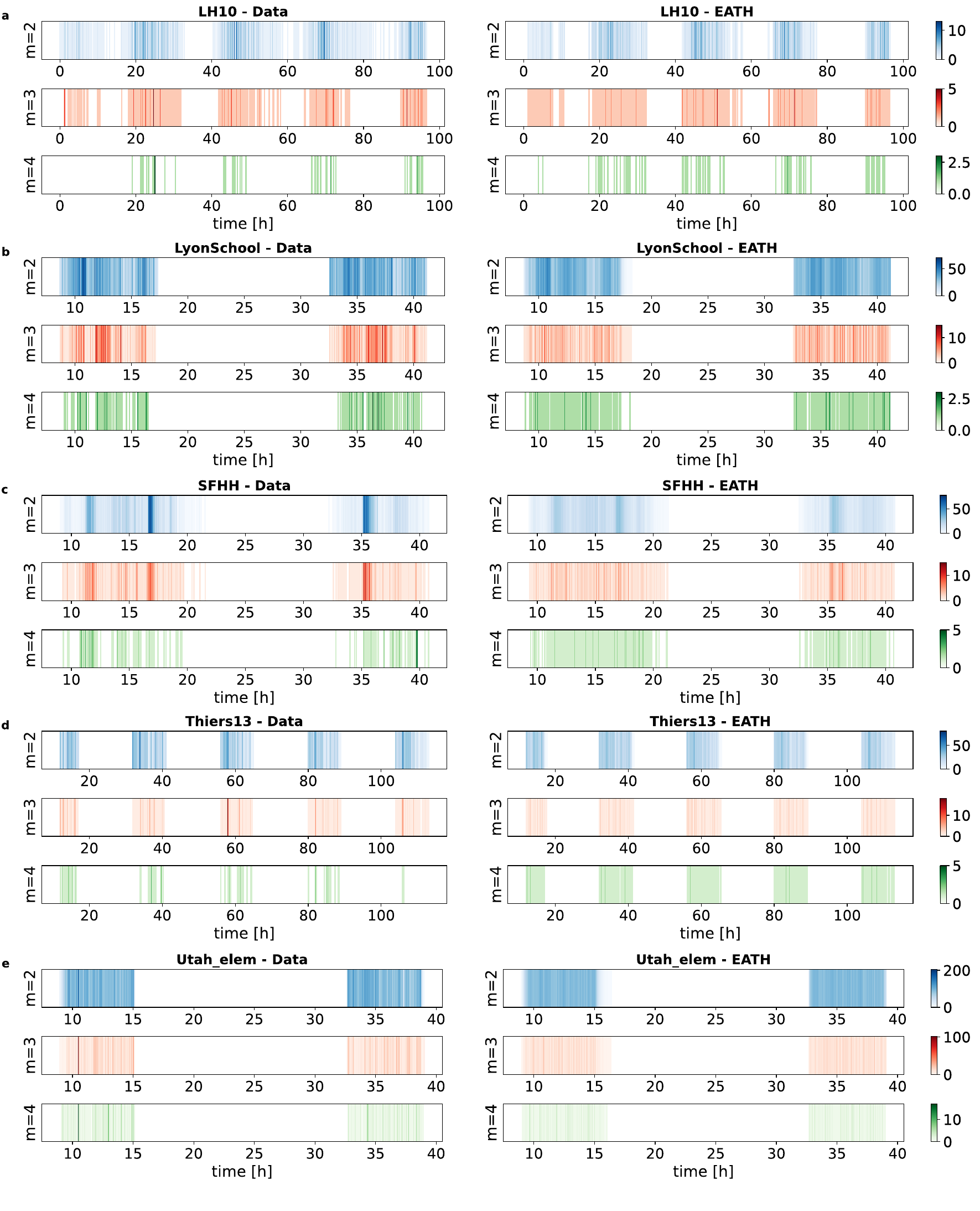}
\caption{\textbf{Overall system activity - II.} In each row we show the number of active hyperedges of a specific size $m \in [2,3,4]$ as a function of time for the empirical hypergraph (first column) and for the EATH model (second column). In each row we consider a different dataset (see title).}
\label{fig:figure2}
\end{figure*}

\begin{figure*}[ht!]
\includegraphics[width=\textwidth]{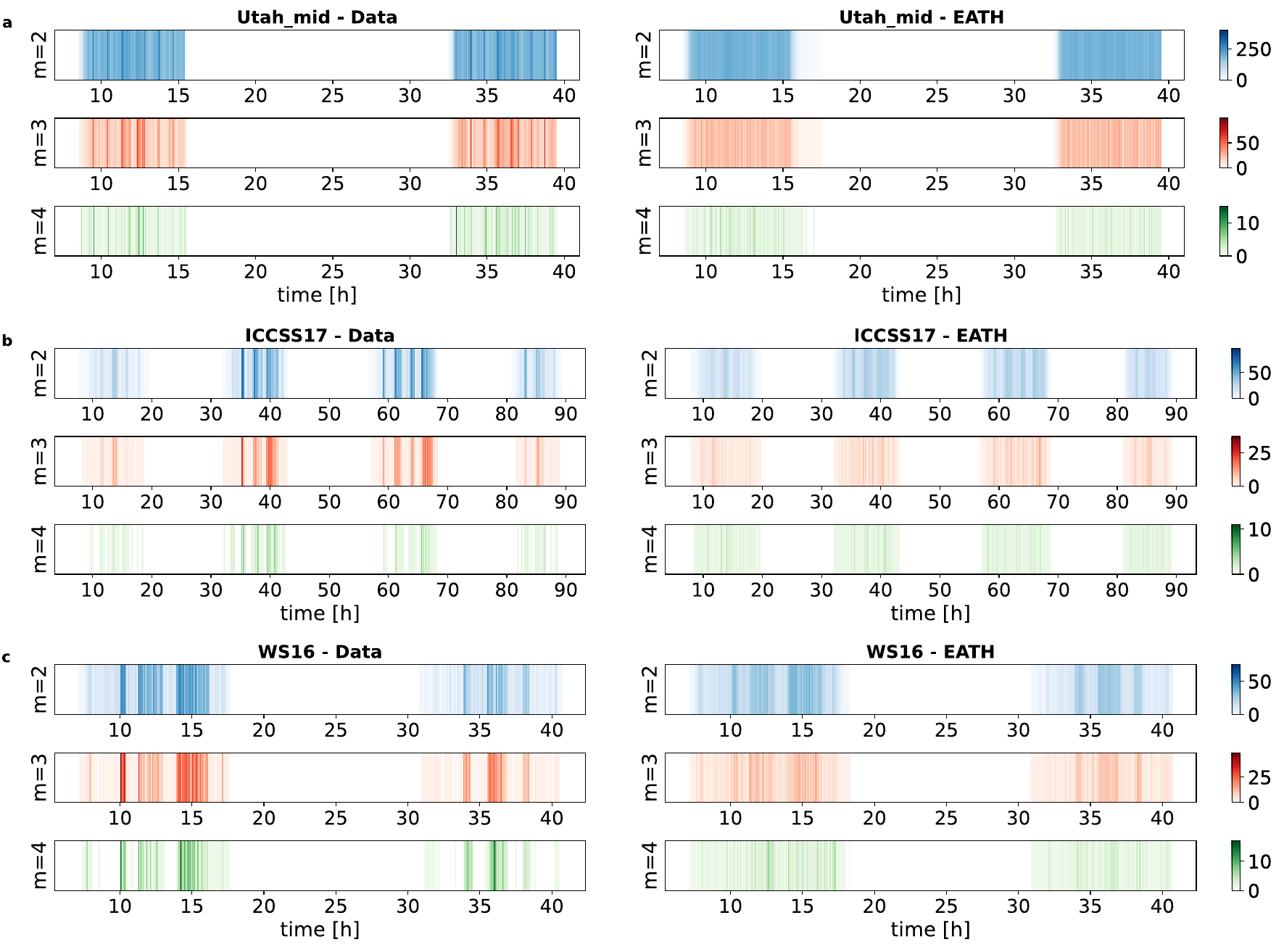}
\caption{\textbf{Overall system activity - III.} In each row we show the number of active hyperedges of a specific size $m \in [2,3,4]$ as a function of time for the empirical hypergraph (first column) and for the EATH model (second column). In each row we consider a different dataset (see title).}
\label{fig:figure2b}
\end{figure*}

\begin{figure*}[ht!]
\includegraphics[width=\textwidth]{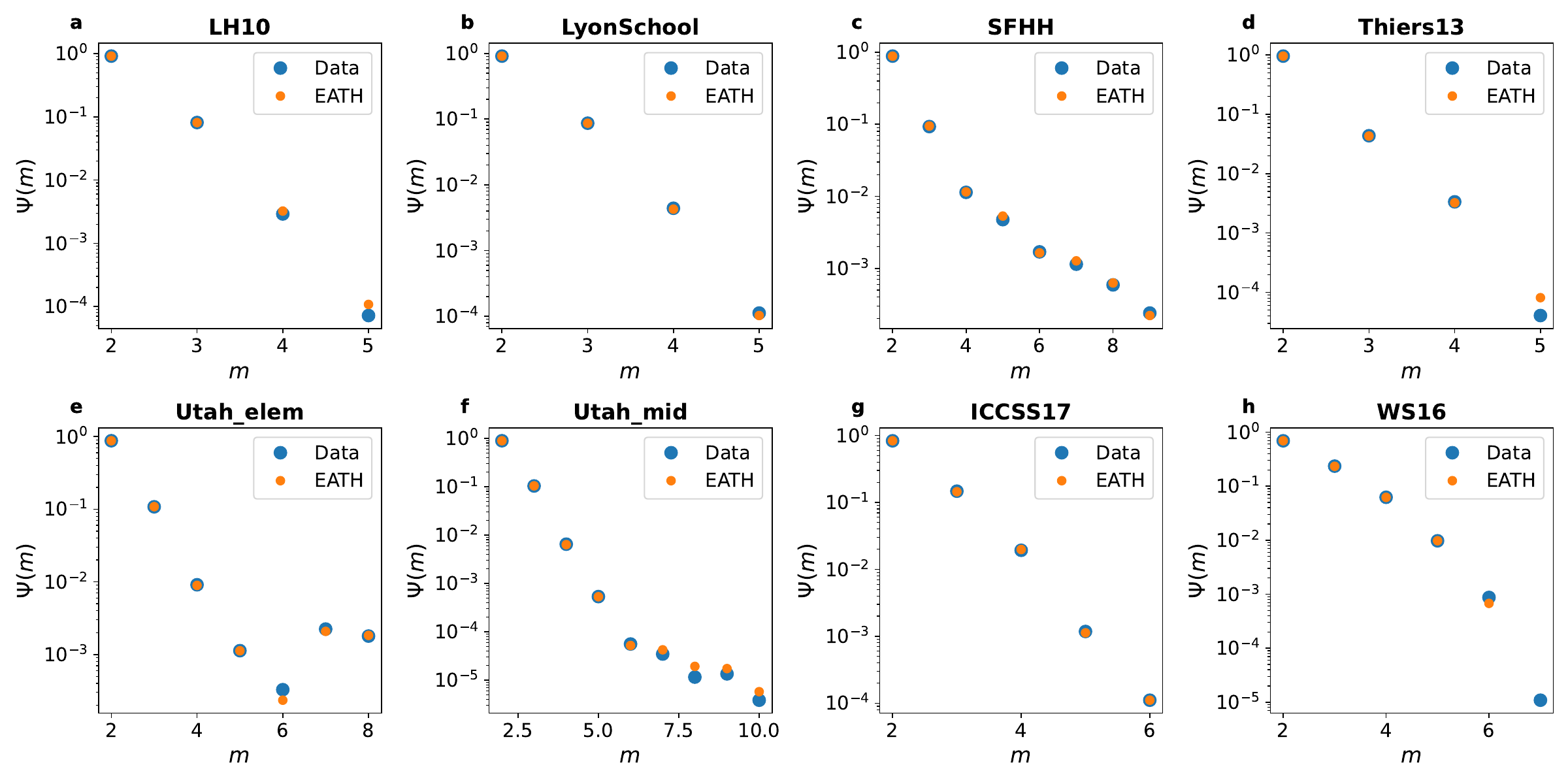}
\caption{\textbf{Hyperedges size distribution.} In each panel we show the hyperedges size distribution $\Psi(m)$ for the empirical hypergraph and for the EATH model. In each panel we consider a different dataset (see title).}
\label{fig:figure3}
\end{figure*}

Here we focus on the system activity and on the empirical parameters. Supplementary Fig. \ref{fig:figure1} shows the evolution of the number of active hyperedges $E_t$ and Supplementary Figs.  \ref{fig:figure2}, \ref{fig:figure2b} show the dynamics of hyperedges activation divided by sizes, for $m \in [2,3,4]$. Supplementary Fig. \ref{fig:figure3} shows the hyperedges size distribution $\Psi(m)$.

\newpage

\newpage

\newpage
In Supplementary Fig. \ref{fig:figure13} we show the values of $g$ and of $\overline{\Delta}$ for the different datasets (see main text). 

The parameter $g$, defining the heterogeneity level for the long-term memory, is tuned for each dataset to reproduce the evolution of the weights distribution and of 
its heterogeneity observed in the empirical projected graph.
We assume that, the denser the empirical projected graph, the lower $g$ should be, since a dense graph requires 
the generation process to explore a wide set of different interactions; 
furthermore, the higher the total number of interactions, the lower $g$ should be, since a high interaction level 
tends to make the heterogeneity grow faster, by frequently reactivating groups. Hence, we set $g=\alpha/(\alpha + \eta E)$ (where $\eta$ is the density of the projected graph and $E$ is the total number of interactions generated), so that if the empirical projected graph is extremely sparse (or with few interactions) $g \xrightarrow{\eta E \to 0} 1$.
Moreover we set a minimum level of memory at $0.2$ so that if $\eta E$ is high, we still keep $g>0$. 

Analogously, the parameter $\overline{\Delta}$, defining the extraction of high- and low-activity phases, 
is tuned for each dataset: also in this case, we assume that a system with higher $\eta E$ features more close and simultaneous events, requiring a lower $\overline{\Delta}$.
We set $\overline{\Delta}= \beta \alpha'/(\alpha'+\eta E)$, so that if the empirical system has few interactions (or a low density) $\overline{\Delta} \xrightarrow{\eta E \to 0} \beta$, where $\beta$ is the maximum allowed time distance for which we can consider two node events to be part of the same high-activity phase. Moreover we set a minimum resolution at $2 min$ so that if $\eta E$ is high still $\overline{\Delta}>0$, i.e. we still have a separation in high- and low- activity phases. 

The obtained values of $g$ and $\overline{\Delta}$ allow to reproduce well
the evolution of the weights distribution $P(w_l)$ and of its heterogeneity and the system time-scales, 
providing an a posteriori validation of these assumptions (see Sections \ref{sez:section1b}, \ref{sez:section1c}). 

\begin{figure*}[ht!]
\includegraphics[width=\textwidth]{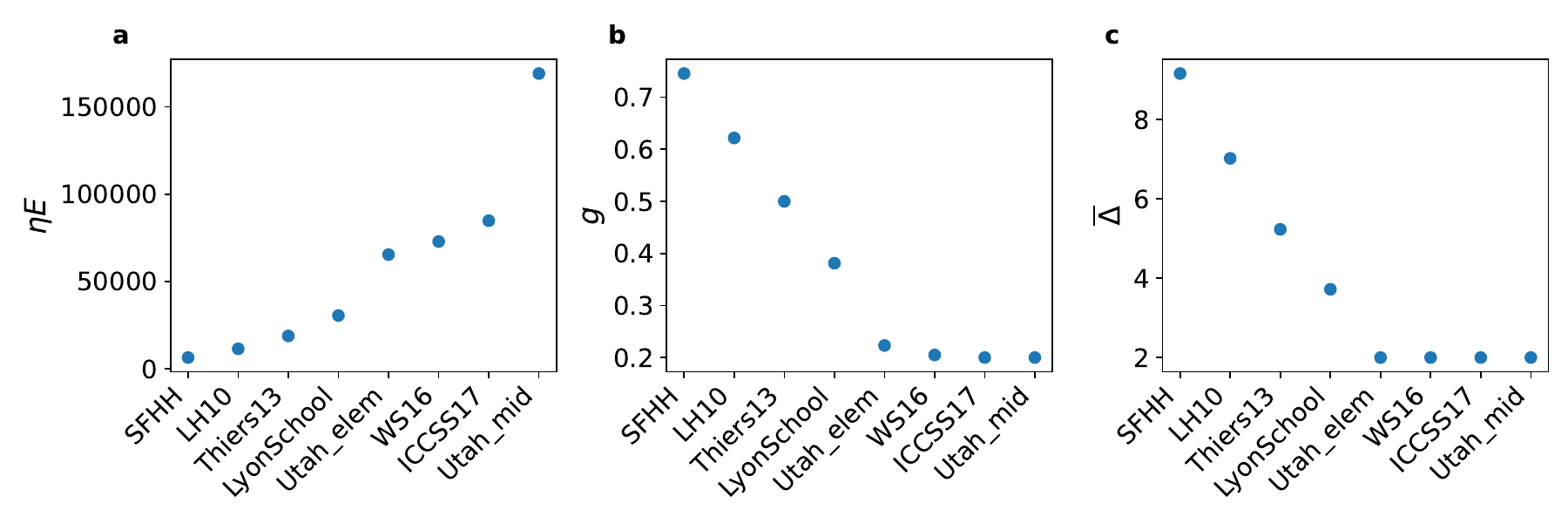}
\caption{\textbf{$\bm{g}$ and $\bm{\overline{\Delta}}$ parameters estimation from data.} 
\textbf{a} $\eta E$ for each dataset; in panel \textbf{b} and \textbf{c} we show respectively the values
of the parameter $g$ and of $\overline{\Delta}$, for the heterogeneity level of the long-term memory and the resolution in defining periods of high- and low- activity. In this case, $g$ has a minimum fixed value of $0.2$, $\overline{\Delta}$ has a minimum value of $2 min$, $\beta=15min$. $\alpha$ and $\alpha'$ are respectively obtained by fixing $g=0.5$ for the Thiers13 dataset and $\overline{\Delta}=2min$ for the Utah\_elem, to reproduce the temporal-topological observed features.}
\label{fig:figure13}
\end{figure*}

\newpage
\subsection{Topological properties}
\label{sez:section1b}
\begin{figure*}[ht!]
\includegraphics[width=\textwidth]{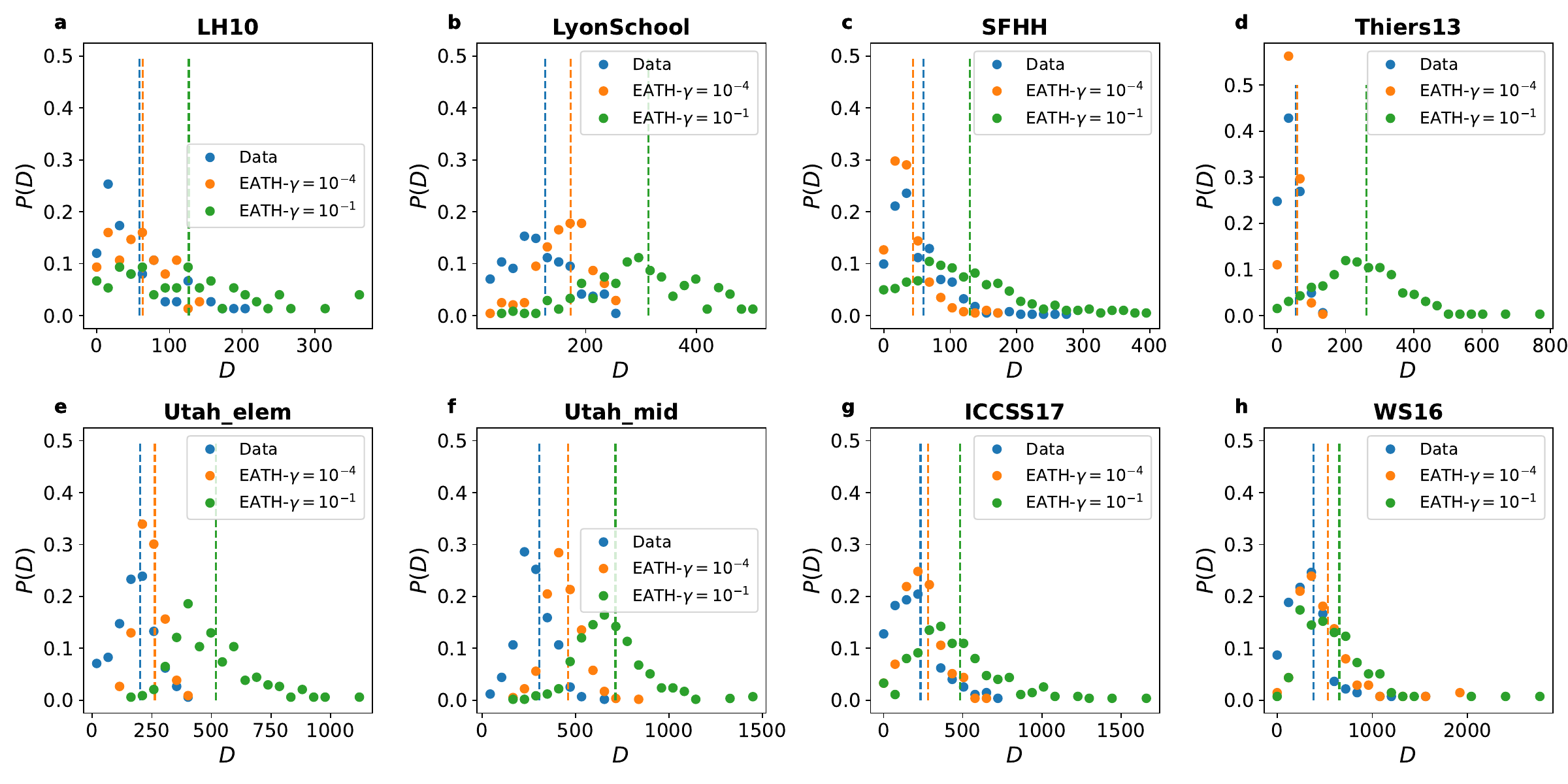}
\caption{\textbf{Higher-order topological properties - I.} In each panel we show the distribution $P(D)$ of nodes total hyperdegree in the aggregated hypergraph $\mathcal{H}$ for the empirical system and the model with (EATH) and without (EATHw) memory. In each panel we consider a different dataset (see title).}
\label{fig:figure4}
\end{figure*}

\begin{figure*}[ht!]
\includegraphics[width=\textwidth]{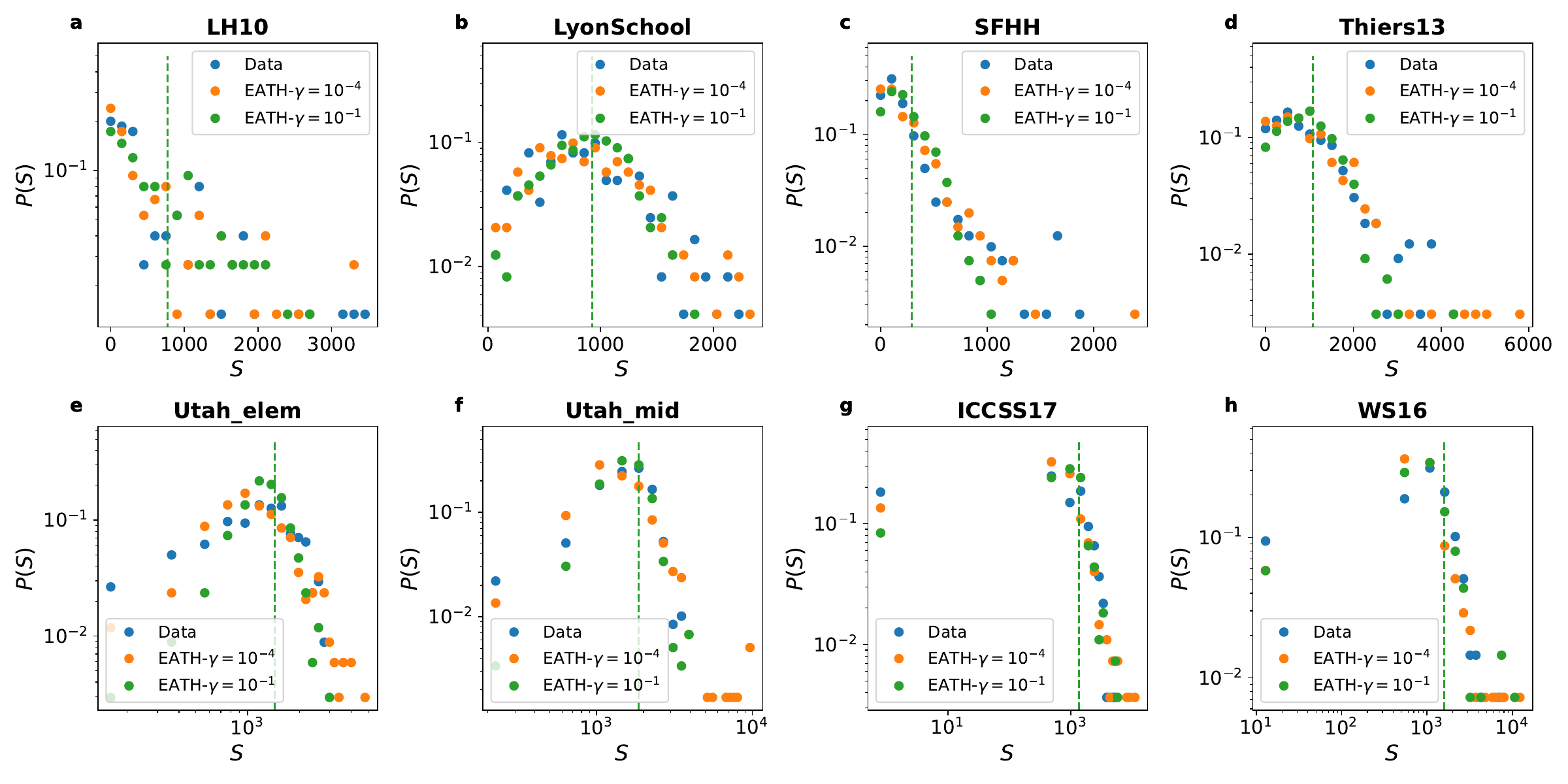}
\caption{\textbf{Higher-order topological properties - II.} In each panel we show the distribution $P(S)$ of nodes hyperstrength in the weighted aggregated hypergraph $\mathcal{H}$ for the empirical system and the model with (EATH) and without (EATHw) memory. In each panel we consider a different dataset (see title).}
\label{fig:figure5}
\end{figure*}

\begin{figure*}[ht!]
\includegraphics[width=\textwidth]{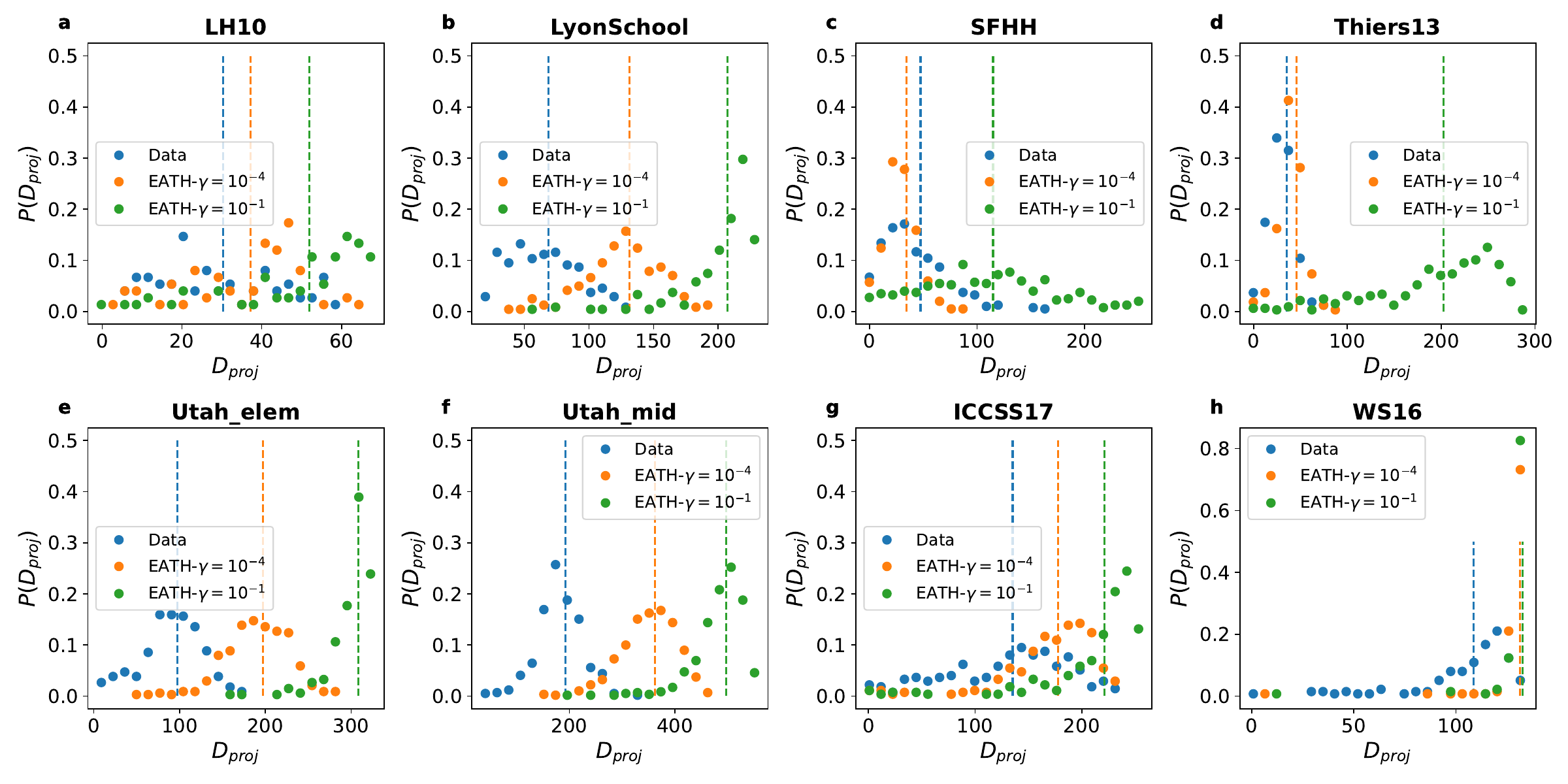}
\caption{\textbf{Pairwise topological properties - I.} In each panel we show the distribution $P(D_{proj})$ of nodes degree in the projected aggregated graph $\mathcal{G}$ for the empirical system and the model with (EATH) and without (EATHw) memory. In each panel we consider a different dataset (see title).}
\label{fig:figure6}
\end{figure*}

\begin{figure*}[ht!]
\includegraphics[width=\textwidth]{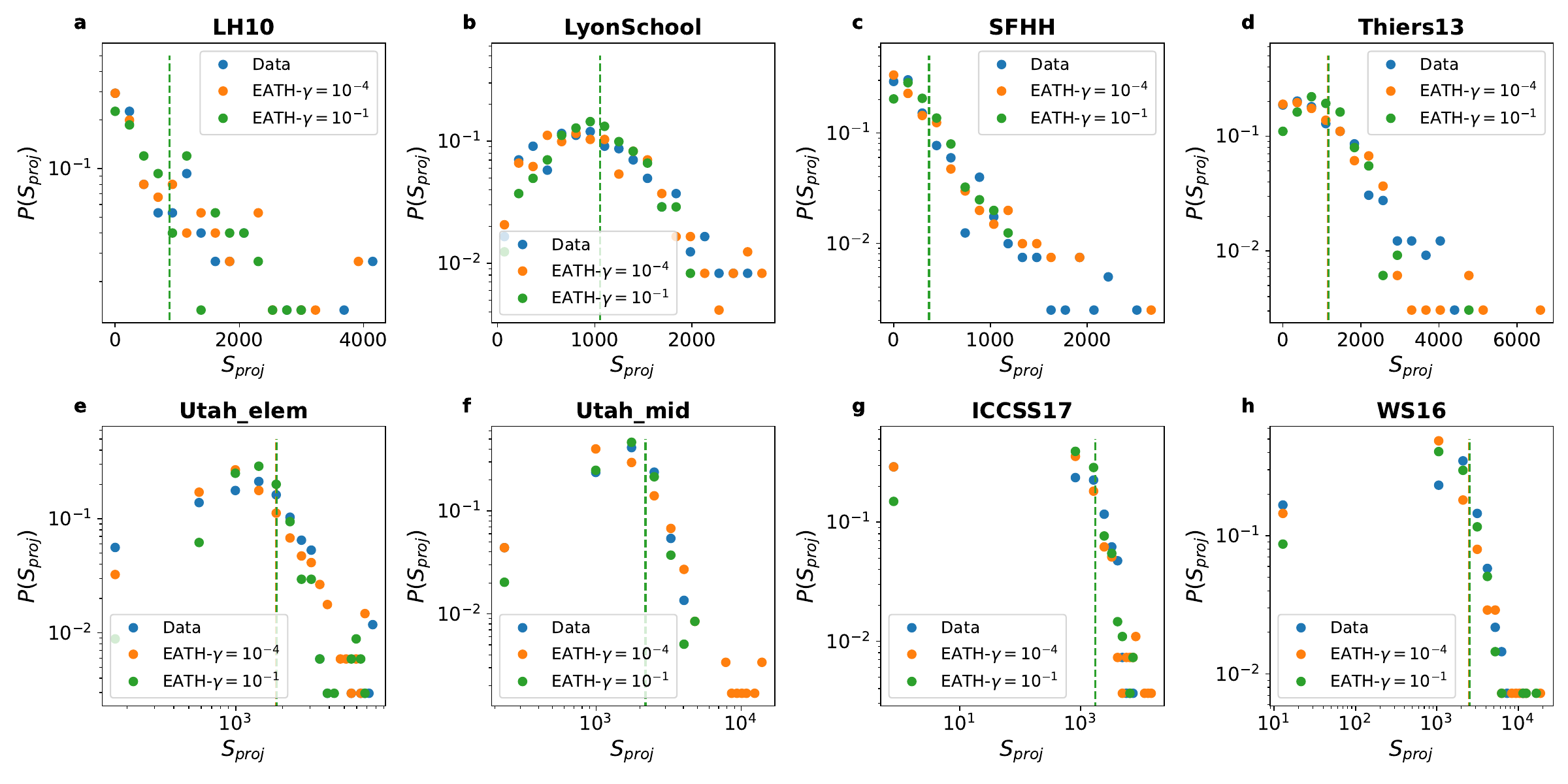}
\caption{\textbf{Pairwise topological properties - II.} In each panel we show the distribution $P(S_{proj})$ of nodes strength in the weighted projected aggregated graph $\mathcal{G}$ for the empirical system and the model with (EATH) and without (EATHw) memory. In each panel we consider a different dataset (see title).}
\label{fig:figure7}
\end{figure*}

\begin{figure*}[ht!]
\includegraphics[width=\textwidth]{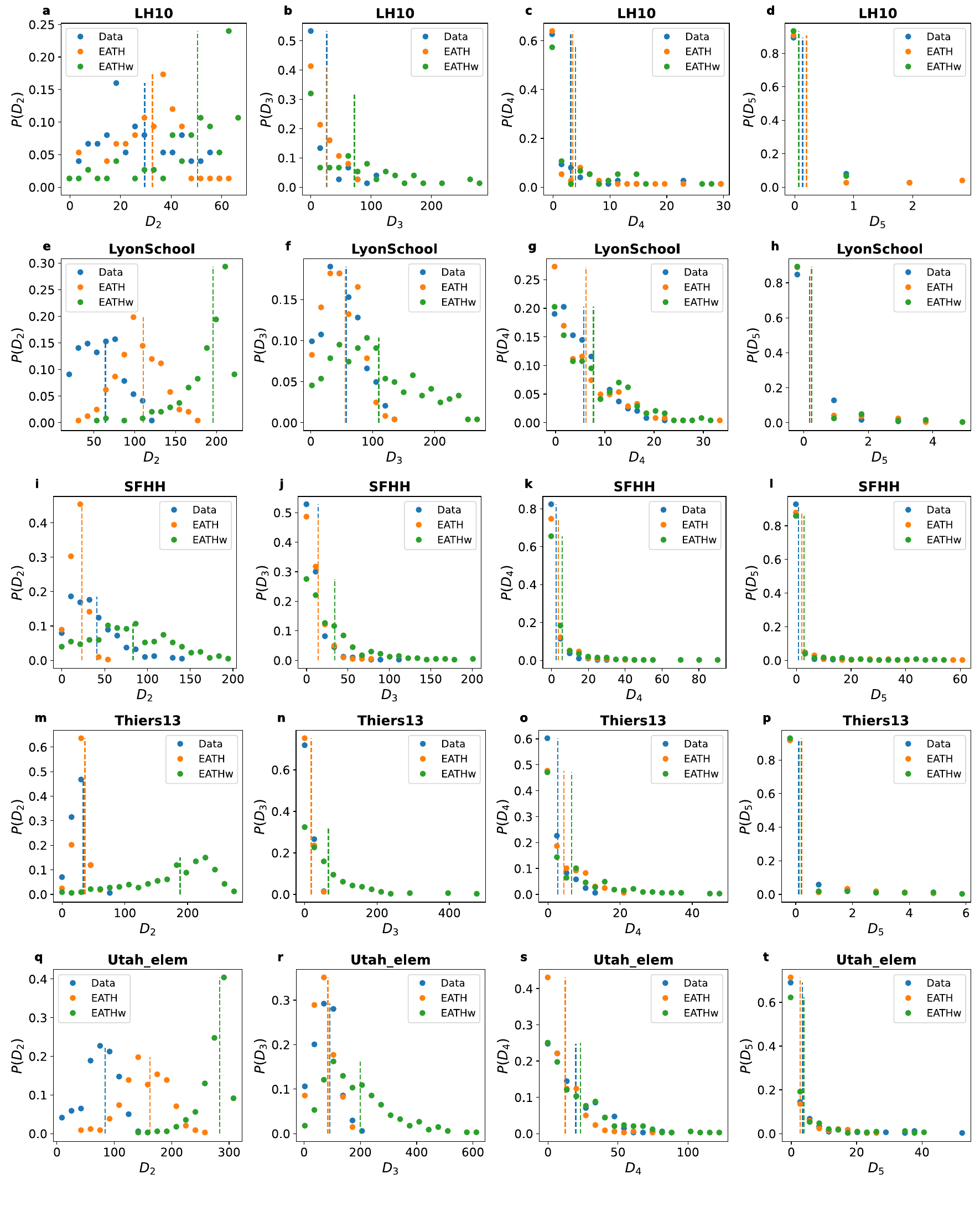}
\caption{\textbf{Higher-order topological properties - III.} In each row we consider a different dataset (see title), and in each panel we show the distribution $P(D_m)$ of nodes hyperdegree at size $m$, for $m \in [2,3,4,5]$, in the aggregated hypergraph $\mathcal{H}$ for the empirical system and the model with (EATH) and without (EATHw) memory. In all panels the dashed vertical lines indicate the average of the corresponding distributions.}
\label{fig:figure8}
\end{figure*}

\begin{figure*}[ht!]
\includegraphics[width=\textwidth]{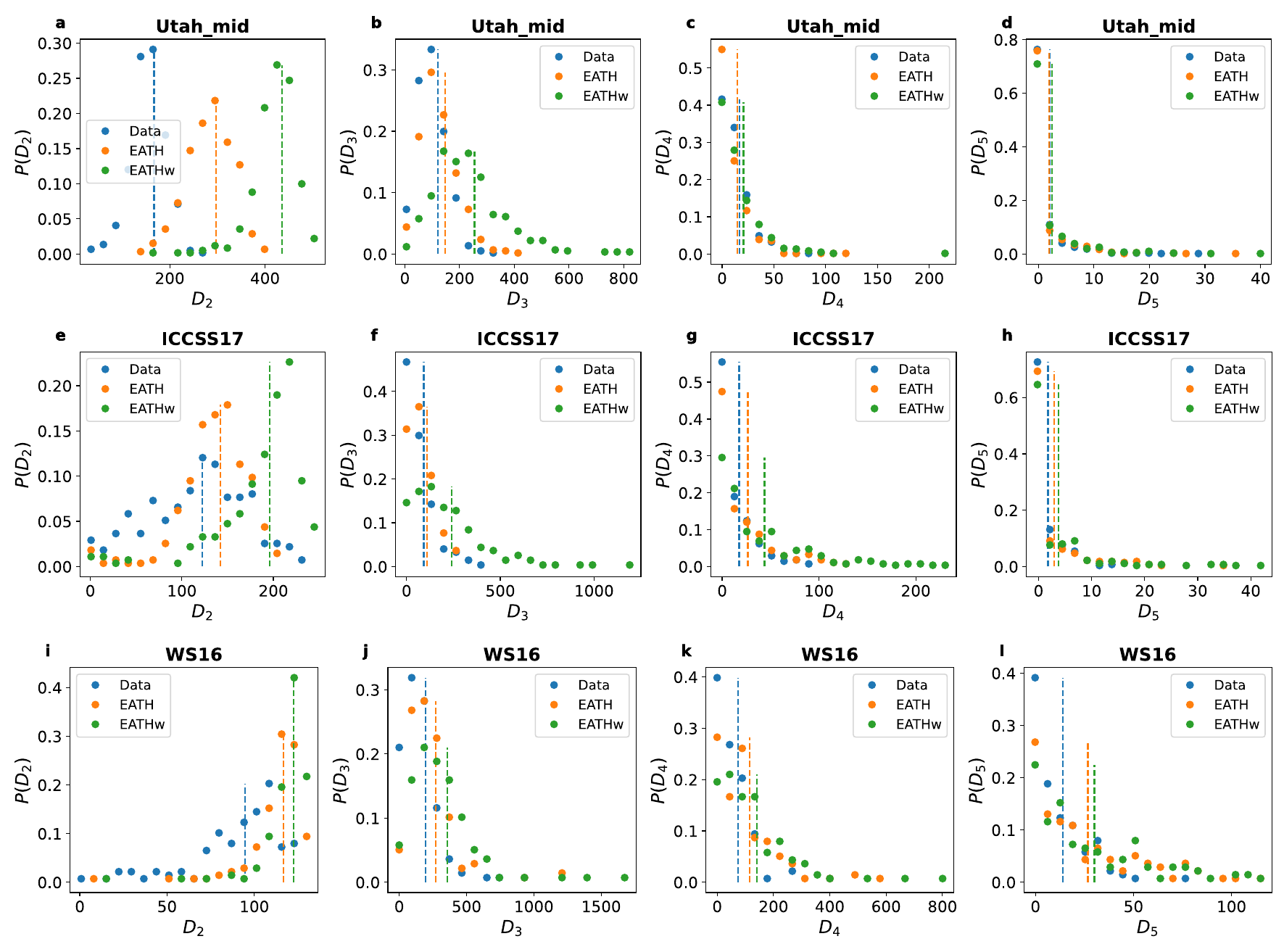}
\caption{\textbf{Higher-order topological properties - IV.} In each row we consider a different dataset (see title), and in each panel we show the distribution $P(D_m)$ of nodes hyperdegree at size $m$, for $m \in [2,3,4,5]$, in the aggregated hypergraph $\mathcal{H}$ for the empirical system and the model with (EATH) and without (EATHw) memory. In all panels the dashed vertical lines indicate the average of the corresponding distributions.}
\label{fig:figure8b}
\end{figure*}

\begin{figure*}[ht!]
\includegraphics[width=0.75\textwidth]{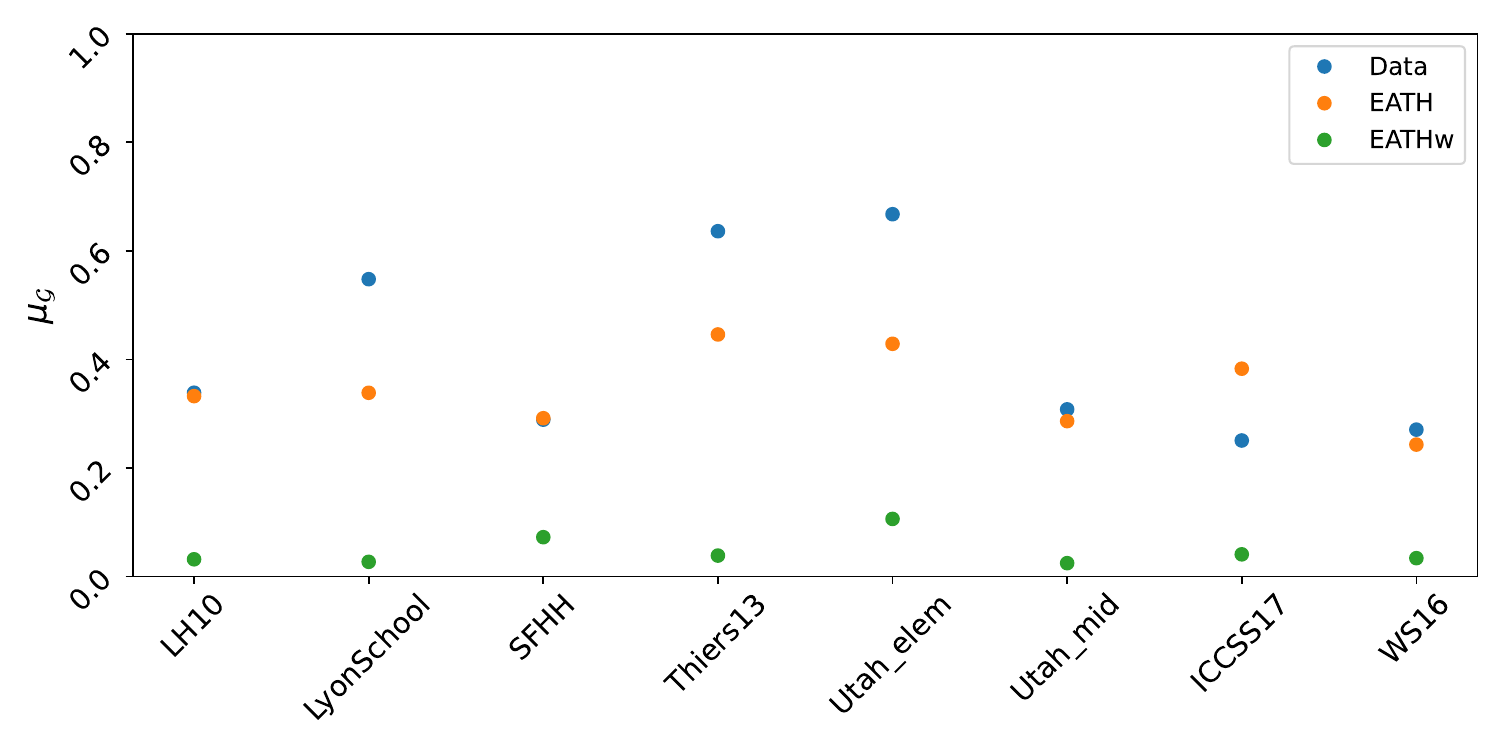}
\caption{\textbf{Projected graph modularity.} For each dataset we show the modularity $\mu_{\mathcal{G}}$ in the weighted projected aggregated graph $\mathcal{G}$ for the empirical system and the model with (EATH) and without (EATHw) memory. $\mu_{\mathcal{G}}$ is obtained by detecting communities through the Louvain Community Detection Algorithm and accounting for the link weights in both the community detection and modularity estimation.}
\label{fig:figure9}
\end{figure*}

Here we focus on topological properties of the empirical and generated hypergraphs. Supplementary Figs. \ref{fig:figure4}-\ref{fig:figure8b} show the topological properties of the systems in the aggregated hypergraph $\mathcal{H}$ and in its pairwise projection $\mathcal{G}$, in terms of the degree and strength distributions. Supplementary Fig. \ref{fig:figure9} shows the modularity value $\mu_{\mathcal{G}}$ in the aggregated projected graph $\mathcal{G}$. In Supplementary Fig. \ref{fig:figure10} we show the evolution of the heterogeneity $\langle w_l^2 \rangle/\langle w_l \rangle^2$ of the weights of the graph obtained projecting the hypergraph aggregated up to time $t$; in Supplementary Figs. \ref{fig:figure11_1}-\ref{fig:figure12} we show the distribution of the weights in the aggregated projected graph, $P(w_l^t)$, for different times $t$ of aggregation, and of the weights in the aggregated hypergraph, $P(w_e)$. 

\newpage

\newpage

\newpage

\newpage
\begin{figure*}[ht!]
\includegraphics[width=\textwidth]{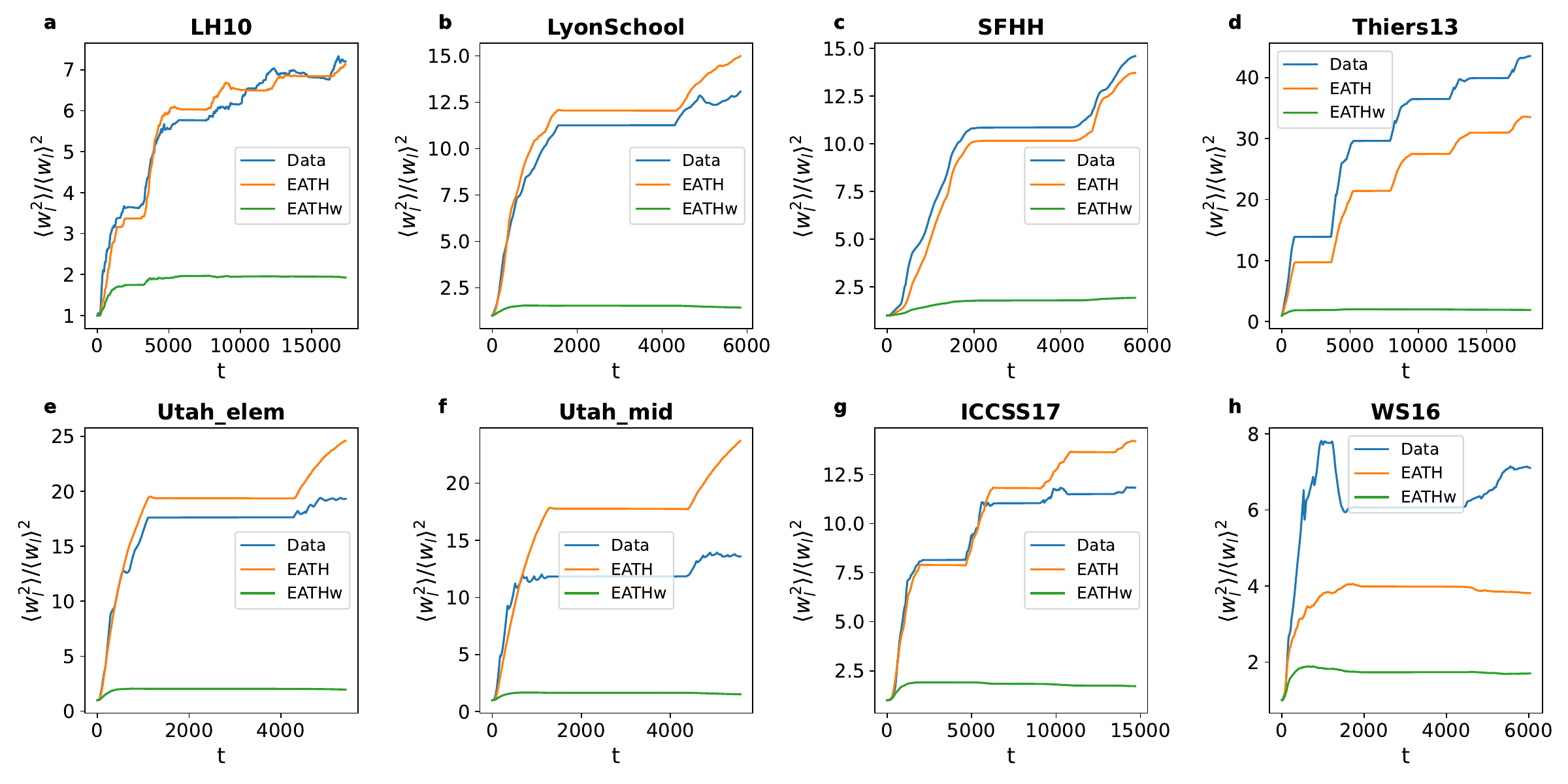}
\caption{\textbf{Weights heterogeneity evolution.} In each panel we show the temporal evolution of the heterogeneity $\langle w_l^2 \rangle/\langle w_l \rangle^2$ of the weights $w_l$ in the aggregated projected weighted graph, obtained by projecting the hypergraph aggregated up to time $t$. In each panel we consider an empirical system (see title) and the model with (EATH) and without (EATHw) memory.}
\label{fig:figure10}
\end{figure*}

\begin{figure*}[ht!]
\includegraphics[width=\textwidth]{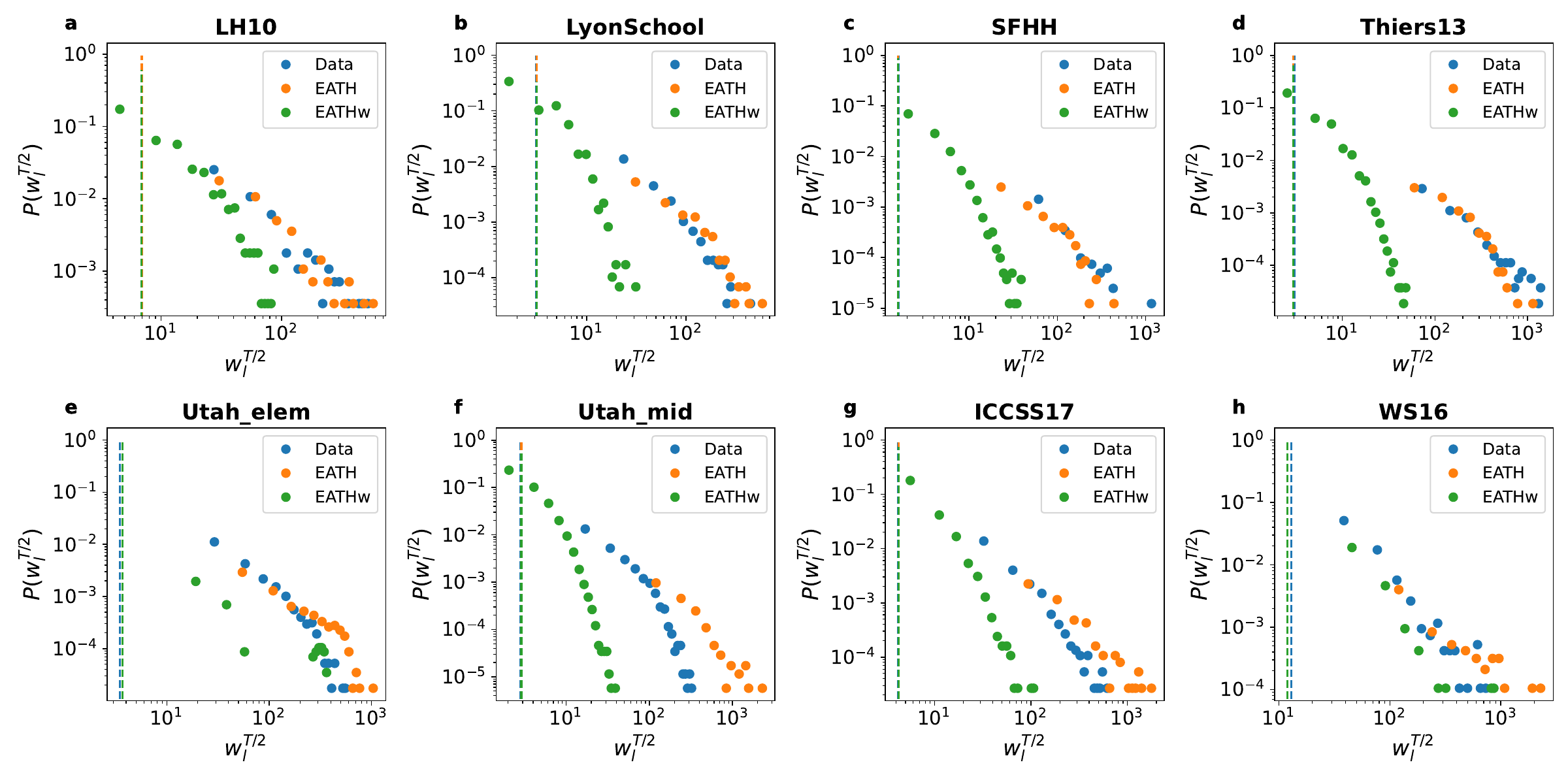}
\caption{\textbf{Weights distribution - I.} In each panel we show the distribution of the weights $w_l^{\mathcal{T}/2}$ in the aggregated projected weighted graph, obtained by projecting the hypergraph aggregated up to time $\mathcal{T}/2$, i.e. at half of the total time span. In each panel we consider an empirical system (see title) and the model with (EATH) and without (EATHw) memory.}
\label{fig:figure11_1}
\end{figure*}

\newpage
\begin{figure*}[ht!]
\includegraphics[width=\textwidth]{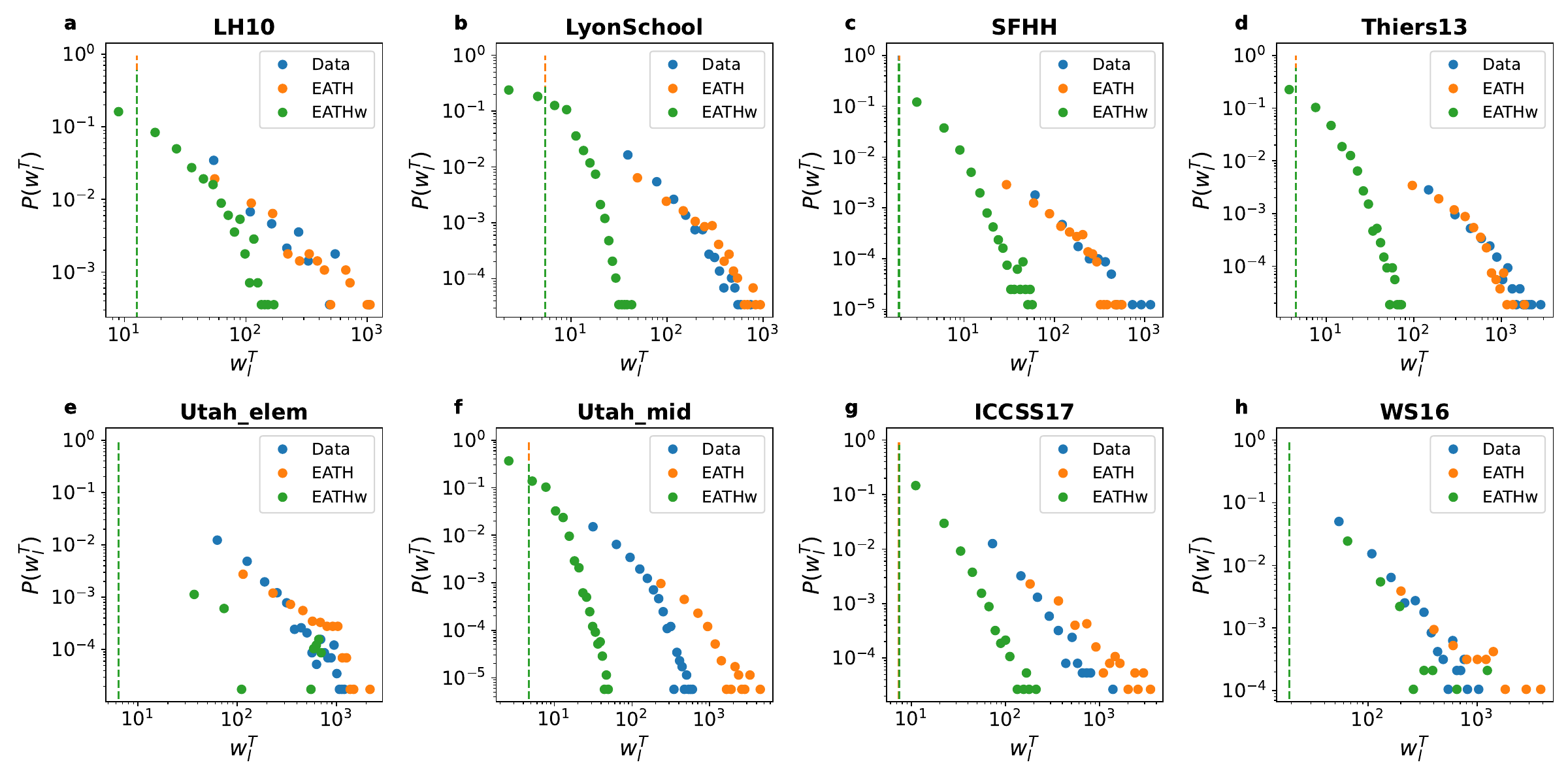}
\caption{\textbf{Weights distribution - II.} In each panel we show the distribution of the weights $w_l^{\mathcal{T}}$ in the aggregated projected weighted graph, obtained by projecting the hypergraph aggregated up to time $\mathcal{T}$, i.e. at the end of the total time span. In each panel we consider an empirical system (see title) and the model with (EATH) and without (EATHw) memory.}
\label{fig:figure11_2}
\end{figure*}

\begin{figure*}[ht!]
\includegraphics[width=\textwidth]{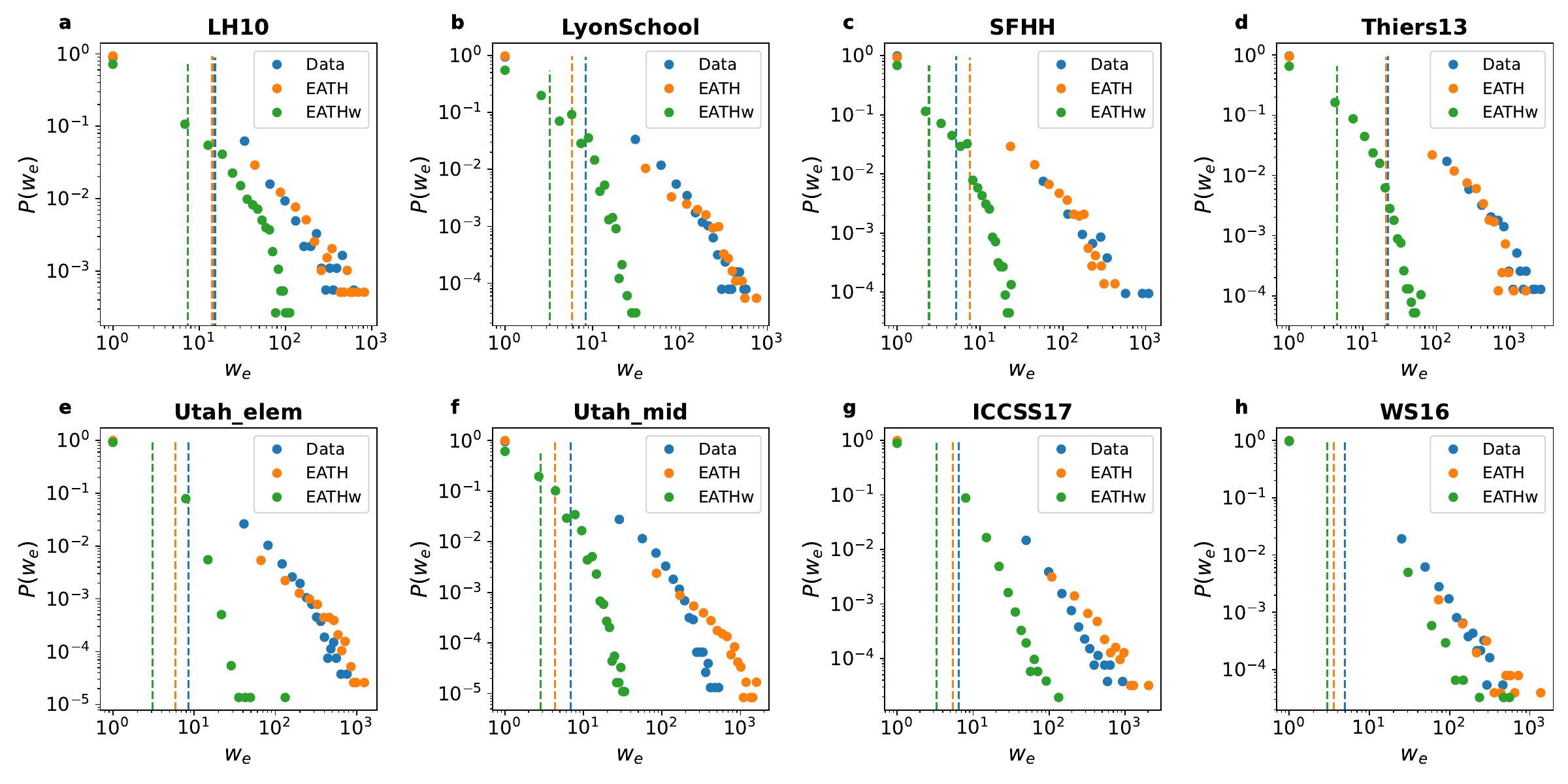}
\caption{\textbf{Weights distribution - III.} In each panel we show the distribution of the weights $w_e$ in the aggregated weighted hypergraph $\mathcal{H}$ at time $\mathcal{T}$, i.e. at the end of the total time span. In each panel we consider an empirical system (see title) and the model with (EATH) and without (EATHw) memory.}
\label{fig:figure12}
\end{figure*}

\newpage
\subsection{Temporal properties}
\label{sez:section1c}
\begin{figure*}[ht!]
\includegraphics[width=\textwidth]{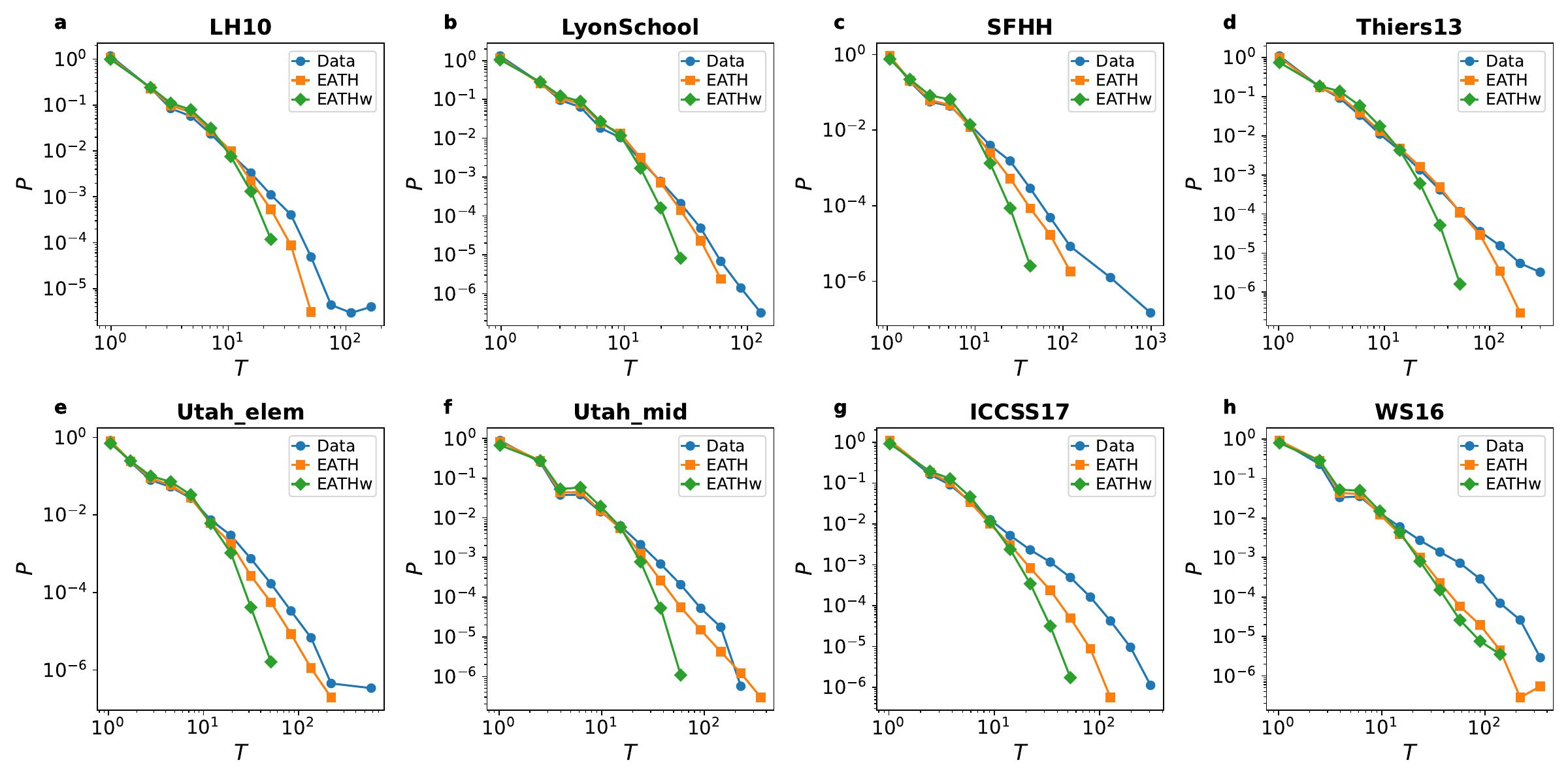}
\caption{\textbf{Events durations distribution - Nodes.} In each panel we show the distribution of events durations for nodes $P(T)$. In each panel we consider an empirical system (see title) and the model with (EATH) and without (EATHw) memory.}
\label{fig:figure14}
\end{figure*}

\begin{figure*}[ht!]
\includegraphics[width=\textwidth]{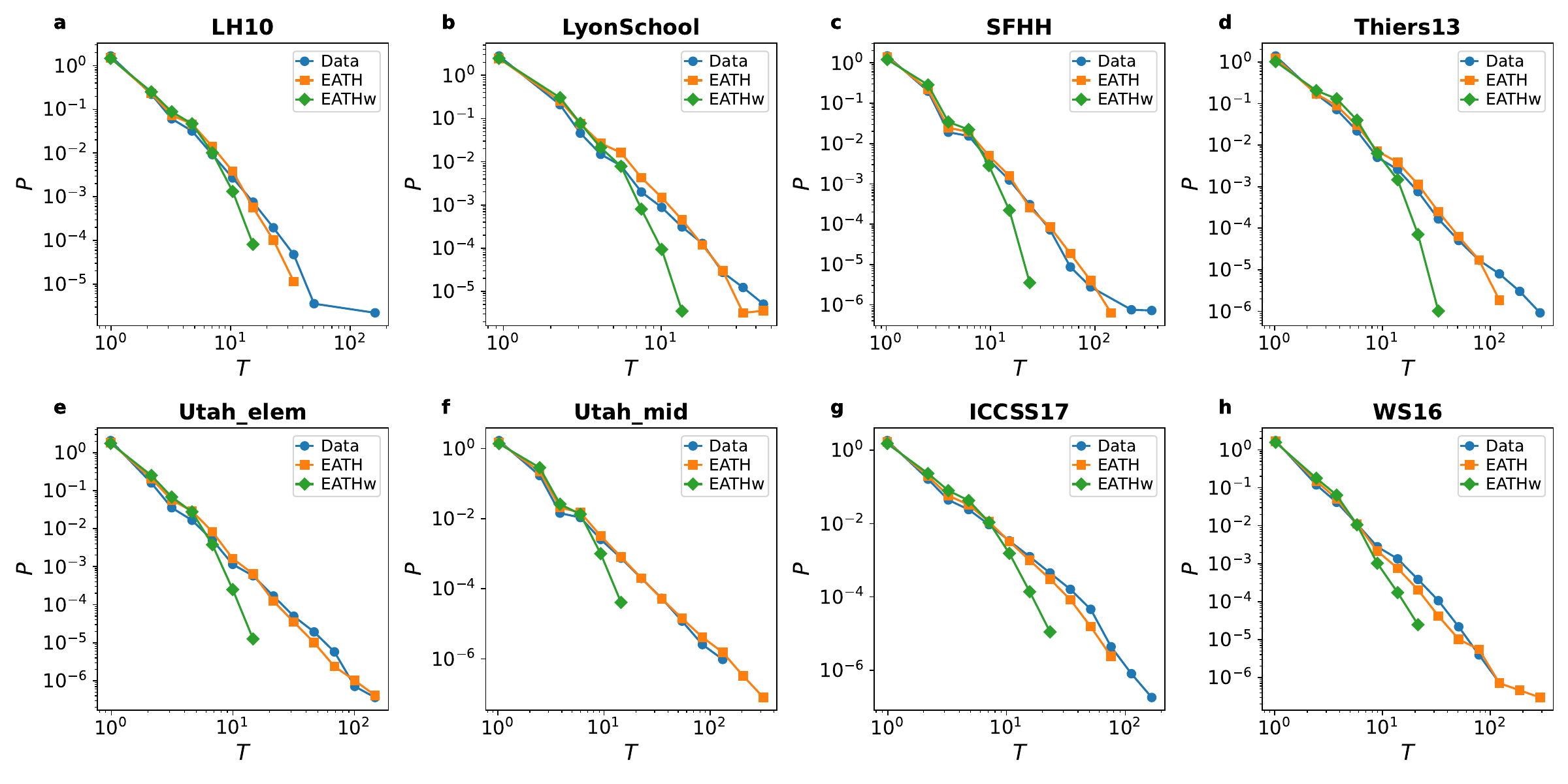}
\caption{\textbf{Events durations distributions - Hyperedges.} In each panel we show the distribution of event durations for hyperedges $P(T)$. In each panel we consider an empirical system (see title) and the model with (EATH) and without (EATHw) memory.}
\label{fig:figure17}
\end{figure*}

\begin{figure*}[ht!]
\includegraphics[width=\textwidth]{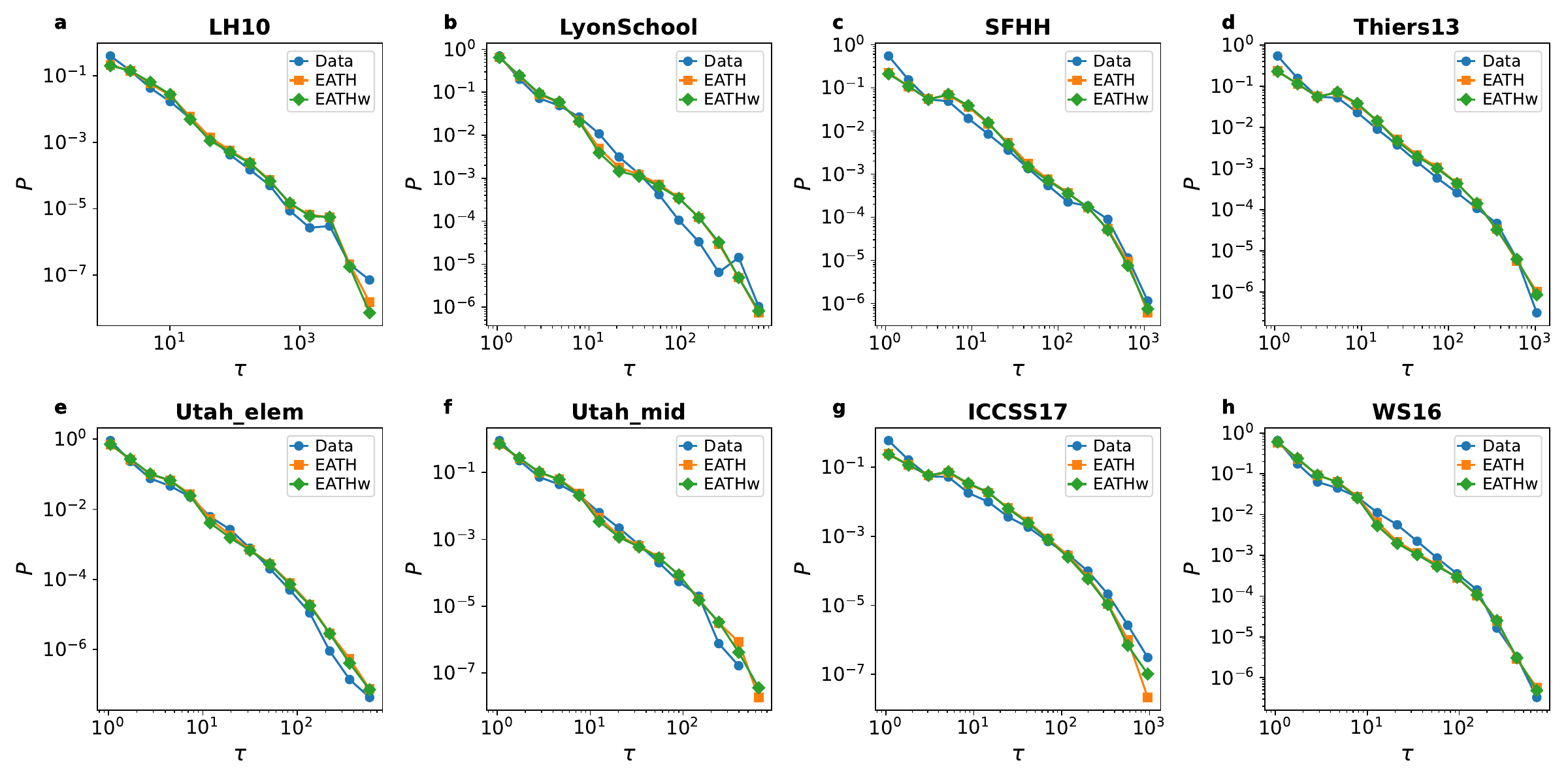}
\caption{\textbf{Inter-event times distributions - Nodes.} In each panel we show the distribution of inter-event times for nodes $P(\tau)$. In each panel we consider an empirical system (see title) and the model with (EATH) and without (EATHw) memory.}
\label{fig:figure15}
\end{figure*}

\begin{figure*}[ht!]
\includegraphics[width=\textwidth]{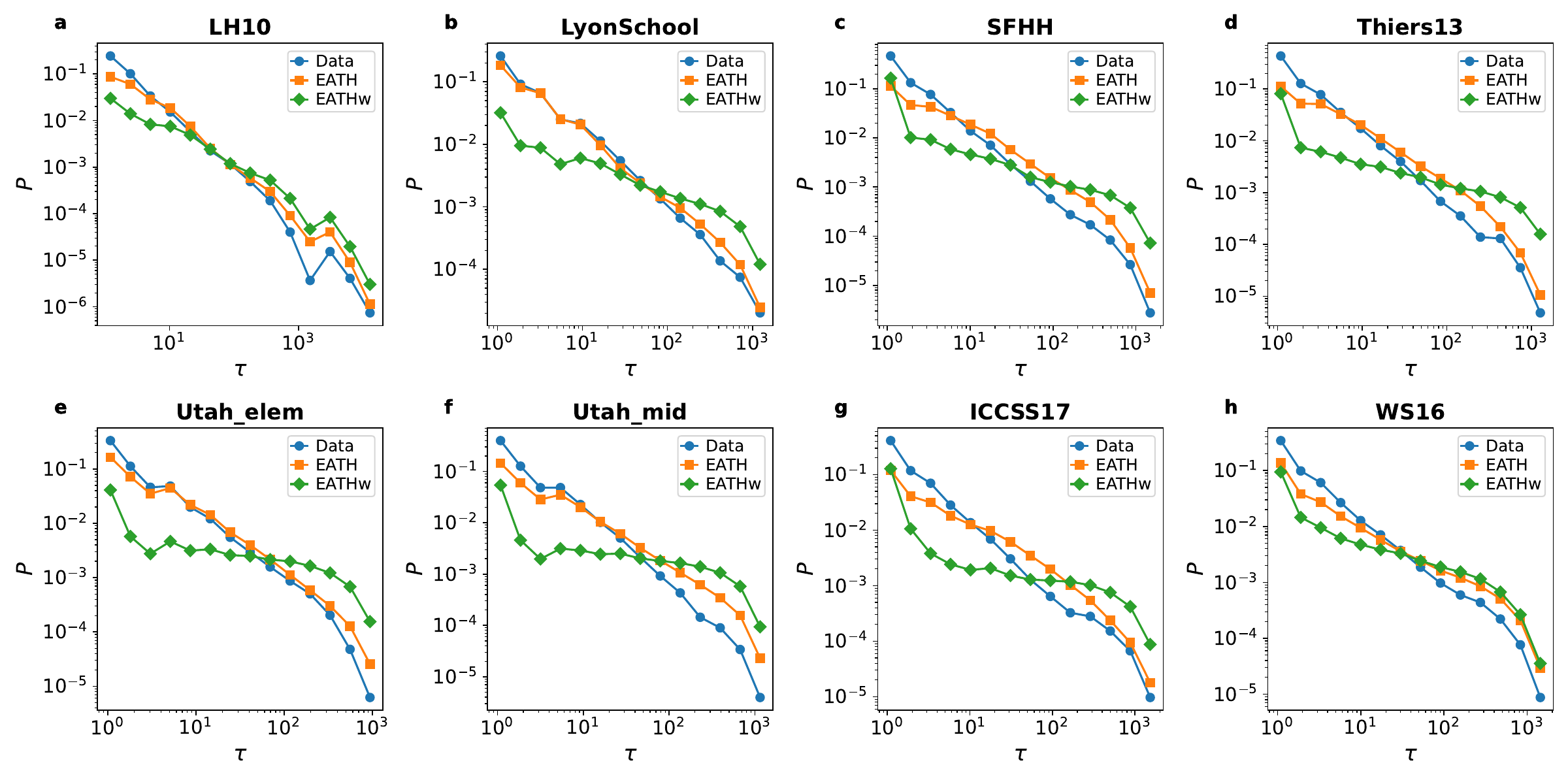}
\caption{\textbf{Inter-event times distributions - Hyperedges.} In each panel we show the distribution of inter-event times for hyperedges $P(\tau)$. In each panel we consider an empirical system (see title) and the model with (EATH) and without (EATHw) memory.}
\label{fig:figure18}
\end{figure*}

\begin{figure*}[ht!]
\includegraphics[width=\textwidth]{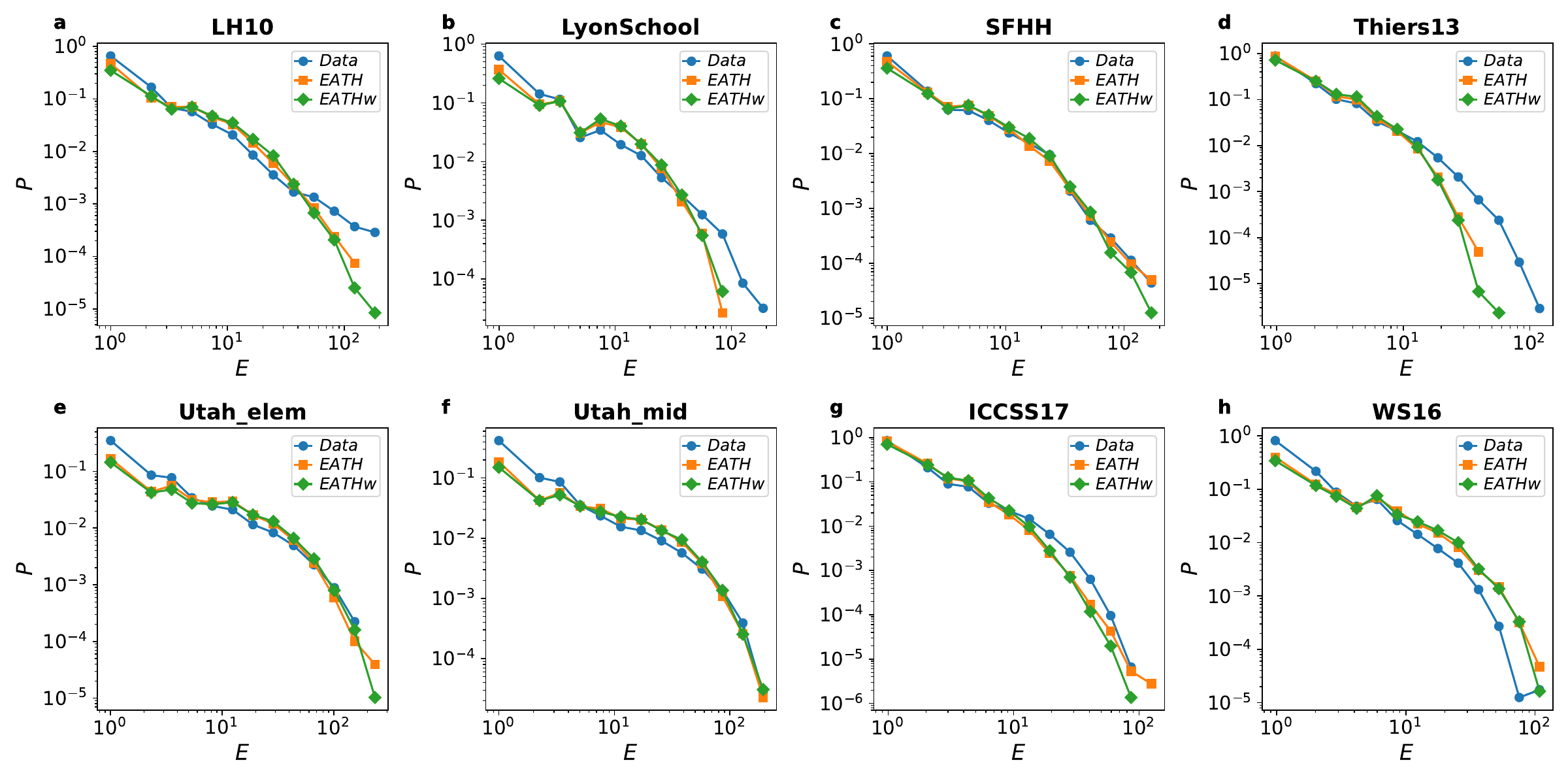}
\caption{\textbf{Distributions of the number of events in a train - Nodes.} In each panel we show the distribution of the number of events in a train of events for nodes $P(E)$, where a train is defined with $\Delta=15 \delta t$ for all the datasets, except for LH10 and SFHH where we consider $\Delta=60 \delta t$. In each panel we consider an empirical system (see title) and the model with (EATH) and without (EATHw) memory.}
\label{fig:figure16}
\end{figure*}

\begin{figure*}[ht!]
\includegraphics[width=\textwidth]{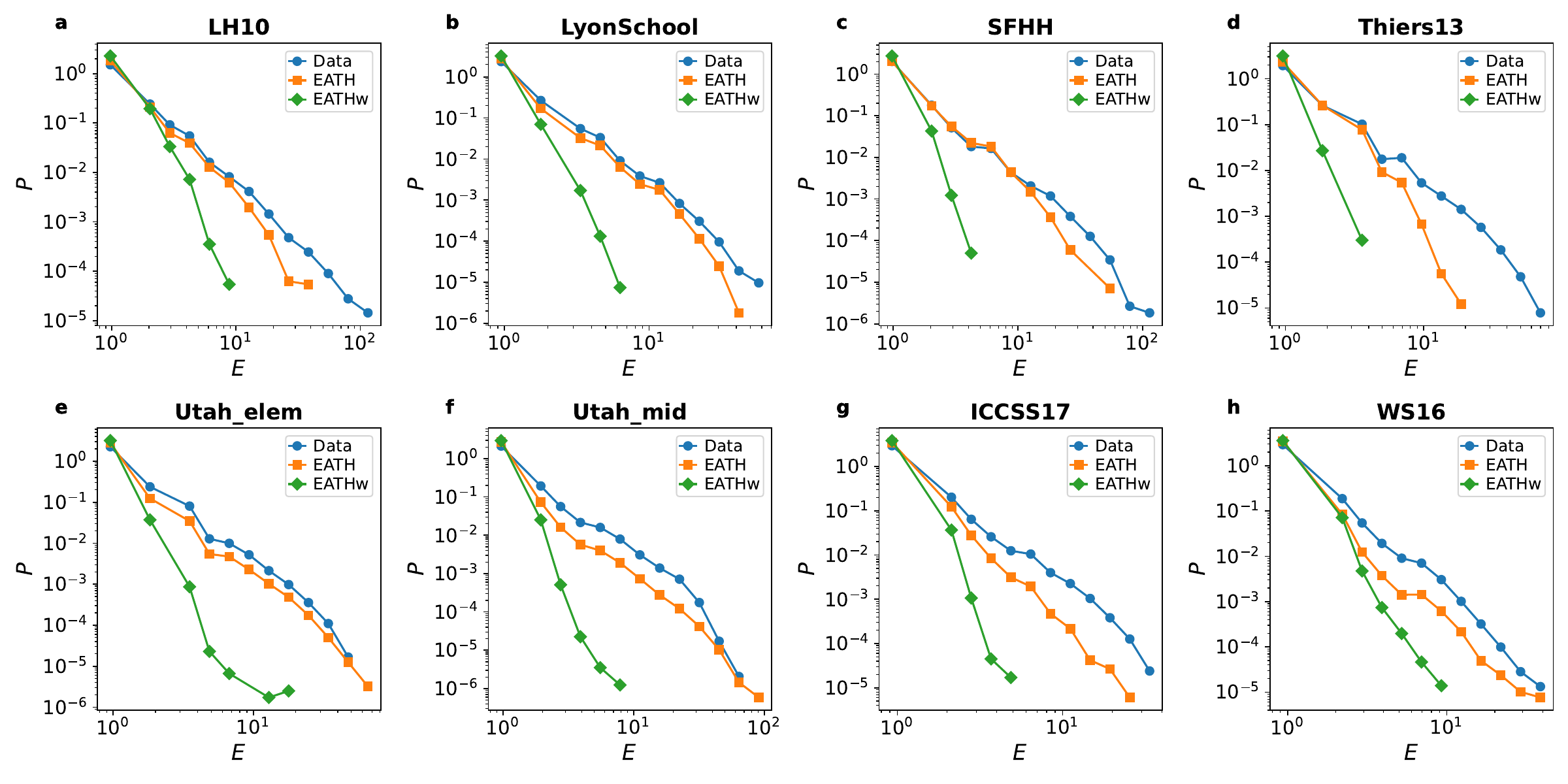}
\caption{\textbf{Distributions of the number of events in a train - Hyperedges.} In each panel we show the distribution of the number of events in a train of events for hyperedges $P(E)$, where a train is defined with $\Delta=15 \delta t$ for all the datasets, except for LH10 and SFHH where we consider $\Delta=60 \delta t$. In each panel we consider an empirical system (see title) and the model with (EATH) and without (EATHw) memory.}
\label{fig:figure19}
\end{figure*}

Here we focus on temporal properties of the empirical and generated hypergraphs. Supplementary Figs. \ref{fig:figure14}-\ref{fig:figure19} show the temporal properties of the hypergraphs, i.e. the distribution of the event duration, $P(T)$, the distribution of the inter-event times, $P(\tau)$, and the distribution of the number of events in a train of events of resolution $\Delta$, $P(E)$, both considering nodes and hyperedges. Supplementary Table \ref{tab:table1} reports the burstiness values for the duration, $\Delta B_T$, and inter-event time distributions, $\Delta B_{\tau}$, for both nodes and hyperedges considering the hypergraphs generated with the EATH model. Analogously Supplementary Table \ref{tab:table2} reports the burstiness values for the temporal distributions obtained with the EATH model without memory (EATHw). In Supplementary Figs. \ref{fig:figure20}-\ref{fig:figure26} we show the distributions of the duration and inter-event times burstiness, $\Delta B_T$, $\Delta B_{\tau}$, for single nodes and single hyperedges, and their correlations. Supplementary Figs. \ref{fig:figure27}, \ref{fig:figure23} show respectively the distribution of the persistence activity, $P(a_T(i))$, and of the instantaneous activity, $P(a_h(i))$, actually measured for each node from the generated hypergraphs and from the empirical datasets.

\begin{table}[h!]
    \begin{tabular}{c|c|c|c|c}
    \hline
    \hline
         & \multicolumn{2}{c}{Nodes} \vline & \multicolumn{2}{c}{Hyperedges} \\ \hline
         & $\Delta B_{T}$ & $\Delta B_{\tau}$ & $\Delta B_{T}$ & $\Delta B_{\tau}$ \\ \hline
         LH10 & 0.23 &0.71 & 0.30 & 0.42 \\
         LyonSchool & 0.23 & 0.58 & 0.38 & 0.32 \\
         SFHH & 0.41 &0.47 & 0.49 & 0.30 \\
         Thiers13 & 0.39 & 0.45 & 0.45 & 0.30 \\
         Utah\_elem & 0.32 & 0.54 & 0.50 & 0.31 \\
         Utah\_mid & 0.41 & 0.57 & 0.61 & 0.26 \\
         ICCSS17 & 0.38 & 0.35 & 0.47 & 0.26 \\
         WS16 & 0.47 & 0.58 & 0.73 & 0.13 \\
    \hline
    \hline
    \end{tabular}
    \caption{\textbf{Burstiness of temporal distributions - EATH.} For each dataset we report the burstiness of the overall duration distribution, $\Delta B_T$, and the burstiness of the overall inter-event times distribution, $\Delta B_{\tau}$, for both nodes and hyperedges, obtained considering the hypergraphs generated with the EATH model with memory (EATH).} 
    \label{tab:table1}
\end{table}

\begin{table}[h!]
    \begin{tabular}{c|c|c|c|c}
    \hline
    \hline
         & \multicolumn{2}{c}{Nodes} \vline & \multicolumn{2}{c}{Hyperedges} \\ \hline
         & $\Delta B_{T}$ & $\Delta B_{\tau}$ & $\Delta B_{T}$ & $\Delta B_{\tau}$ \\ \hline
         LH10 & 0.14 &0.70 & 0.21 & 0.22 \\
         LyonSchool & 0.13 & 0.60 & 0.25 & -0.03 \\
         SFHH & 0.17 &0.48 & 0.22 & 0.01 \\
         Thiers13 & 0.14 & 0.46 & 0.17 & -0.05 \\
         Utah\_elem & 0.15 & 0.55 & 0.25 & -0.09 \\
         Utah\_mid & 0.13 & 0.58 & 0.24 & -0.1 \\
         ICCSS17 & 0.19 & 0.36 & 0.25 & -0.06 \\
         WS16 & 0.29 & 0.59 & 0.32 & 0.02 \\
    \hline
    \hline
    \end{tabular}
    \caption{\textbf{Burstiness values of temporal distributions - EATHw.} For each dataset we report the burstiness of the overall duration distribution, $\Delta B_T$, and the burstiness of the overall inter-event times distribution, $\Delta B_{\tau}$, for both nodes and hyperedges, obtained considering the hypergraphs generated with the EATH model without memory (EATHw).}
    \label{tab:table2}
\end{table}

\newpage

\newpage

\newpage

\newpage
\begin{figure*}[ht!]
\includegraphics[width=\textwidth]{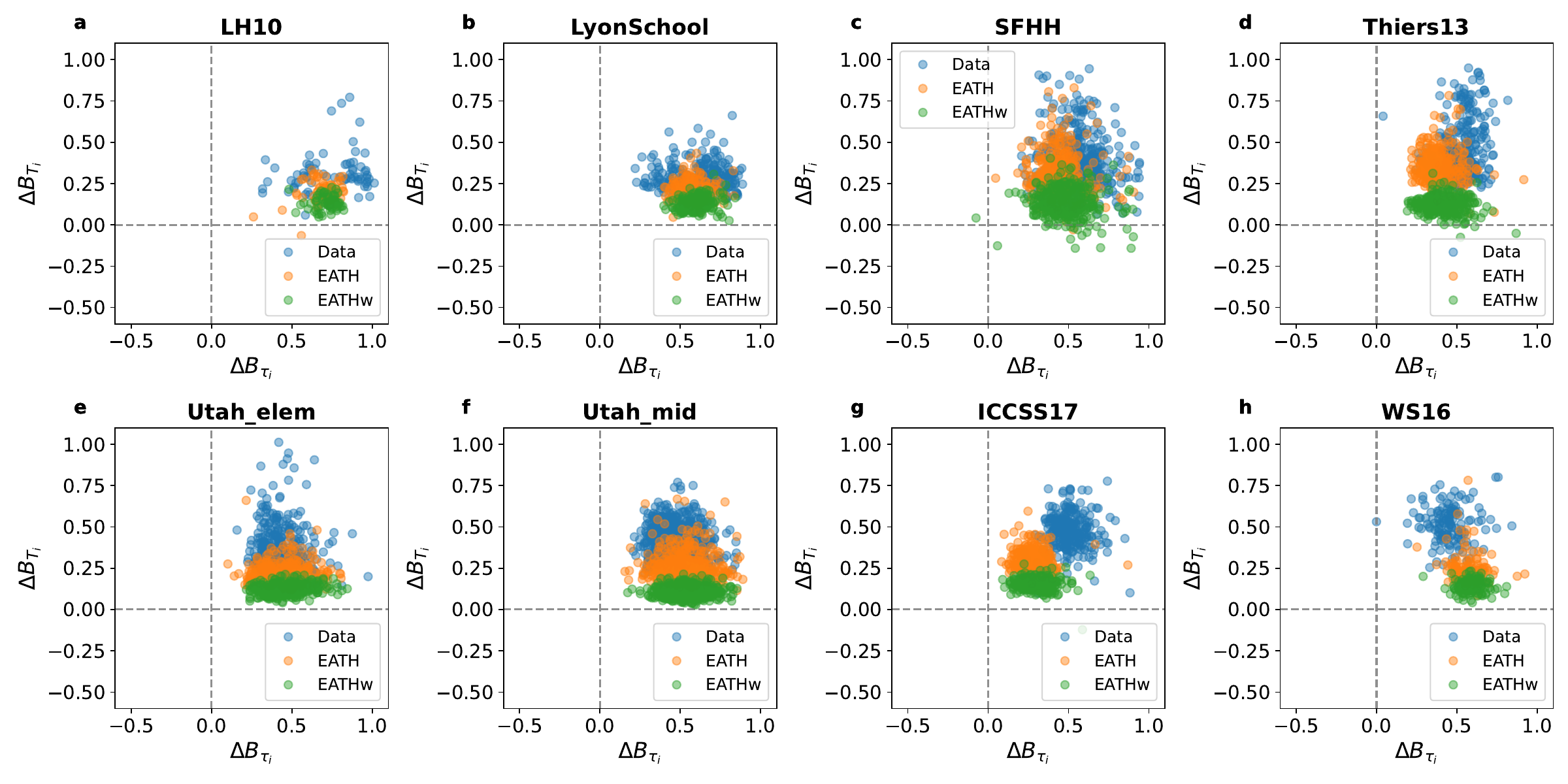}
\caption{\textbf{Nodes temporal heterogeneity - I.} For each node we evaluate the burstiness $\Delta B_{\tau_i}$ and $\Delta B_{T_i}$ of their inter-event times and durations distributions. In each panel, each point corresponds to a node and we show the correlations between $\Delta B_{\tau_i}$ and $\Delta B_{T_i}$. We consider only the nodes with at least $10$ activation events. In each panel we consider an empirical system (see title) and the model with (EATH) and without (EATHw) memory. The gray dashed lines are a reference for $\Delta B=0$.}
\label{fig:figure20}
\end{figure*}

\begin{figure*}[ht!]
\includegraphics[width=\textwidth]{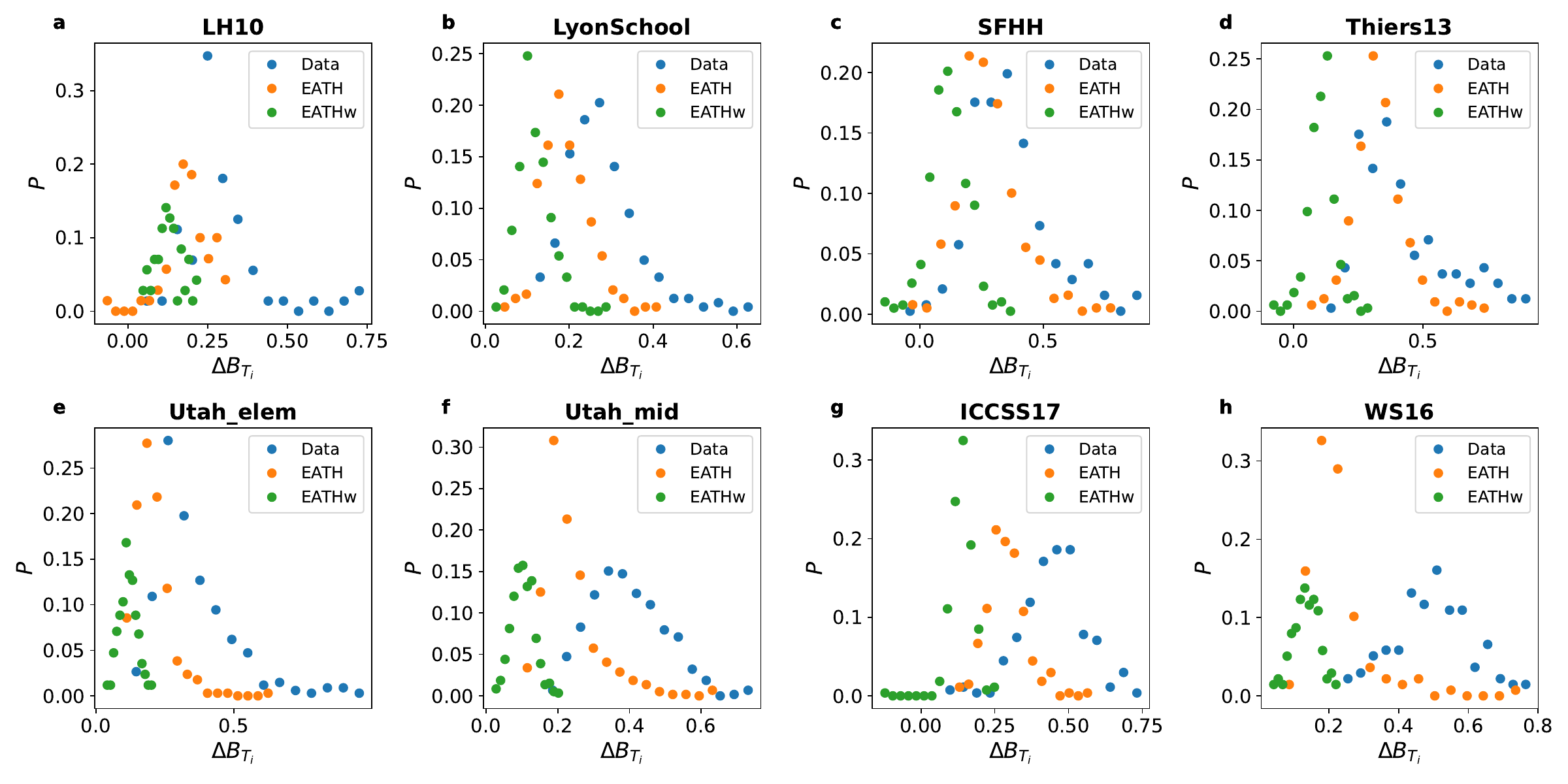}
\caption{\textbf{Nodes temporal heterogeneity - II.} For each node we evaluate the burstiness of their durations distribution $\Delta B_{T_i}$: in each panel we show the distribution of this node burstiness, $P(\Delta B_{T_i})$. We consider only the nodes with at least $10$ activation events. In each panel we consider an empirical system (see title) and the model with (EATH) and without (EATHw) memory.}
\label{fig:figure21}
\end{figure*}

\newpage
\begin{figure*}[ht!]
\includegraphics[width=\textwidth]{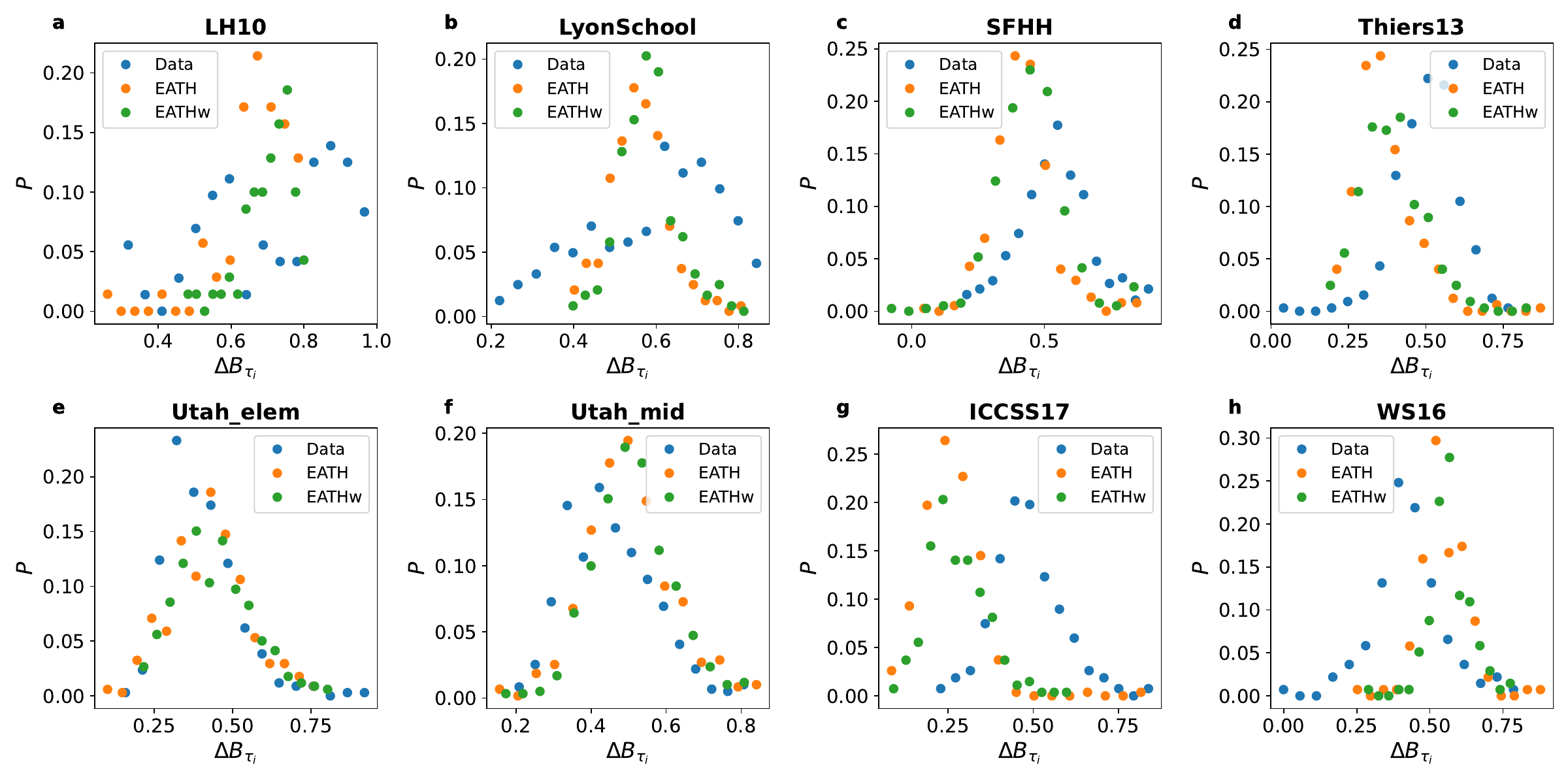}
\caption{\textbf{Nodes temporal heterogeneity - III.} For each node we evaluate the burstiness of their inter-event times distribution $\Delta B_{\tau_i}$: in each panel we show the distribution of this node burstiness, $P(\Delta B_{\tau_i})$. We consider only the nodes with at least $10$ activation events. In each panel we consider an empirical system (see title) and the model with (EATH) and without (EATHw) memory.}
\label{fig:figure22}
\end{figure*}

\begin{figure*}[ht!]
\includegraphics[width=\textwidth]{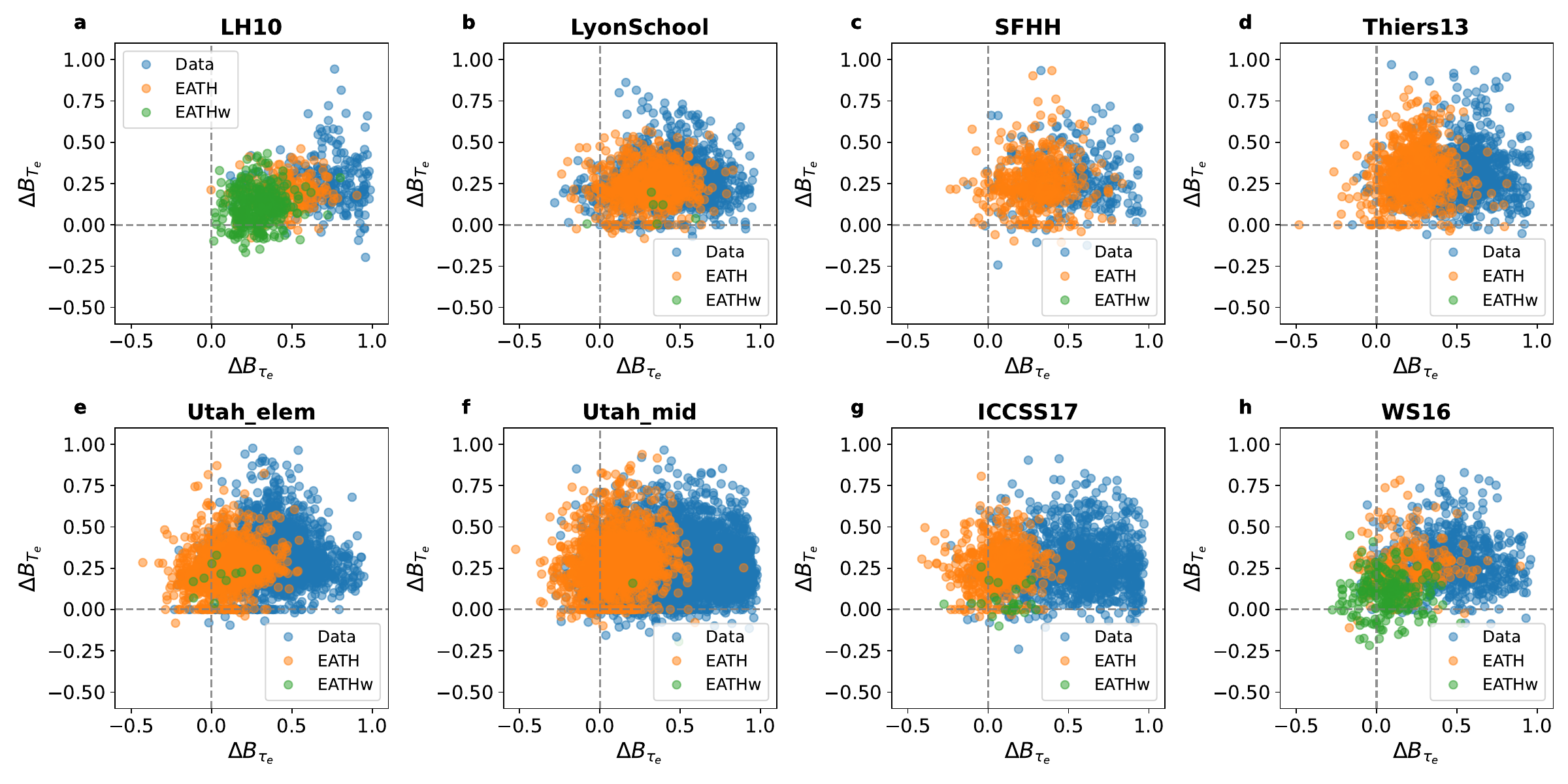}
\caption{\textbf{Hyperedges temporal heterogeneity - I.} For each hyperedge we evaluate the burstiness $\Delta B_{\tau_e}$ and $\Delta B_{T_e}$ of their inter-event time and duration distributions. In each panel, each point corresponds to a hyperedge and we show the correlations between $\Delta B_{\tau_e}$ and $\Delta B_{T_e}$. We consider only the hyperedges with at least $10$ activation events. In each panel we consider an empirical system (see title) and the model with (EATH) and without (EATHw) memory. Note that in some datasets there are no results for the EATHw case, since no group activated enough times to define its burstiness parameters. The gray dashed lines are a reference for $\Delta B=0$.}
\label{fig:figure24}
\end{figure*}

\newpage
\begin{figure*}[ht!]
\includegraphics[width=\textwidth]{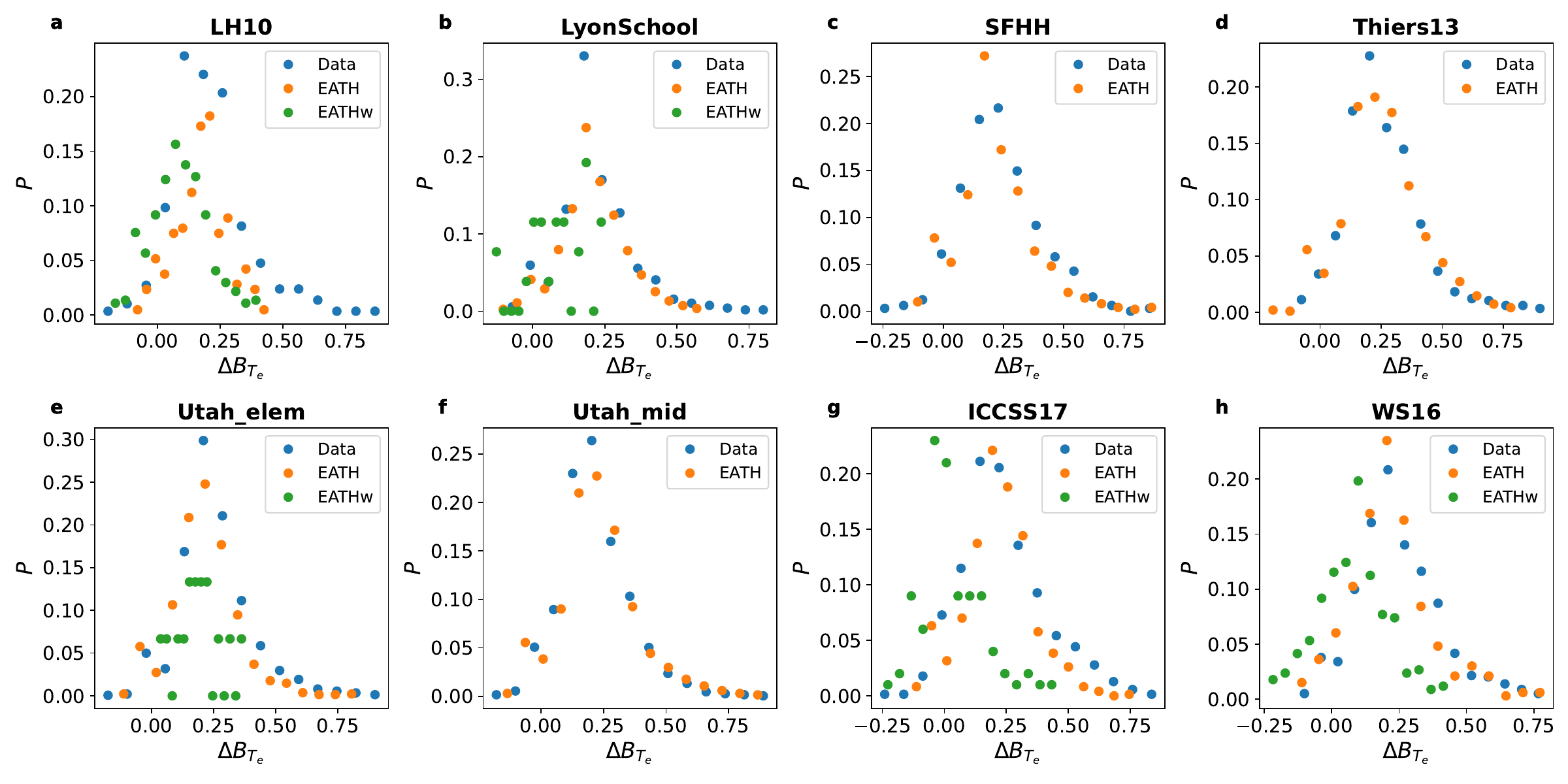}
\caption{\textbf{Hyperedges temporal heterogeneity - II.} For each hyperedge we evaluate the burstiness of their durations distribution $\Delta B_{T_e}$: in each panel we show the distribution of this hyperedge burstiness, $P(\Delta B_{T_e})$. We consider only the hyperedges with at least $10$ activation events. In each panel we consider an empirical system (see title) and the model with (EATH) and without (EATHw) memory. Note that in some datasets there are no points for the EATHw case, since no group activated enough times to define its burstiness parameters.}
\label{fig:figure25}
\end{figure*}

\begin{figure*}[ht!]
\includegraphics[width=\textwidth]{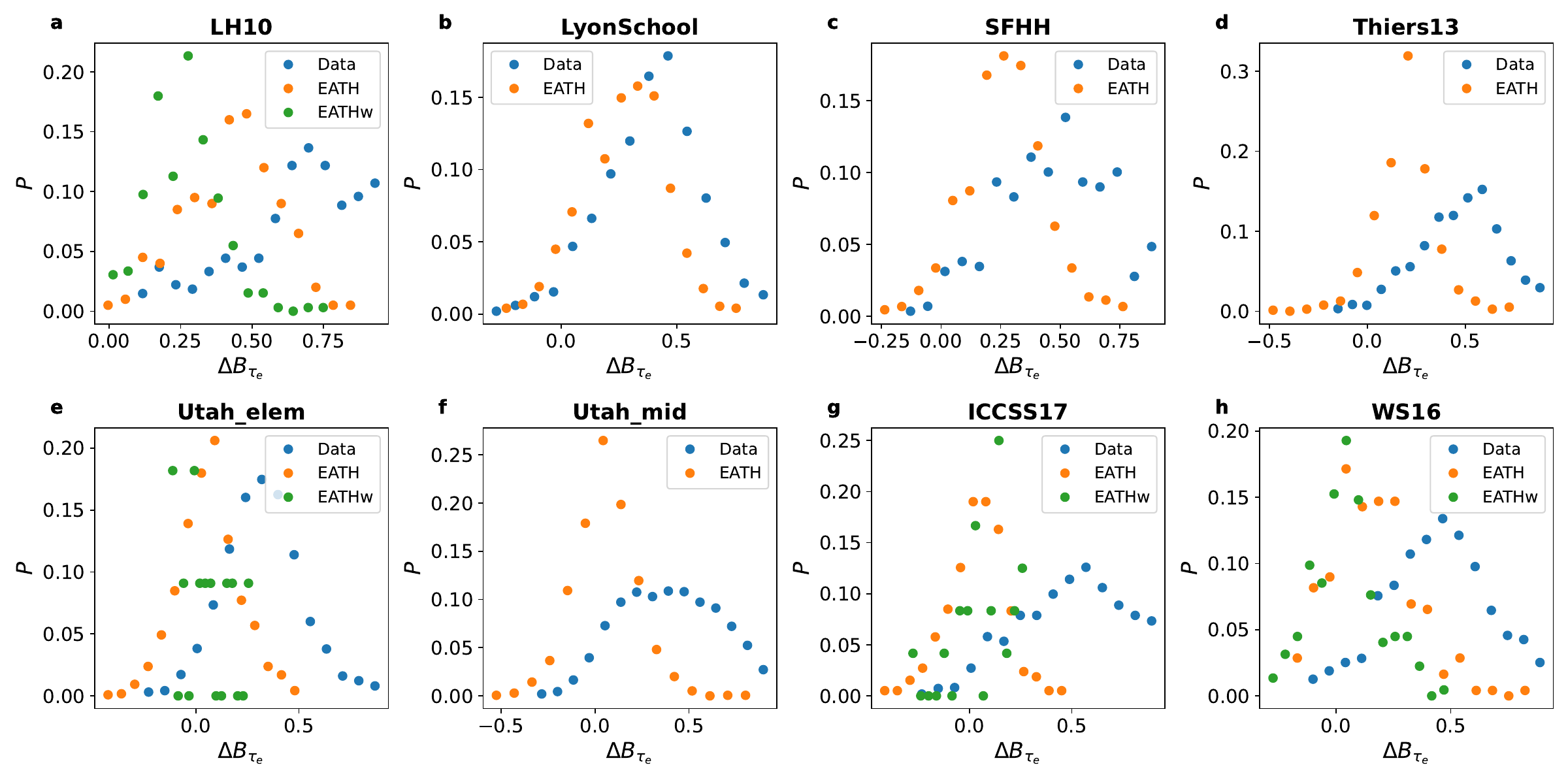}
\caption{\textbf{Hyperedges temporal heterogeneity - III.} For each hyperedge we evaluate the burstiness of their inter-event times distribution $\Delta B_{\tau_e}$: in each panel we show the distribution of this hyperedge burstiness, $P(\Delta B_{\tau_e})$. We consider only the hyperedges with at least $10$ activation events. In each panel we consider an empirical system (see title) and the model with (EATH) and without (EATHw) memory. Note that in some datasets there are no points for the EATHw case, since no group activated enough times to define its burstiness parameters.}
\label{fig:figure26}
\end{figure*}

\newpage
\begin{figure*}[ht!]
\includegraphics[width=\textwidth]{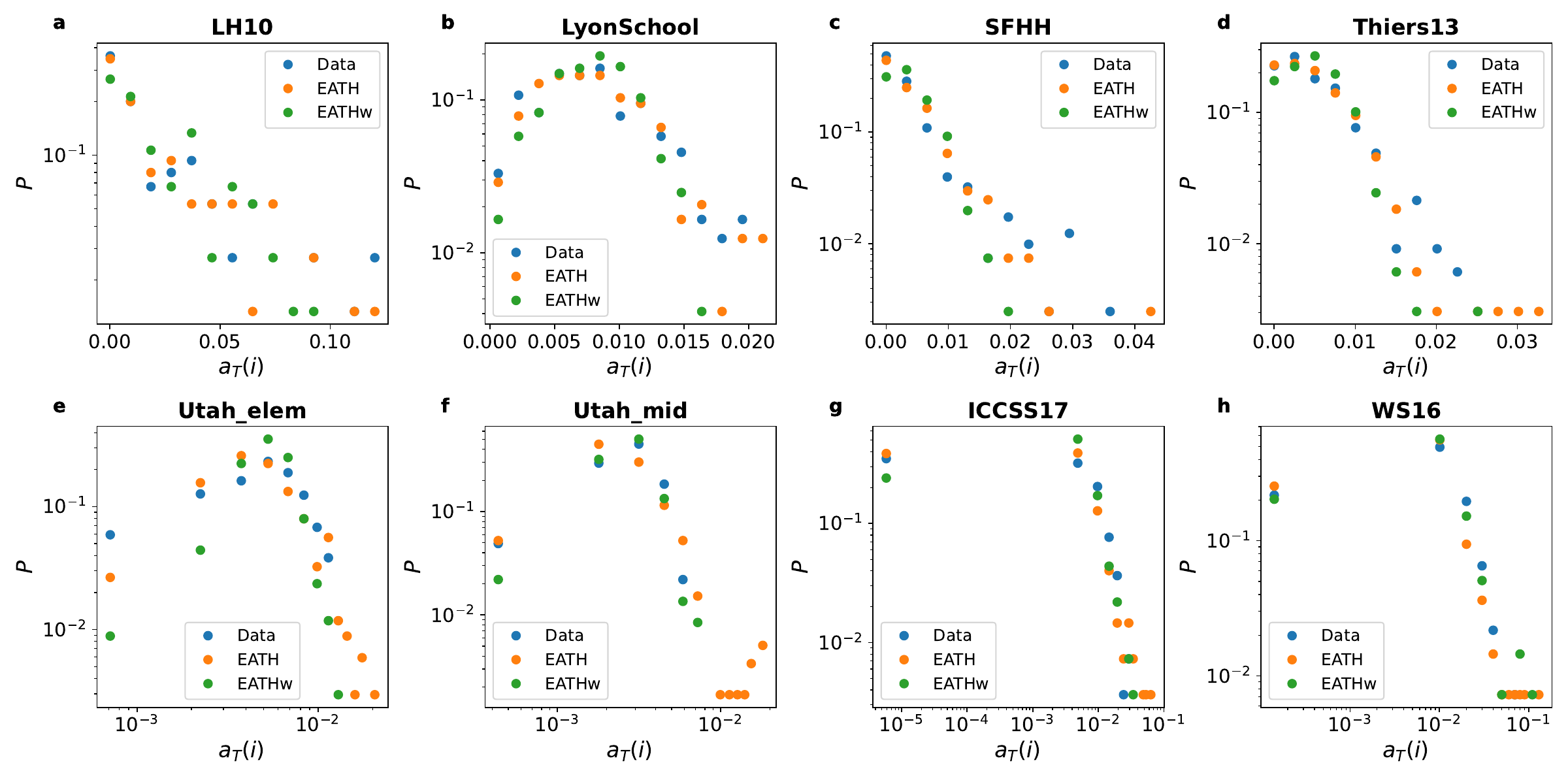}
\caption{\textbf{Nodes activity heterogeneity - I.} In each panel we show the distribution of nodes persistence activity $P(a_T(i))$, where the activity of each node is measured as described in the main text, in the empirical and generated hypergraphs. In each panel we consider an empirical system (see title) and the model with (EATH) and without (EATHw) memory.}
\label{fig:figure27}
\end{figure*}

\begin{figure*}[ht!]
\includegraphics[width=\textwidth]{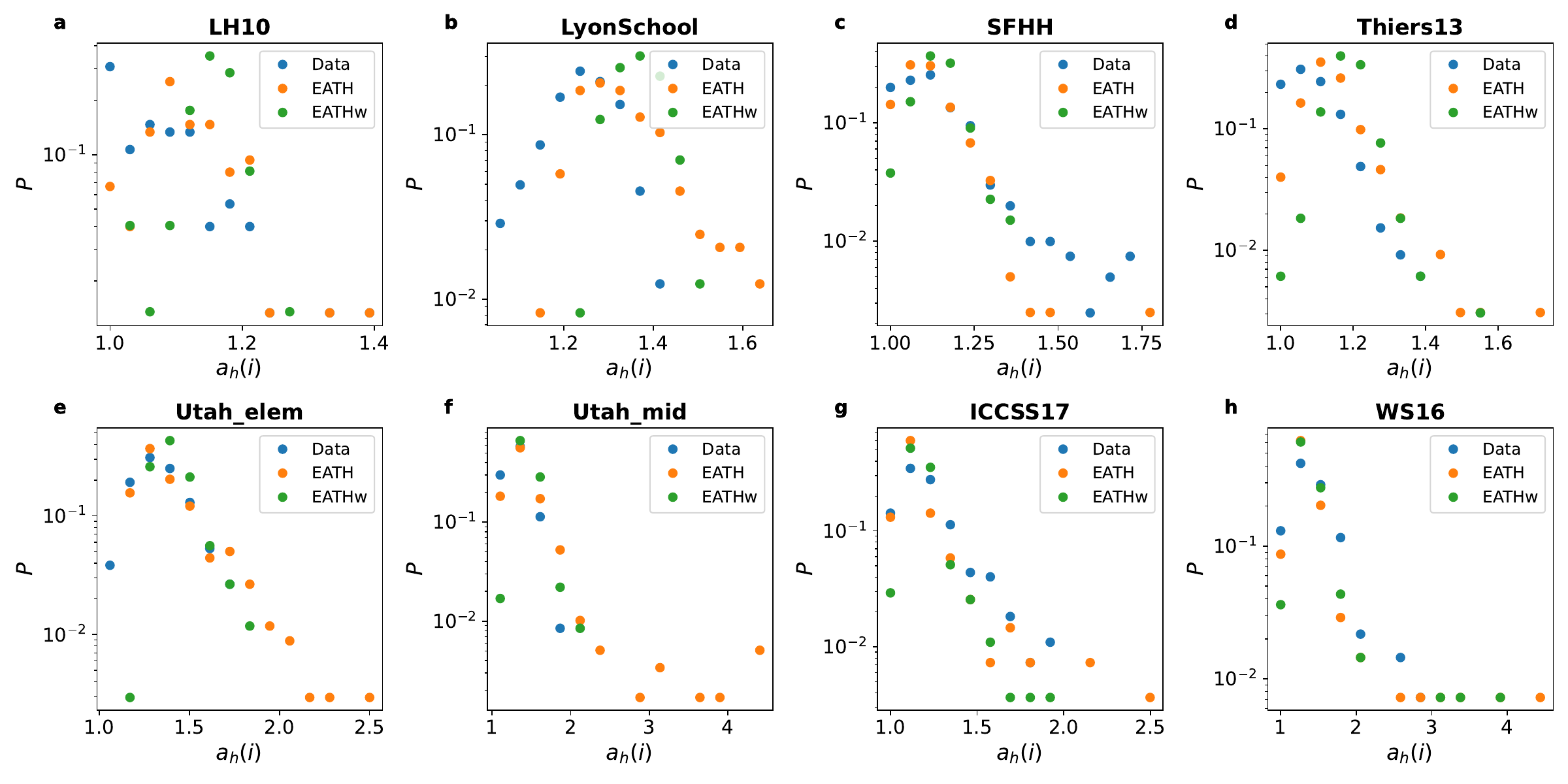}
\caption{\textbf{Nodes activity heterogeneity - II.} In each panel we show the distribution of nodes instantaneous activity $P(a_h(i))$, where the activity of each node is measured as described in the main text, in the empirical and generated hypergraphs. In each panel we consider an empirical system (see title) and the model with (EATH) and without (EATHw) memory.}
\label{fig:figure23}
\end{figure*}

\newpage
\subsection{Nodes participation at different orders}
\label{sez:section1d}
\begin{figure*}[ht!]
\includegraphics[width=\textwidth]{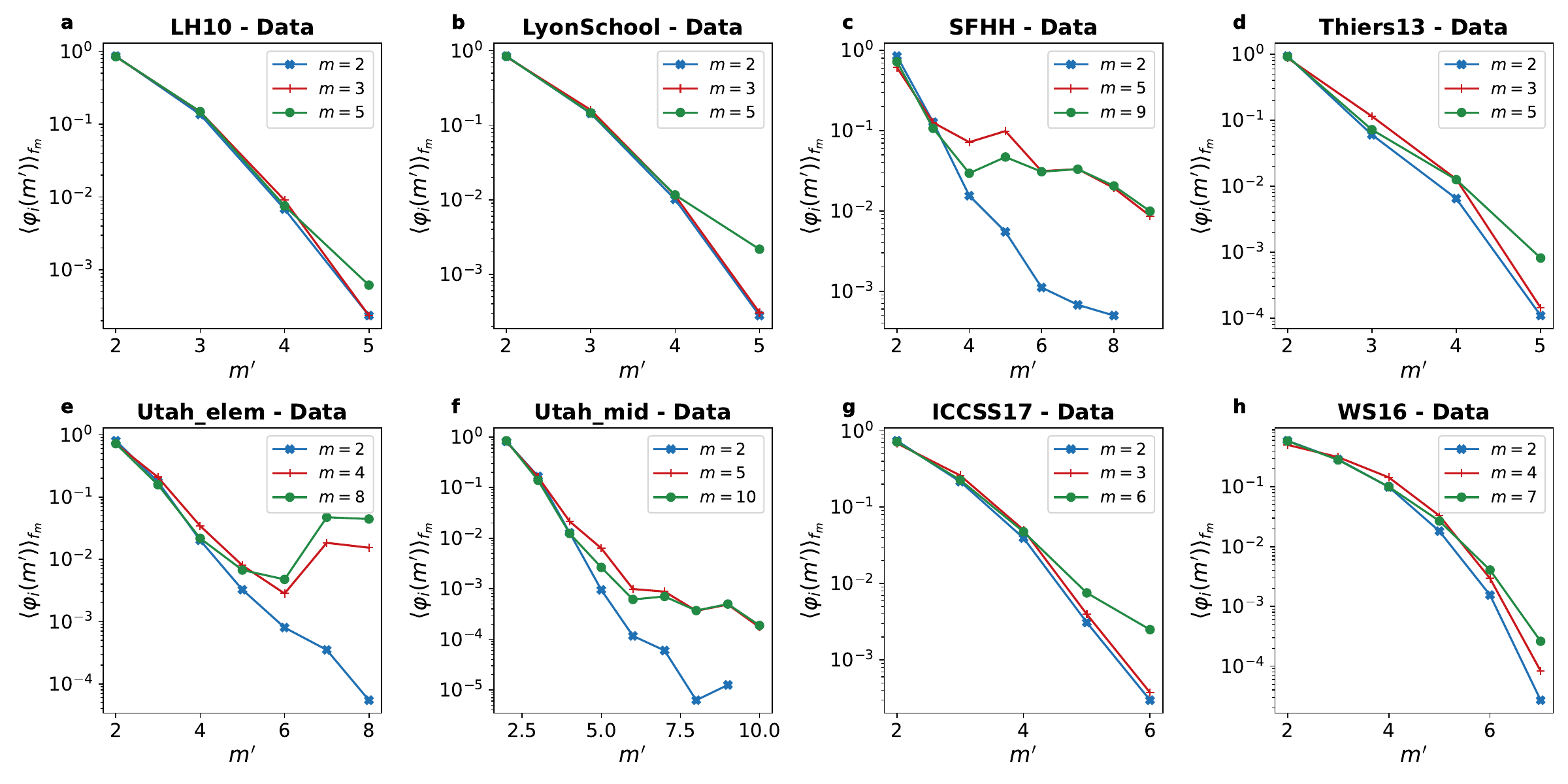}
\caption{\textbf{Order propensity - Data.} In each panel we show the average order propensity $\langle \varphi_i(m') \rangle_{f_m}$ as a function of $m'$, averaged over the nodes occupying the top $f N$ positions of node rankings obtained considering the time spent by nodes interacting at order $m$, for different $m$ (see legend). In this case, we consider $f=0.1$ and we consider the empirical systems (see title).}
\label{fig:figure31_0}
\end{figure*}

\begin{figure*}[ht!]
\includegraphics[width=\textwidth]{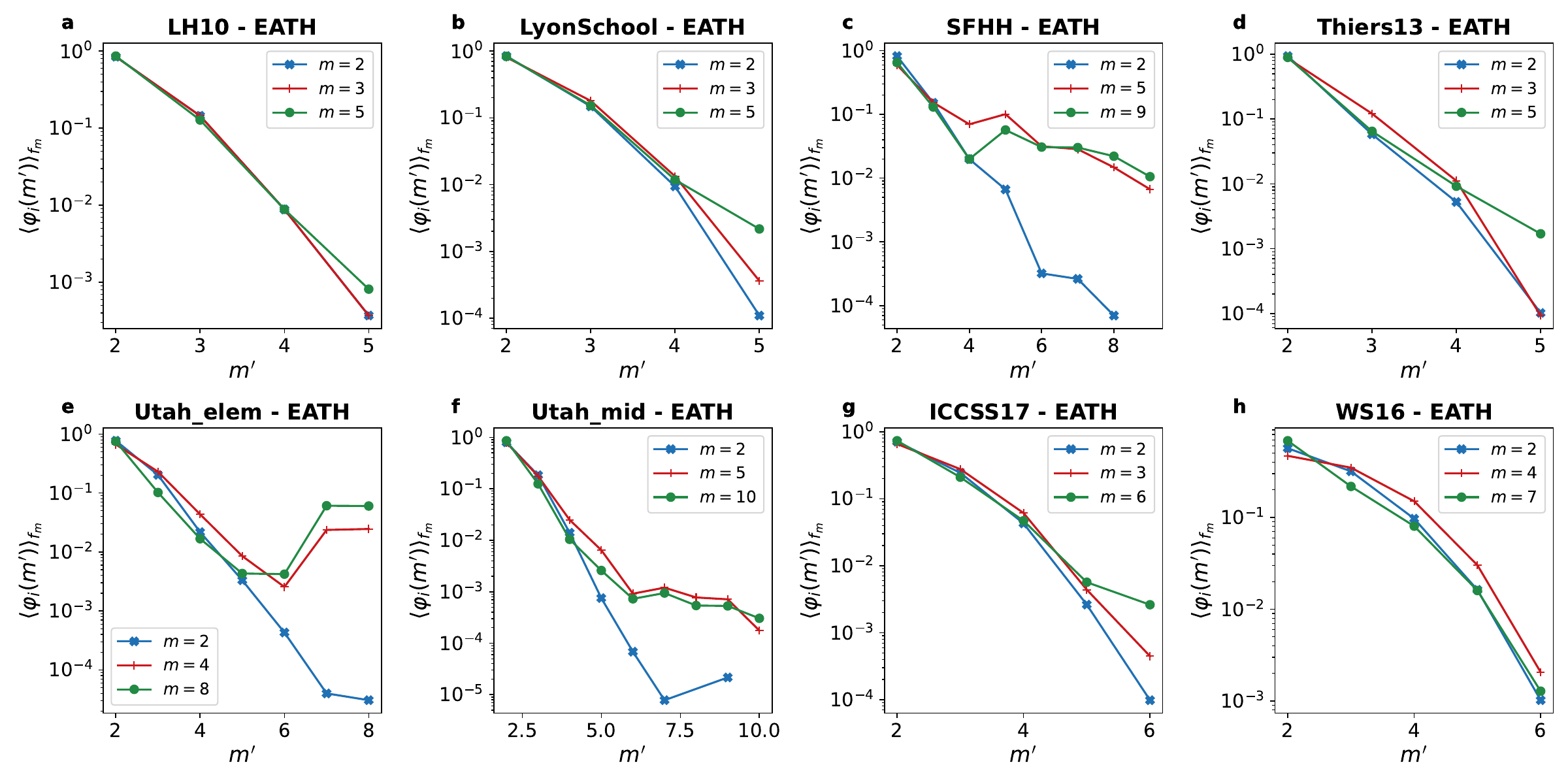}
\caption{\textbf{Order propensity - EATH.} In each panel we show the average order propensity $\langle \varphi_i(m') \rangle_{f_m}$ as a function of $m'$, averaged over the nodes occupying the top $f N$ positions of node rankings obtained considering the time spent by nodes interacting at order $m$, for different $m$ (see legend). In this case, we consider $f=0.1$ and we consider the hypergraphs generated with the EATH model (see title).}
\label{fig:figure31_1}
\end{figure*}

\begin{figure*}[ht!]
\includegraphics[width=\textwidth]{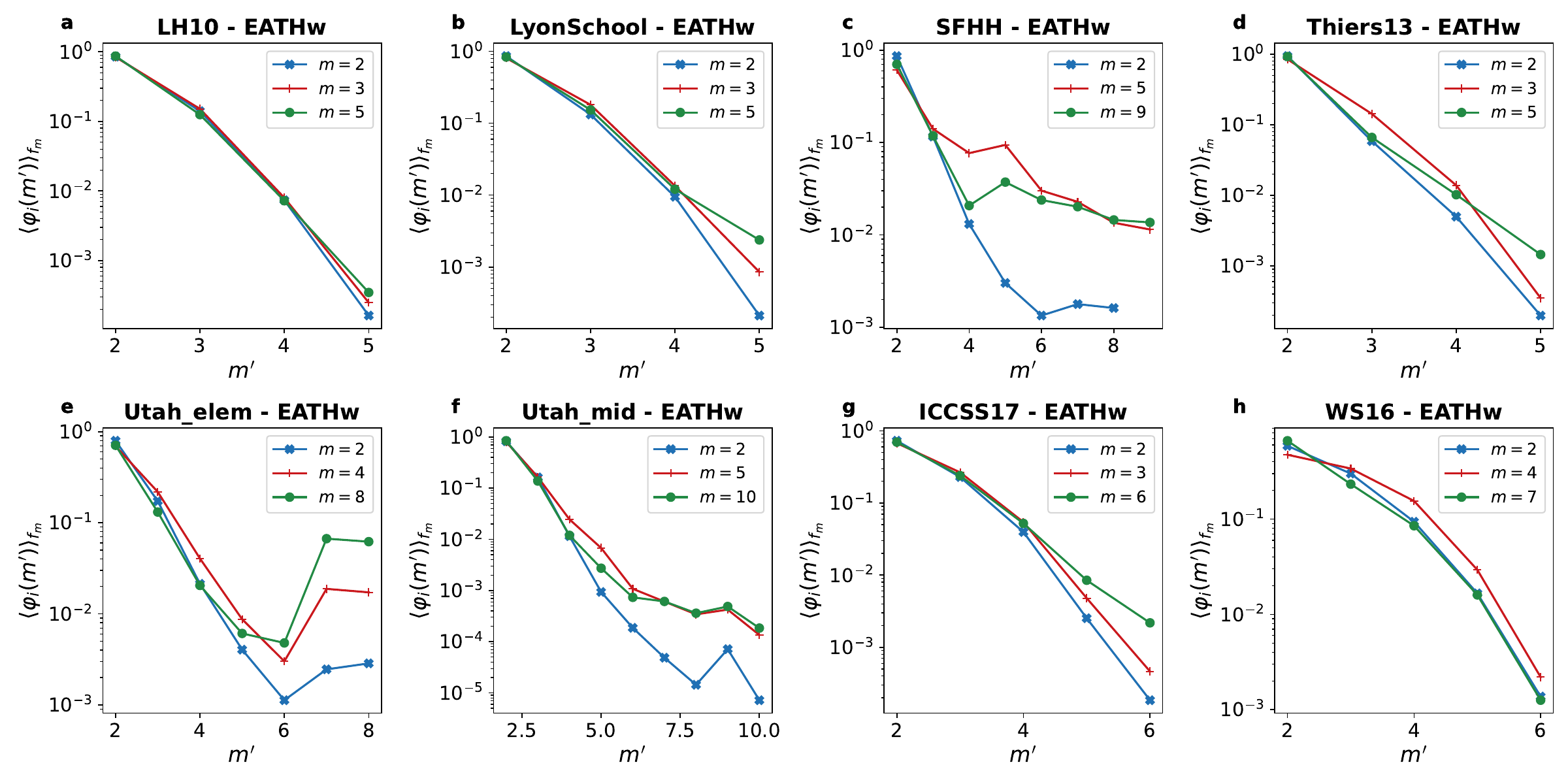}
\caption{\textbf{Order propensity - EATHw.} In each panel we show the average order propensity $\langle \varphi_i(m') \rangle_{f_m}$ as a function of $m'$, averaged over the nodes occupying the top $f N$ positions of node rankings obtained considering the time spent by nodes interacting at order $m$, for different $m$ (see legend). In this case, we consider $f=0.1$ and we consider the hypergraphs generated with the EATH model without memory (EATHw) (see title).}
\label{fig:figure31_2}
\end{figure*}

\begin{figure*}[ht!]
\includegraphics[width=\textwidth]{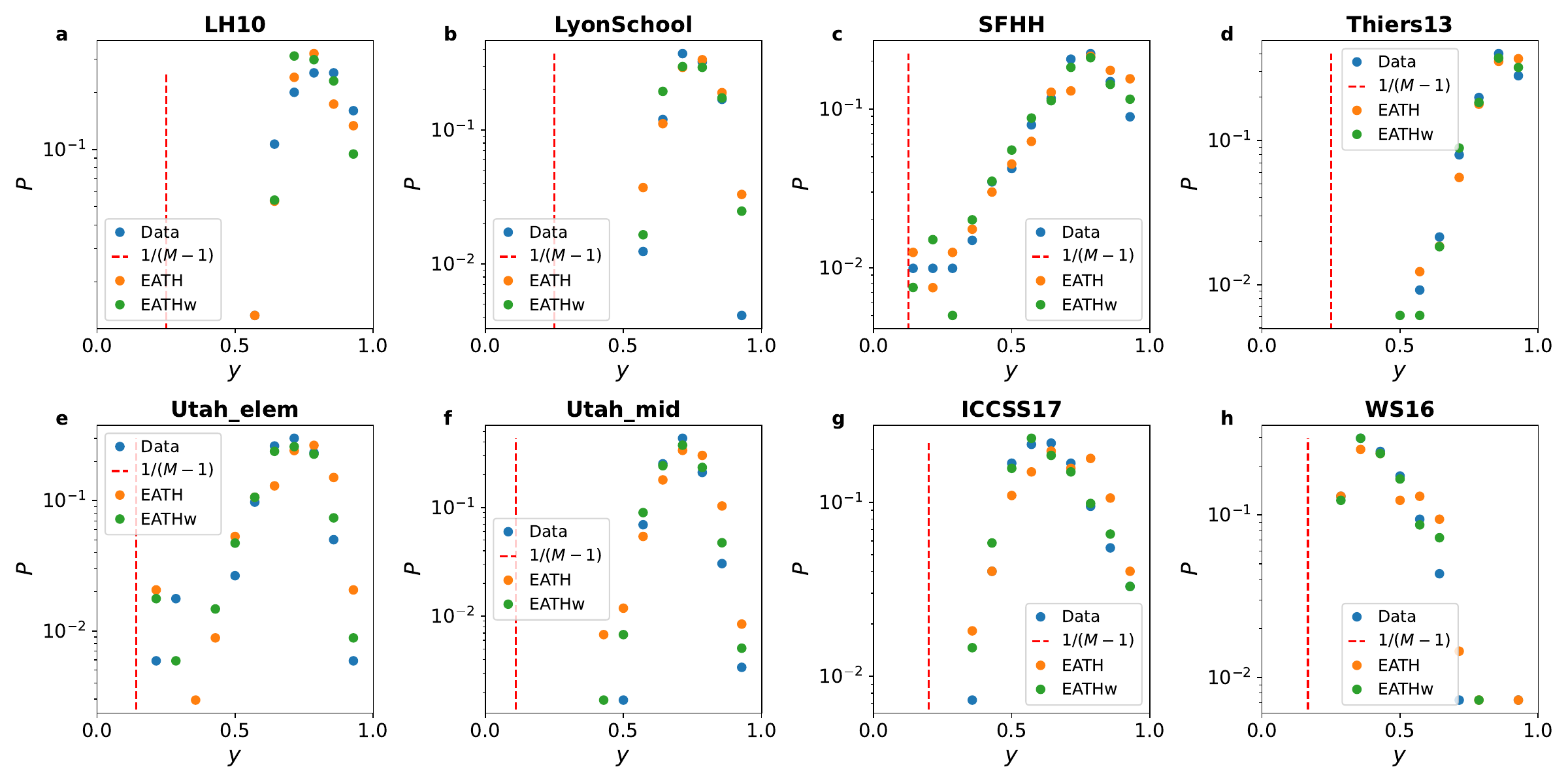}
\caption{\textbf{Nodes participation ratio.} In each panel we show the distribution of the nodes participation ratio, $P(y)$, where $y(i)=\sum_{m=2}^M \varphi_i(m)^2$. The red dashed line correspond to $y=1/(M-1)$. In each panel we consider an empirical system (see title) and the model with (EATH) and without (EATHw) memory.}
\label{fig:figure30}
\end{figure*}

Here we focus on the patterns of nodes participation at different orders of interactions. Supplementary Figs. \ref{fig:figure31_0}-\ref{fig:figure31_2} shows the average order propensity $\langle \varphi_i(m') \rangle_{f_m}$ as a function of $m'$, averaged over different groups of nodes (see caption). Supplementary Fig. \ref{fig:figure30} show the distribution of the nodes participation ratio $P(y)$, where $y(i)=\sum_{m=2}^M \varphi_i(m)^2$. In Supplementary Figs. \ref{fig:figure28}, \ref{fig:figure29} we show the correlation $\rho(m,m')$ between the nodes rankings obtained at orders $m$ and $m'$, by considering the time the nodes spent interacting at that order. 

\newpage

\newpage
\begin{figure*}[ht!]
\includegraphics[width=\textwidth]{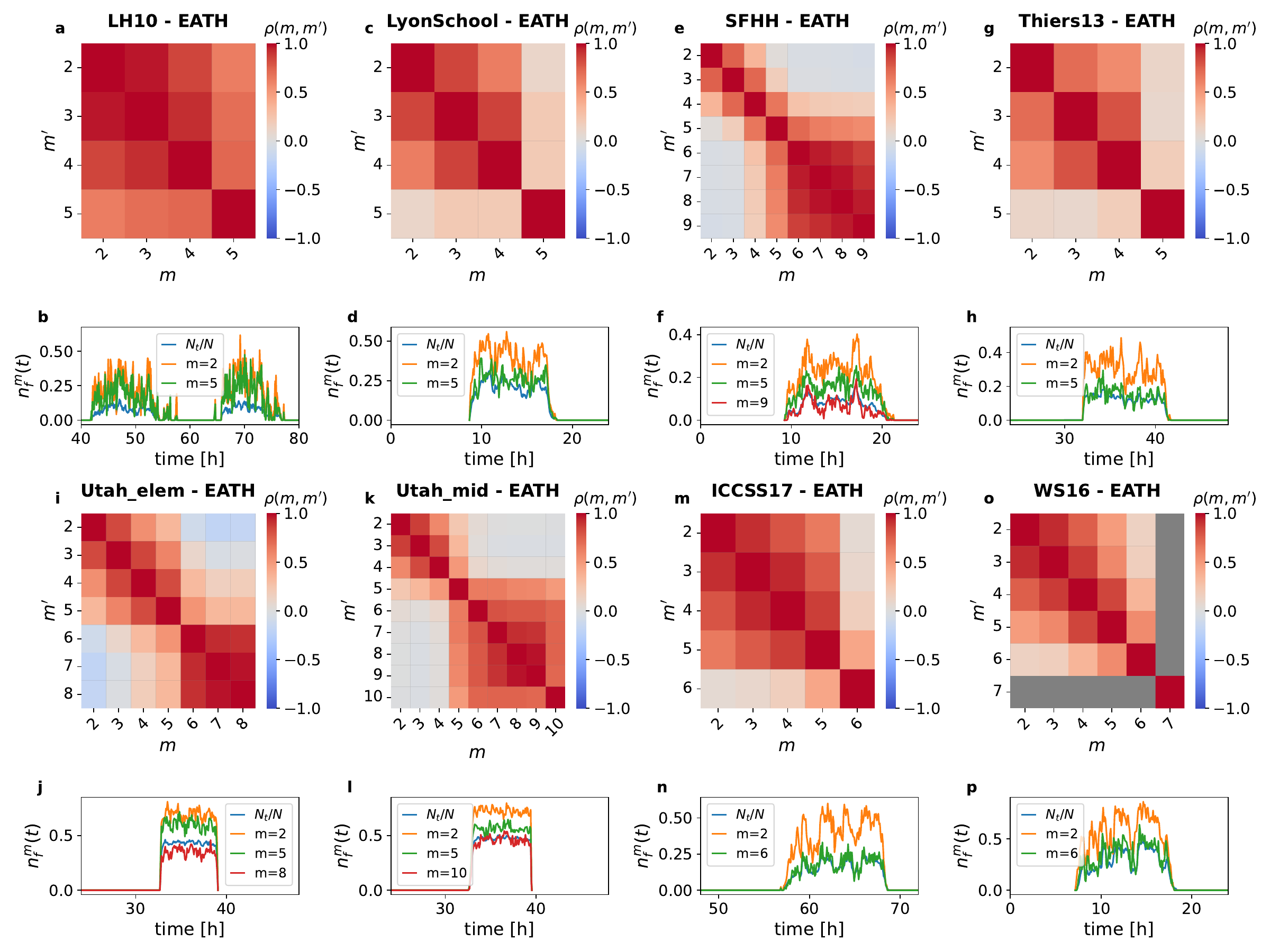}
\caption{\textbf{Participation of nodes at different interaction orders - EATH.} For each dataset (see title), at each order $m$ we rank the nodes based on the time they spend interacting at size $m$, and we estimate the fraction $n_f^m(t)$ of nodes active at time $t$ (in hyperedges of any size), among the nodes occupying the top $f N$ positions of the nodes ranking at order $m$. We show the Pearson's correlation coefficient $\rho(m,m')$ between the rankings obtained at order $m$ and $m'$ and the temporal evolution of $n_f^m(t)$ for different $m$ (see legend), fixing $f=0.1$; we also plot the total fraction $N_t/N$ of active nodes in the population at time $t$. In this case we consider the hypergraphs generated with the EATH model with memory.}
\label{fig:figure28}
\end{figure*}

\newpage
\begin{figure*}[ht!]
\includegraphics[width=\textwidth]{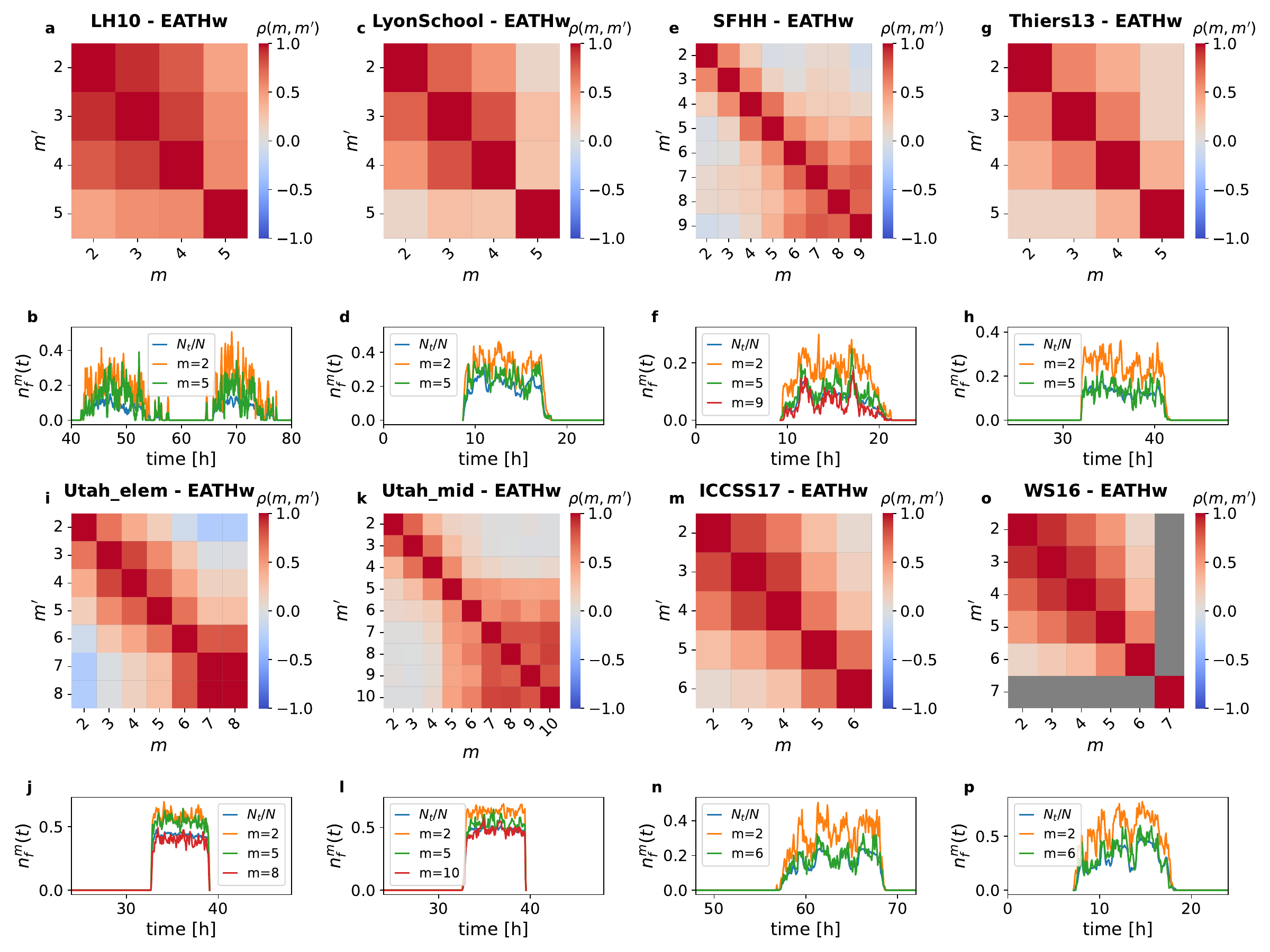}
\caption{\textbf{Participation of nodes at different interaction orders - EATHw.} For each dataset (see title), at each order $m$ we rank the nodes based on the time they spend interacting at size $m$, and we estimate the fraction $n_f^m(t)$ of nodes active at time $t$ (in hyperedges of any size), among the nodes occupying the top $f N$ positions of the nodes ranking at order $m$. We show the Pearson's correlation coefficient $\rho(m,m')$ between the rankings obtained at order $m$ and $m'$ and the temporal evolution of $n_f^m(t)$ for different $m$ (see legend), fixing $f=0.1$; we also plot the total fraction $N_t/N$ of active nodes in the population at time $t$. In this case we consider the hypergraphs generated with the EATH model without memory (EATHw).}
\label{fig:figure29}
\end{figure*}

\newpage
\subsection{Groups aggregation and disaggregation dynamics}
\label{sez:section1e}
\begin{figure*}[ht!]
\includegraphics[width=0.8\textwidth]{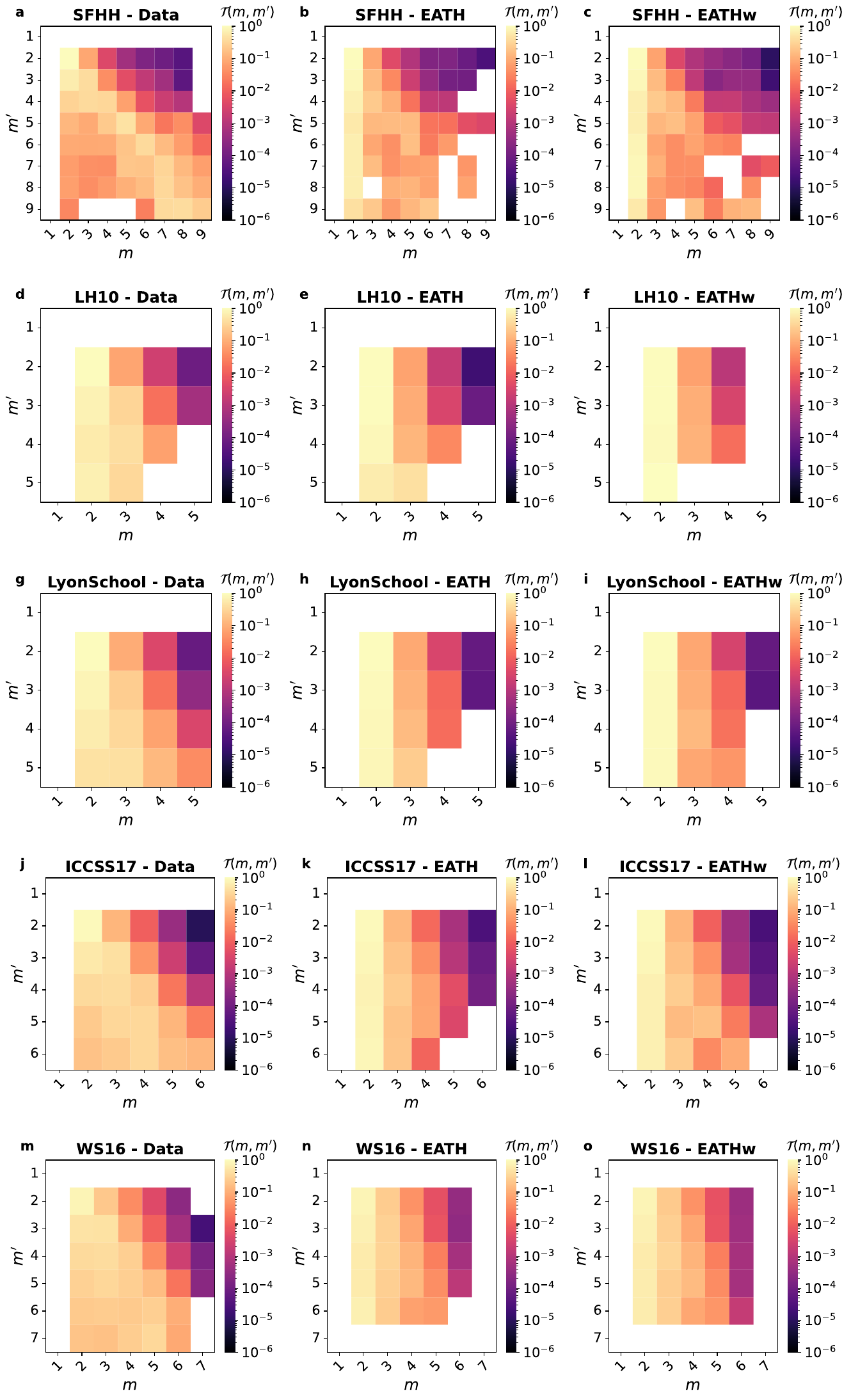}
\caption{\textbf{Transition matrix.} In each panel we show the transition matrix $\mathcal{T}(m,m')$, where each element is the conditional probability that a node that is member of a group of size $m'$ at time $t$ is next member of a different group of size $m$ at time $t + 1$ - given that it undergoes a group change between $t$ and $t + 1$. Each row correspond to a dataset (see title) and we consider the empirical system and the model with (EATH) and without (EATHw) memory.}
\label{fig:figure32_0}
\end{figure*}

\begin{figure*}[ht!]
\includegraphics[width=0.8\textwidth]{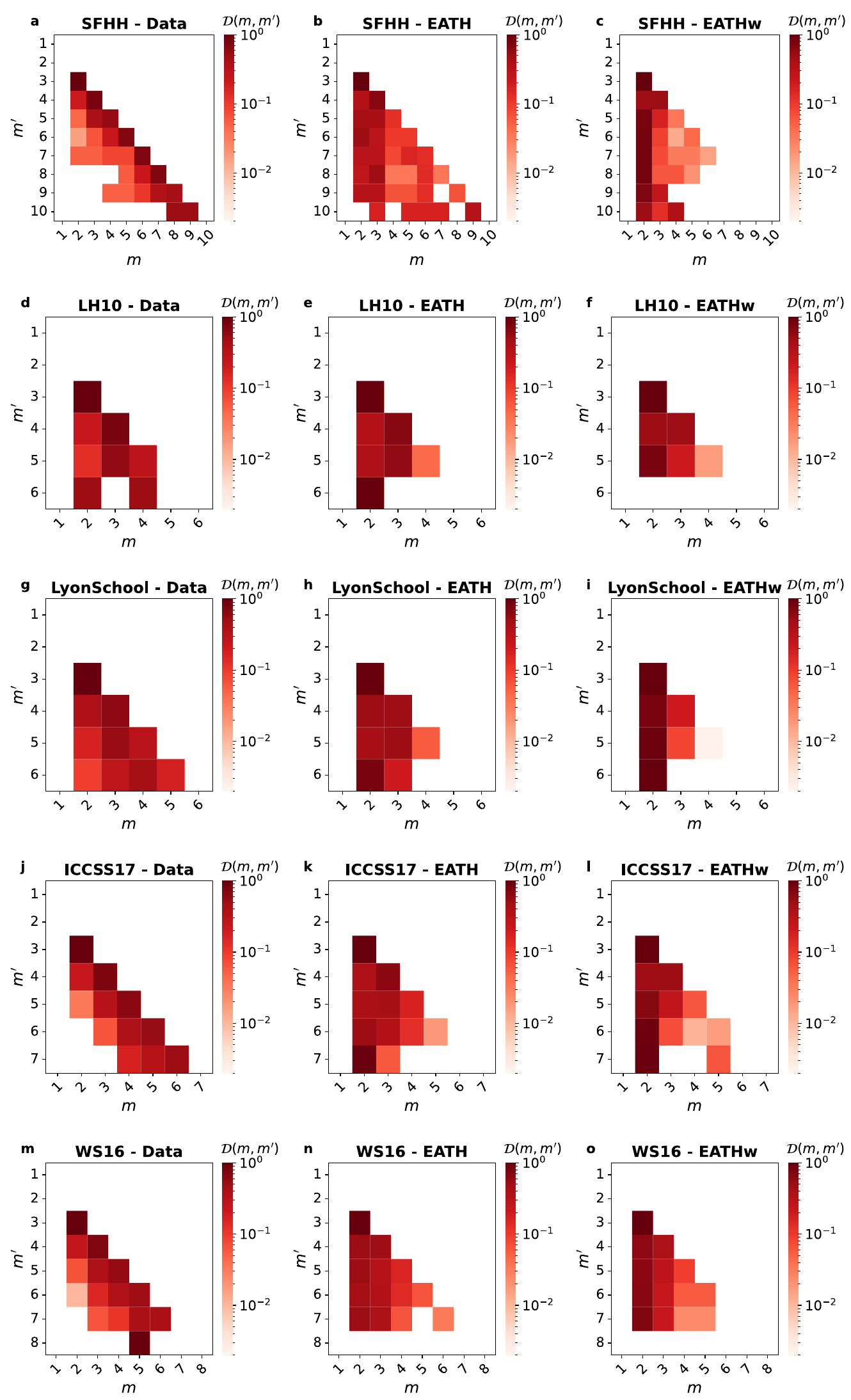}
\caption{\textbf{Disaggregation matrix.} In each panel we show the disaggregation matrix $\mathcal{D}(m,m')$, where each element is the probability that the largest sub-group leaving a group of size $m'$ is of size $m$. Each row correspond to a dataset (see title) and we consider the empirical system and the model with (EATH) and without (EATHw) memory.}
\label{fig:figure32_1}
\end{figure*}

\begin{figure*}[ht!]
\includegraphics[width=0.8\textwidth]{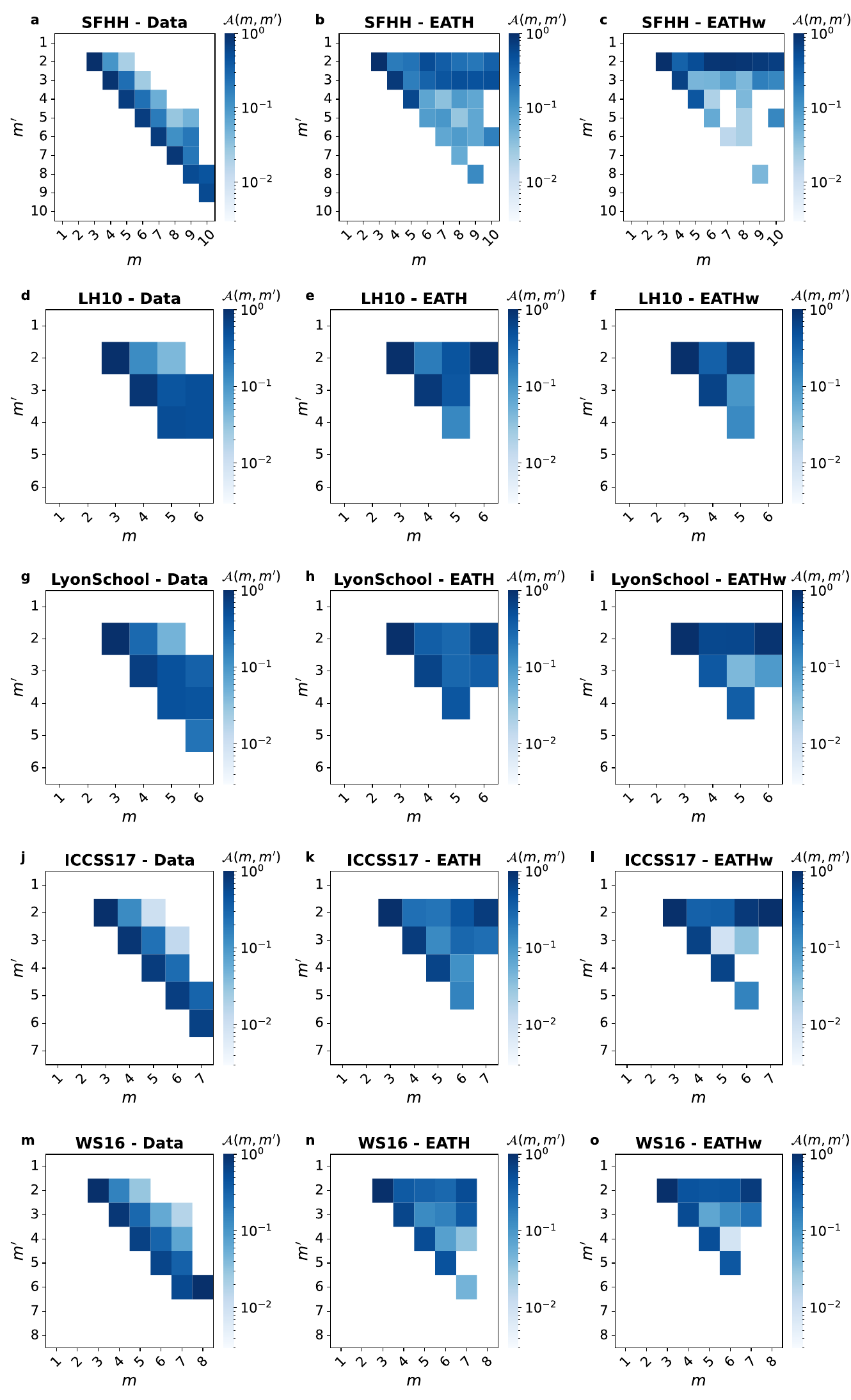}
\caption{\textbf{Aggregation matrix.} In each panel we show the aggregation matrix $\mathcal{A}(m,m')$, where each element is the probability that the largest sub-group joining a group of size $m'$ is of size $m$. Each row correspond to a dataset (see title) and we consider the empirical system and the model with (EATH) and without (EATHw) memory.}
\label{fig:figure32_2}
\end{figure*}

Here we focus on node movements across groups, and on the dynamics of group aggregation and disaggregation \cite{Iacopini2024}, in the empirical and generated hypergraphs. Supplementary Fig. \ref{fig:figure32_0} shows the transition matrix $\mathcal{T}(m,m')$, i.e. the conditional probability that a node that is member of a group of size $m'$ at time $t$ is next member of a different group of size $m$ at time $t + 1$ - given that it undergoes a group change between $t$ and $t + 1$. Supplementary Fig. \ref{fig:figure32_1} shows the disaggregation matrix $\mathcal{D}(m,m')$, i.e. the probability that the largest sub-group leaving a group of size $m'$ is of size $m$. Supplementary Fig. \ref{fig:figure32_2} shows the aggregation matrix $\mathcal{A}(m,m')$, i.e. the probability that the largest sub-group joining a group of size $m'$ is of size $m$ (see Supplementary Ref. \cite{Iacopini2024} for more details).

\newpage

\newpage

\newpage

\newpage
\subsection{Higher-order SIR dynamics}
\label{sez:section1f}
Here we focus on the results of numerical simulations of the higher-order SIR dynamical process \cite{St-Onge2022} (see Methods) for different datasets than the one reported in the main (see Supplementary Figs. \ref{fig:figure33_0}-\ref{fig:figure33_3}).

\begin{figure*}[ht!]
\includegraphics[width=\textwidth]{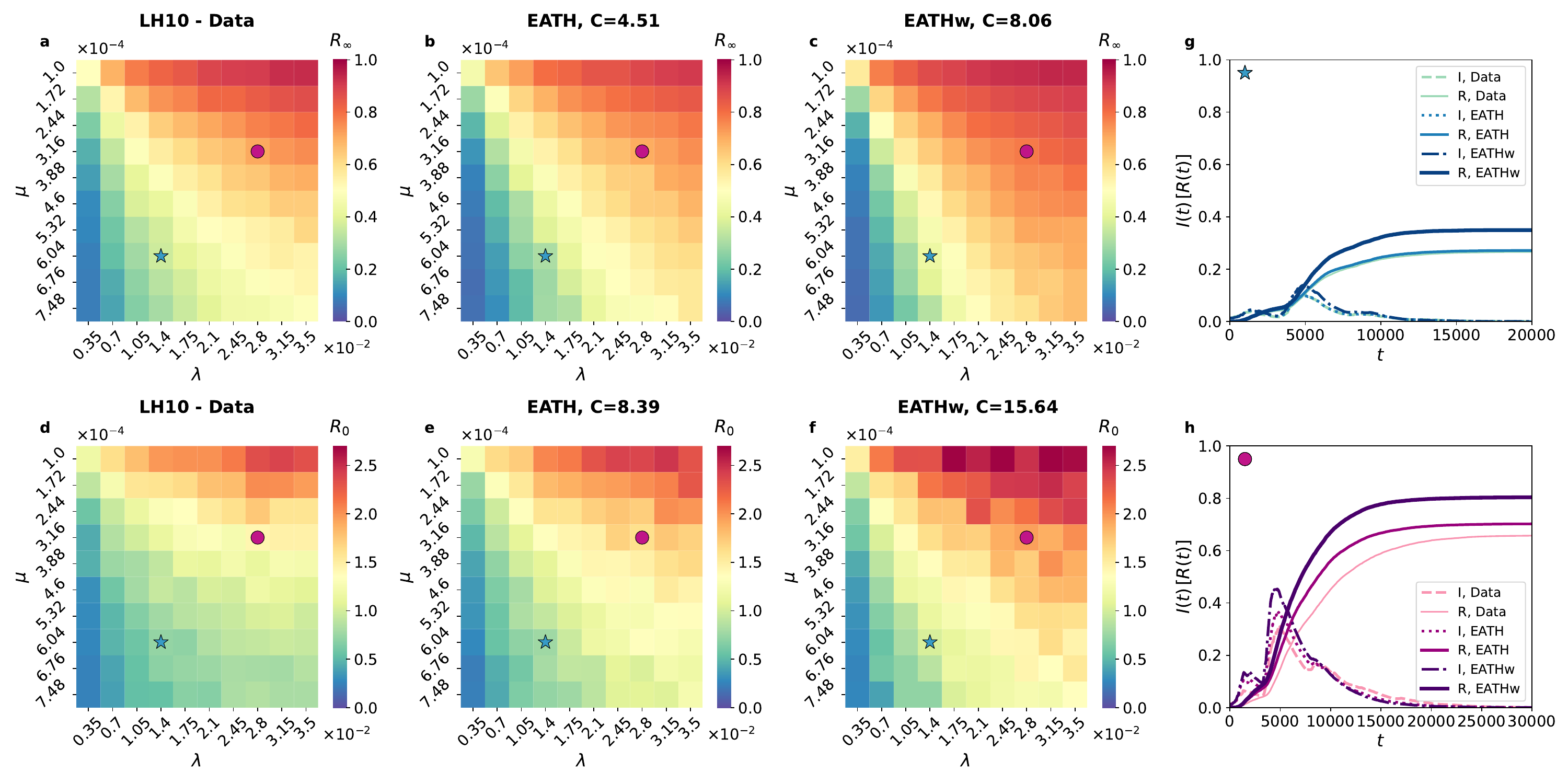}
\caption{\textbf{Higher-order SIR dynamics - LH10.} Panels \textbf{a}-\textbf{c} and \textbf{d}-\textbf{f} show respectively the epidemic final-size $R_{\infty}$ and the basic reproduction number $R_0$ as a function of the epidemiological parameters $(\lambda,\mu)$. The results are obtained by averaging over 400 simulations, fixing $\nu=4$, and considering the empirical hypergraph (Data) and the hypergraphs generated using the model with (EATH) and without memory (EATHw). $C$ indicates the Canberra distance between the empirical and synthetic matrices. In panels \textbf{g},\textbf{h} we show the fraction of infected $I(t)$ and recovered $R(t)$ nodes as a function of time, when the initial seed is infected at $t_0=0$ and averaging the curves over 200 realizations: panel \textbf{g}, $(\lambda,\mu)=(1.4 \, 10^{-2},6.04 \, 10^{-4})$ (see the blue star in panels \textbf{a}-\textbf{f}); panel \textbf{h}, $(\lambda,\mu)=(2.8 \, 10^{-2},3.16 \, 10^{-4})$ (see the purple circle in panels \textbf{a}-\textbf{f}).}
\label{fig:figure33_0}
\end{figure*}

\begin{figure*}[ht!]
\includegraphics[width=\textwidth]{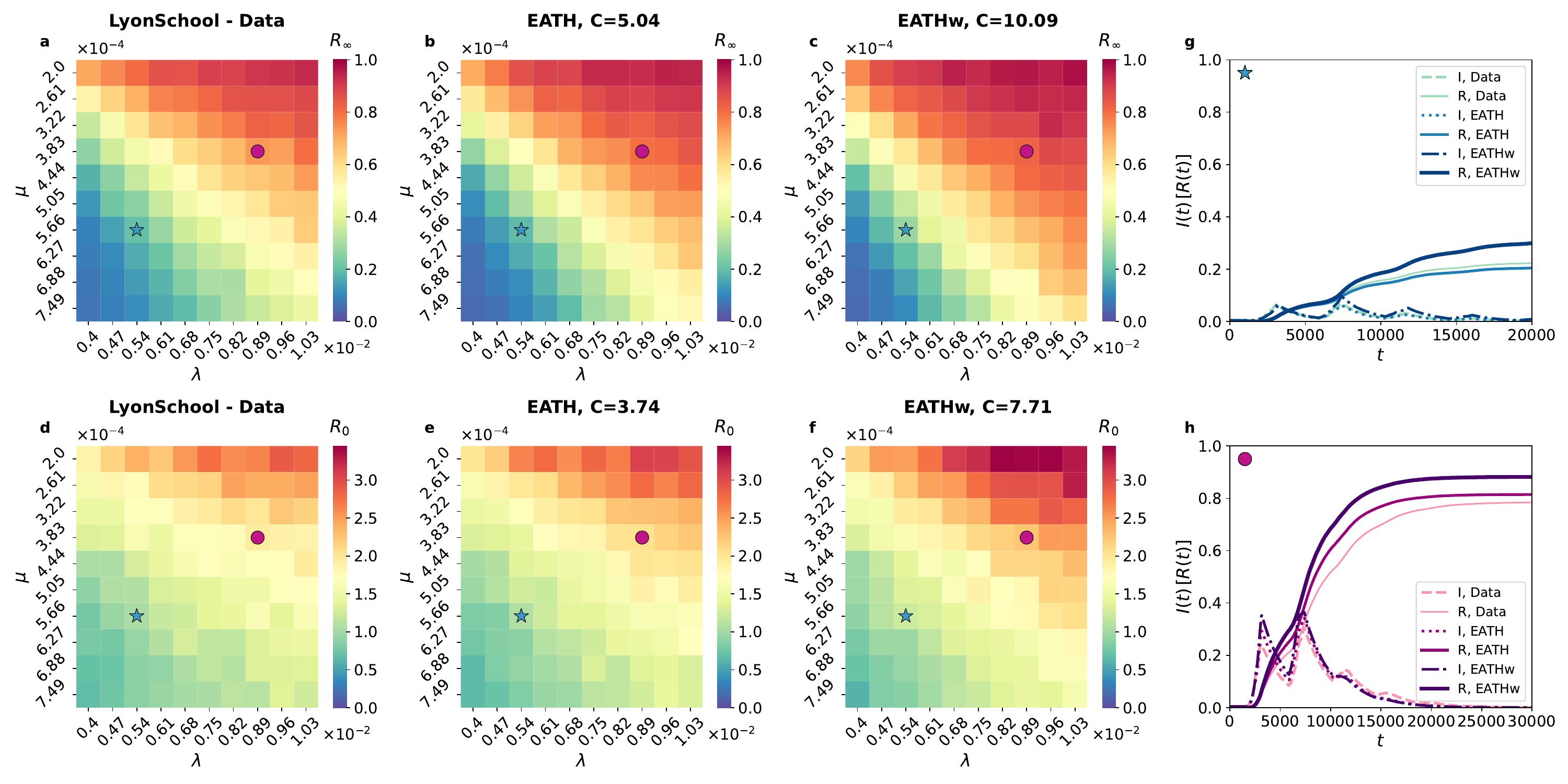}
\caption{\textbf{Higher-order SIR dynamics - LyonSchool.} The results are obtained as described in Supplementary Fig. \ref{fig:figure33_0}, but in this case in panel \textbf{g}, $(\lambda,\mu)=(0.54 \, 10^{-2},5.66 \, 10^{-4})$ (see the blue star in panels \textbf{a}-\textbf{f}); panel \textbf{h}, $(\lambda,\mu)=(0.89 \, 10^{-2},3.83 \, 10^{-4})$ (see the purple circle in panels \textbf{a}-\textbf{f}).}
\label{fig:figure33_1}
\end{figure*}

\newpage
\begin{figure*}[ht!]
\includegraphics[width=\textwidth]{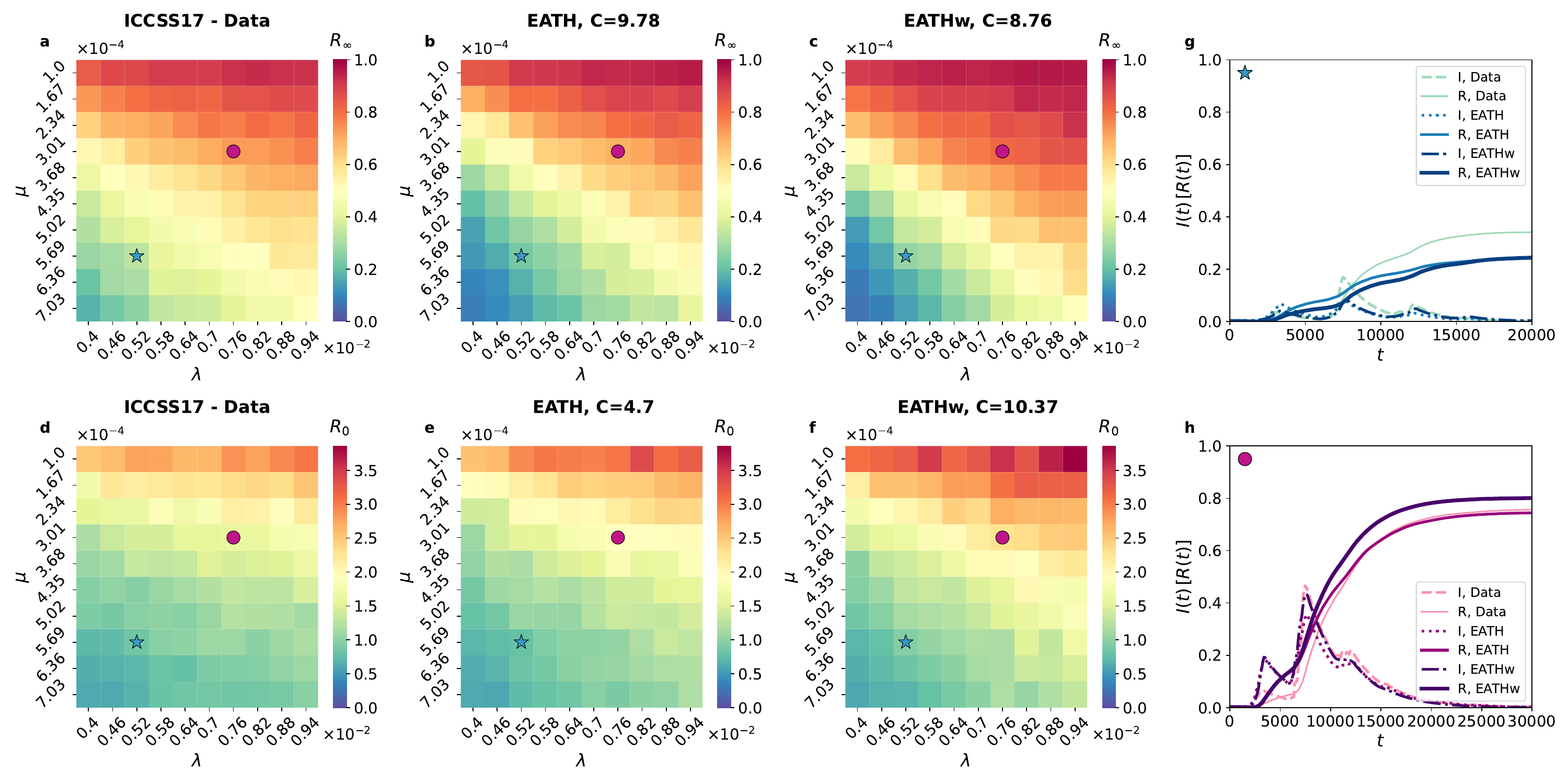}
\caption{\textbf{Higher-order SIR dynamics - ICCSS17.} The results are obtained as described in Supplementary Fig. \ref{fig:figure33_0}, but in this case in panel \textbf{g}, $(\lambda,\mu)=(0.52 \, 10^{-2},5.69 \, 10^{-4})$ (see the blue star in panels \textbf{a}-\textbf{f}); panel \textbf{h}, $(\lambda,\mu)=(0.76 \, 10^{-2},3.01 \, 10^{-4})$ (see the purple circle in panels \textbf{a}-\textbf{f}).}
\label{fig:figure33_2}
\end{figure*}

\begin{figure*}[ht!]
\includegraphics[width=\textwidth]{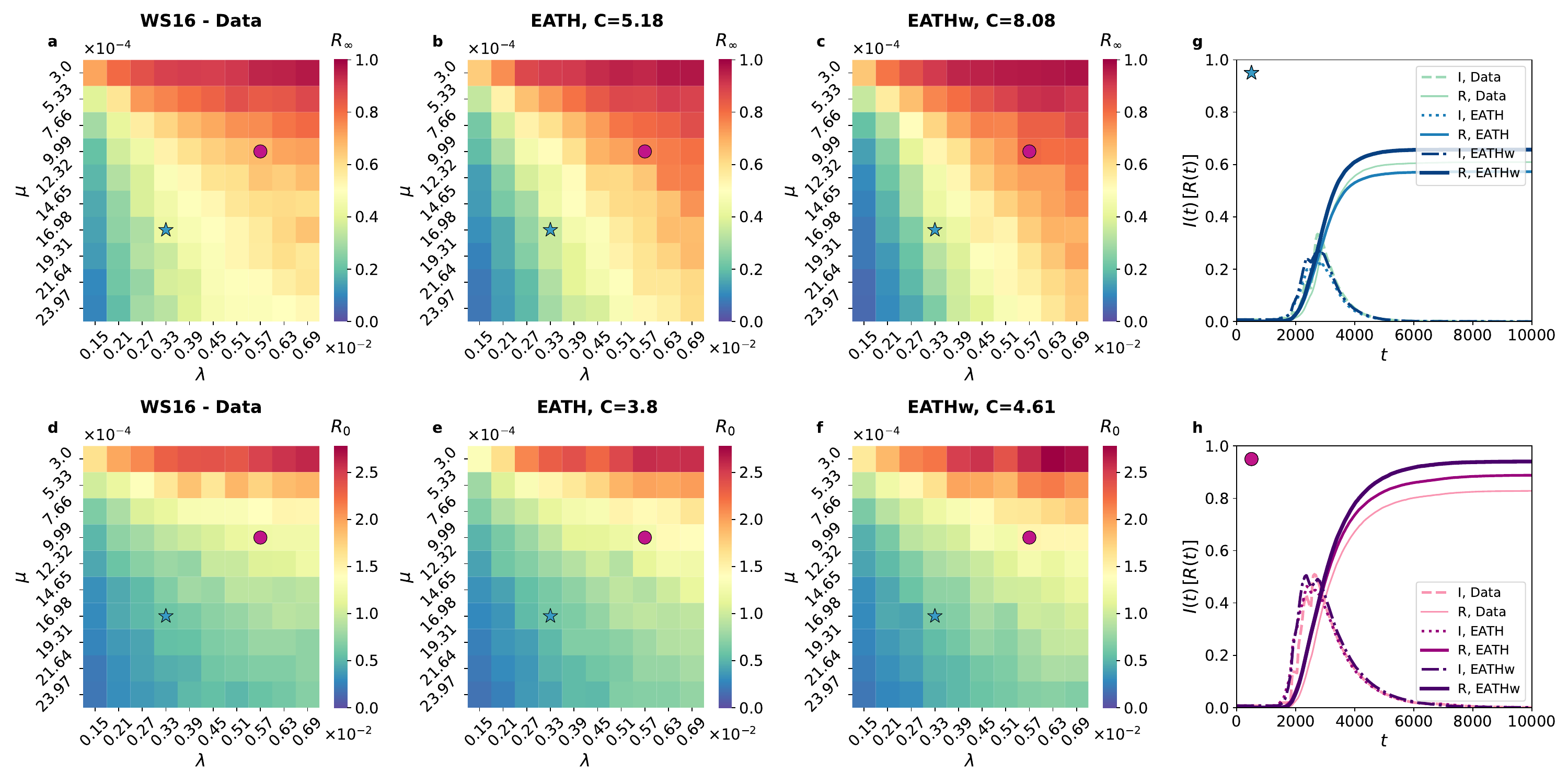}
\caption{\textbf{Higher-order SIR dynamics - WS16.} The results are obtained as described in Supplementary Fig. \ref{fig:figure33_0}, but in this case in panel \textbf{g}, $(\lambda,\mu)=(0.33 \, 10^{-2},16.98 \, 10^{-4})$ (see the blue star in panels \textbf{a}-\textbf{f}); panel \textbf{h}, $(\lambda,\mu)=(0.57 \, 10^{-2},9.99 \, 10^{-4})$ (see the purple circle in panels \textbf{a}-\textbf{f}).}
\label{fig:figure33_3}
\end{figure*}

\newpage
\section{Hybrid temporal hypergraphs generation}
\label{sez:section2}
We present further results on the generation of hybrid hypergraphs: in particular, we consider both the \texttt{Hybrid\_sub} and \texttt{Hybrid\_uni} generation procedures, considering the hospital, LH10, and the conference datasets, WS16, both as source of the population and system features (see captions of Supplementary Figs. \ref{fig:figure34a}, \ref{fig:figure34b}). In Supplementary Table \ref{tab:table3} we report the burstiness values, $\Delta B$, for the temporal distributions of the hybrid systems generated (and also for the hybrid hypergraphs shown in the main text).

\begin{table}[h!]
    \begin{tabular}{c|c|c|c|c}
    \hline
    \hline
         & \multicolumn{2}{c}{Nodes} \vline & \multicolumn{2}{c}{Hyperedges} \\ \hline
         & $\Delta B_{T}$ & $\Delta B_{\tau}$ & $\Delta B_{T}$ & $\Delta B_{\tau}$ \\ \hline
         Hybrid\_sub - WS16 ($d_1$) + Utah\_elem ($d_2$) & 0.35 &0.45 & 0.48 & 0.20 \\
         Hybrid\_uni - WS16 ($d_1$) + Utah\_elem ($d_2$) & 0.47 & 0.57 & 0.61 & 0.21 \\
         Hybrid\_sub - WS16 ($d_1$) + LH10 ($d_2$) & 0.36 &0.64 & 0.43 & 0.19 \\
         Hybrid\_uni - WS16 ($d_1$) + LH10 ($d_2$) & 0.46 & 0.57 & 0.58 & 0.15 \\
         Hybrid\_sub - LH10 ($d_1$) + WS16 ($d_2$) & 0.27 &0.64 & 0.32 & 0.48 \\
         Hybrid\_uni - LH10 ($d_1$) + WS16 ($d_2$) & 0.33 & 0.66 & 0.42 & 0.45 \\
    \hline
    \hline
    \end{tabular}
    \caption{\textbf{Burstiness values of temporal distributions - Hybrid models.} For each hybrid hypergraph generated we report the burstiness of the overall duration distribution, $\Delta B_T$, and the burstiness of the overall inter-event times distribution, $\Delta B_{\tau}$, for both nodes and hyperedges.}
    \label{tab:table3}
\end{table}

\newpage
\begin{figure*}[ht!]
\includegraphics[width=0.95\textwidth]{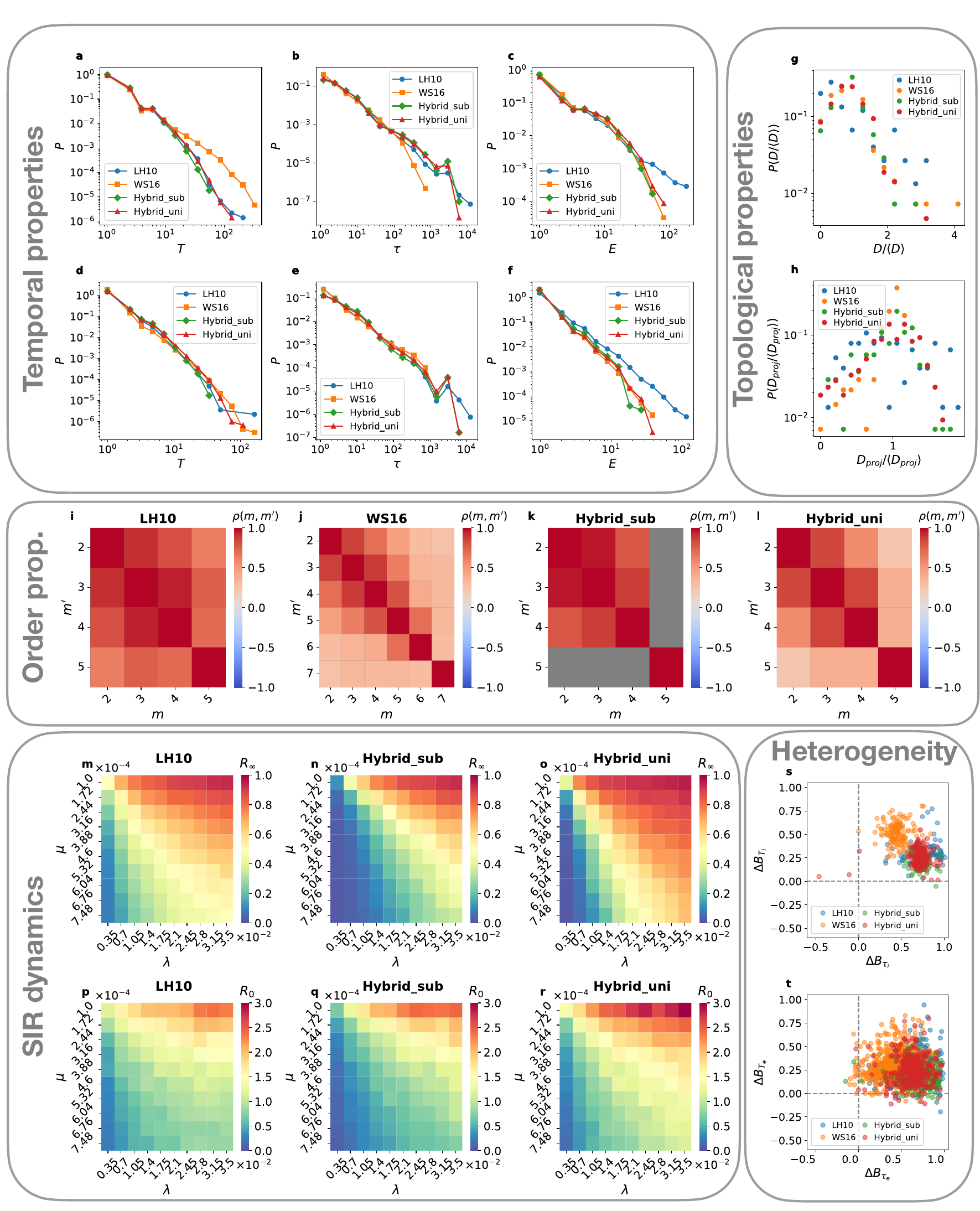}
\caption{\textbf{Hybrid hypergraphs - LH10 (system) + WS16 (population).} We consider two hybrid hypergraphs (\texttt{Hybrid\_sub} and \texttt{Hybrid\_uni}) generated with the EATH model considering the properties of the datasets LH10 ($d_1$ - system properties) and WS16 ($d_2$ - population properties). \textbf{a}-\textbf{f}: distributions of inter-event times, durations and train of events for nodes (\textbf{a}-\textbf{c}) and hyperedges (\textbf{d}-\textbf{f}); \textbf{g},\textbf{h}: hyperdegree distribution, $P(D/\langle D \rangle)$, and degree distribution in the projected graph, $P(D_{proj}/\langle D_{proj} \rangle)$; \textbf{i}-\textbf{l}: Pearson's correlation coefficient $\rho(m,m')$ between the node rankings at size $m$ and $m'$ obtained considering the time spent by nodes interacting at each order; \textbf{m}-\textbf{r}: the epidemic final-size $R_{\infty}$ and the basic reproduction number $R_0$ as a function of $(\lambda, \mu)$, obtained in the same conditions as Supplementary Fig. \ref{fig:figure33_0}; \textbf{s} (\textbf{t}): correlations between the burstiness of inter-event times distribution $\Delta B_{\tau}$ and of duration distribution $\Delta B_{T}$ for single nodes (hyperedges), we consider only the nodes and hyperedges with at least 10 events.} 
\label{fig:figure34a}
\end{figure*}

\newpage
\begin{figure*}[ht!]
\includegraphics[width=0.95\textwidth]{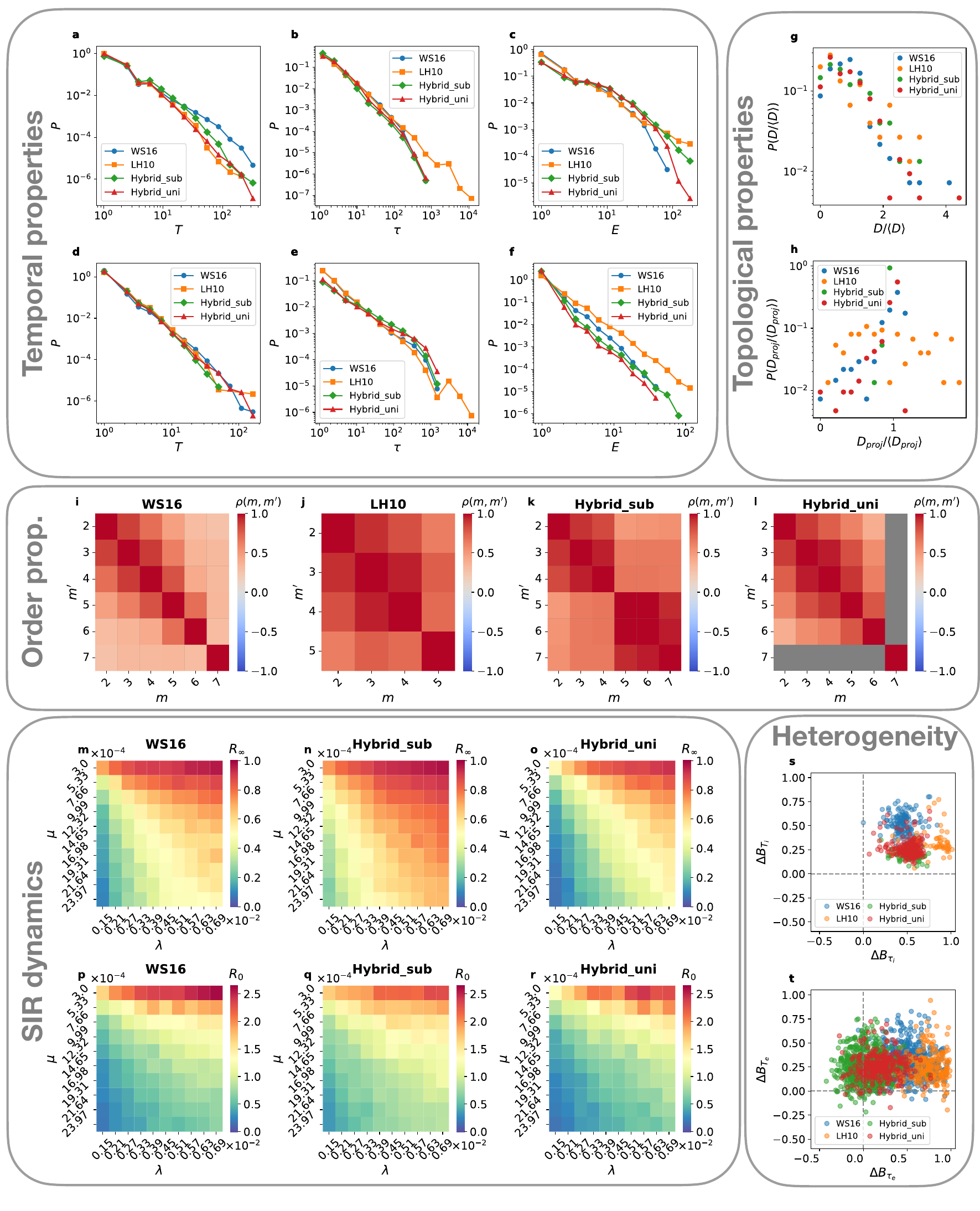}
\caption{\textbf{Hybrid hypergraphs - WS16 (system) + LH10 (population).} We consider two hybrid hypergraphs (\texttt{Hybrid\_sub} and \texttt{Hybrid\_uni}) generated with the EATH model considering the properties of the datasets WS16 ($d_1$ - system properties) and LH10 ($d_2$ - population properties). \textbf{a}-\textbf{f}: distributions of inter-event times, durations and train of events for nodes (\textbf{a}-\textbf{c}) and hyperedges (\textbf{d}-\textbf{f}); \textbf{g},\textbf{h}: hyperdegree distribution, $P(D/\langle D \rangle)$, and degree distribution in the projected graph, $P(D_{proj}/\langle D_{proj} \rangle)$; \textbf{i}-\textbf{l}: Pearson's correlation coefficient $\rho(m,m')$ between the node rankings at size $m$ and $m'$ obtained considering the time spent by nodes interacting at each order; \textbf{m}-\textbf{r}: the epidemic final-size $R_{\infty}$ and the basic reproduction number $R_0$ as a function of $(\lambda, \mu)$, obtained in the same conditions as Supplementary Fig. \ref{fig:figure33_0}; \textbf{s} (\textbf{t}): correlations between the burstiness of inter-event times distribution $\Delta B_{\tau}$ and of duration distribution $\Delta B_{T}$ for single nodes (hyperedges), we consider only the nodes and hyperedges with at least 10 events.} 
\label{fig:figure34b}
\end{figure*}

\newpage
\section{Artificial hypergraphs generation}
\label{sez:section3}
\begin{figure*}[ht!]
\includegraphics[width=\textwidth]{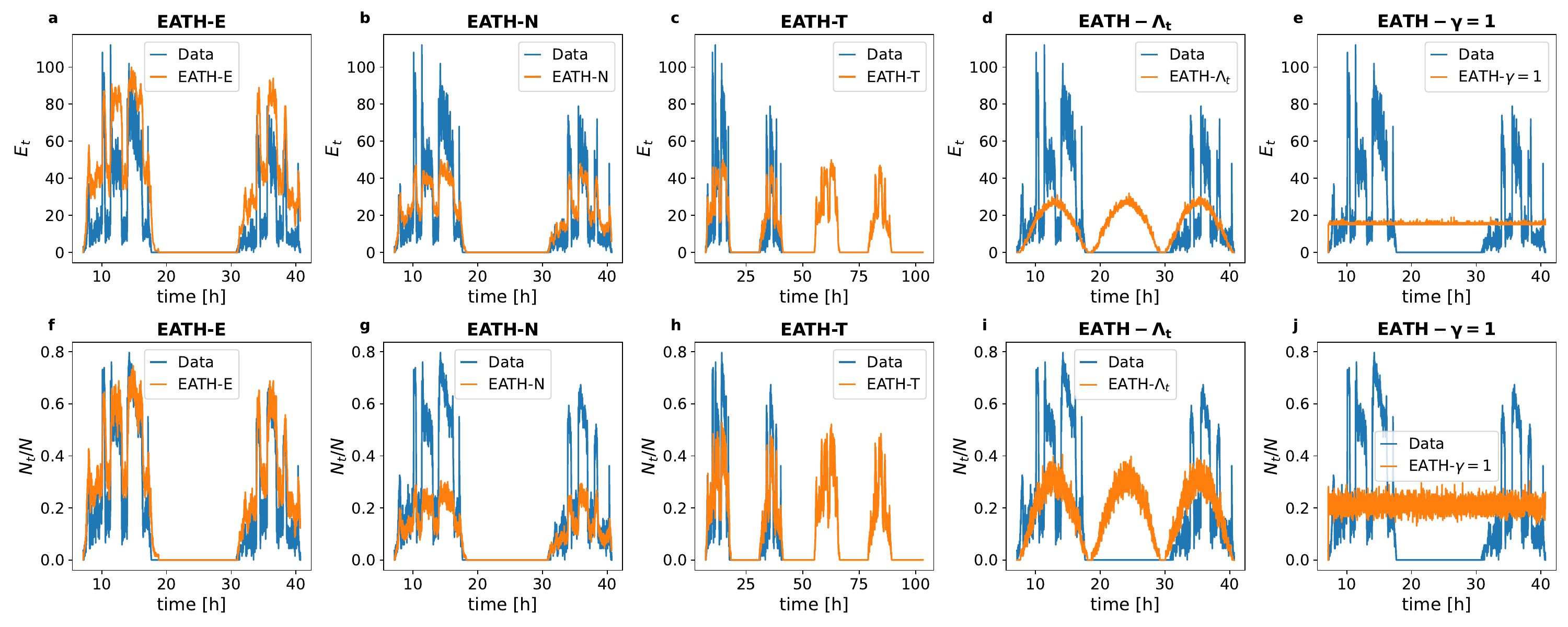}
\caption{\textbf{System activity - I.} In the first row we show the temporal evolution of the number of active hyperedges $E_t$ (of any size). In the second row we show the evolution of the fraction of active nodes $N_t/N$. In all panels we consider the WS16 dataset and the corresponding synthetic hypergraphs: EATH-E, EATH-N, EATH-T, EATH-$\Lambda_t$, EATH-$\gamma=1$ (see title).}
\label{fig:figure35}
\end{figure*}

\begin{figure*}[ht!]
\includegraphics[width=\textwidth]{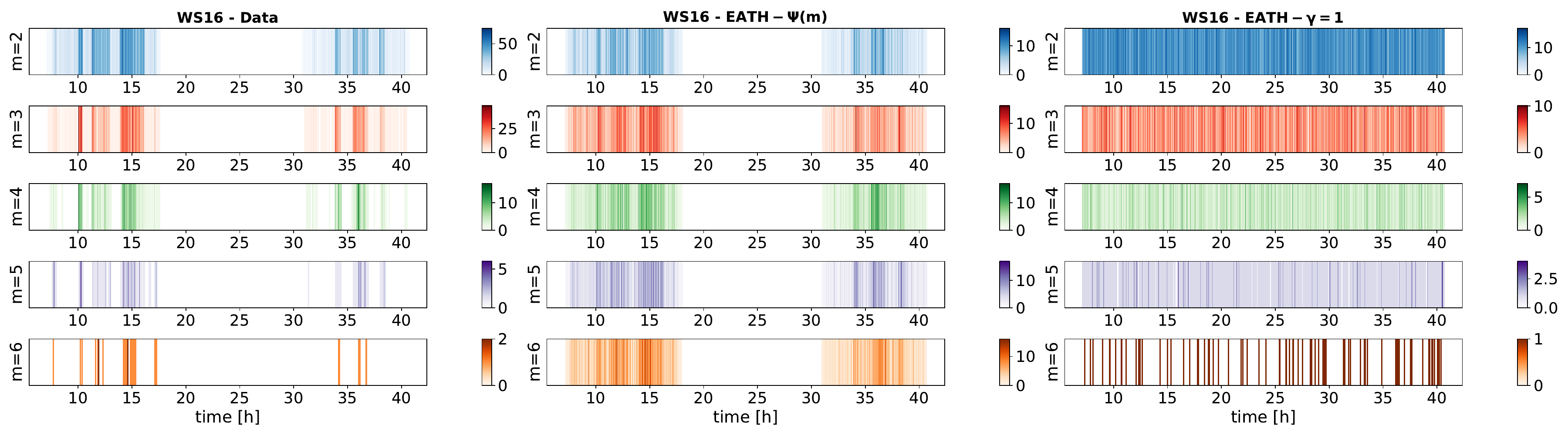}
\caption{\textbf{System activity - II.} We show the temporal evolution of the number of active hyperedges of size $m$, for $m \in [2,3,4,5,6]$, by considering the WS16 dataset and the corresponding synthetic hypergraphs: EATH-$\Psi(m)$, EATH-$\gamma=1$ (see title).}
\label{fig:figure36}
\end{figure*}

\begin{figure*}[ht!]
\includegraphics[width=\textwidth]{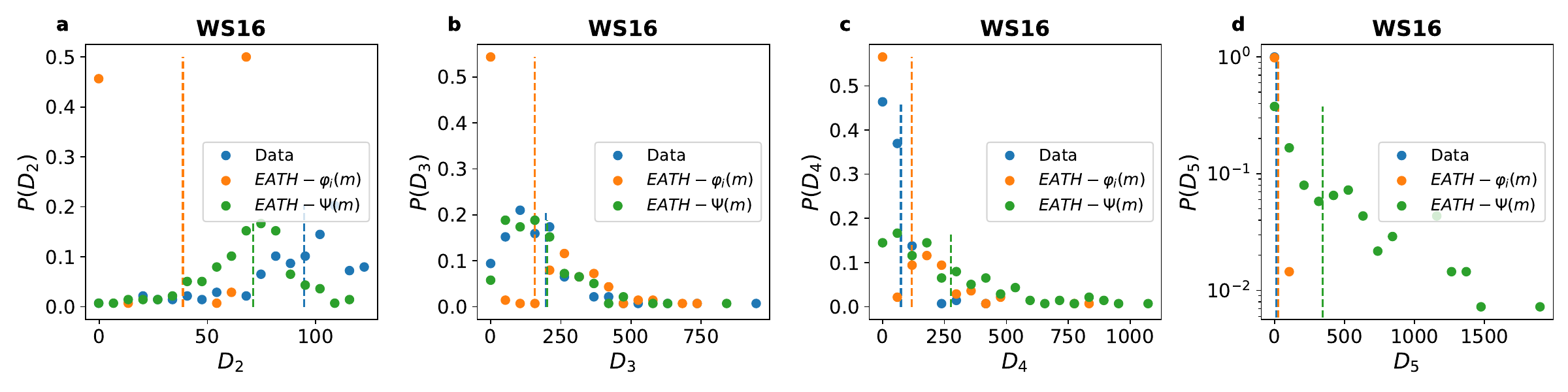}
\caption{\textbf{Higher-order topological properties.} In each panel we show the distribution $P(D_m)$ of nodes hyperdegree at size $m$, for $m \in [2,3,4,5]$, in the aggregated hypergraph $\mathcal{H}$ for the WS16 dataset and the corresponding synthetic hypergraphs: EATH-$\varphi_i(m)$, EATH-$\Psi(m)$.}
\label{fig:figure37}
\end{figure*}

\begin{figure*}[ht!]
\includegraphics[width=\textwidth]{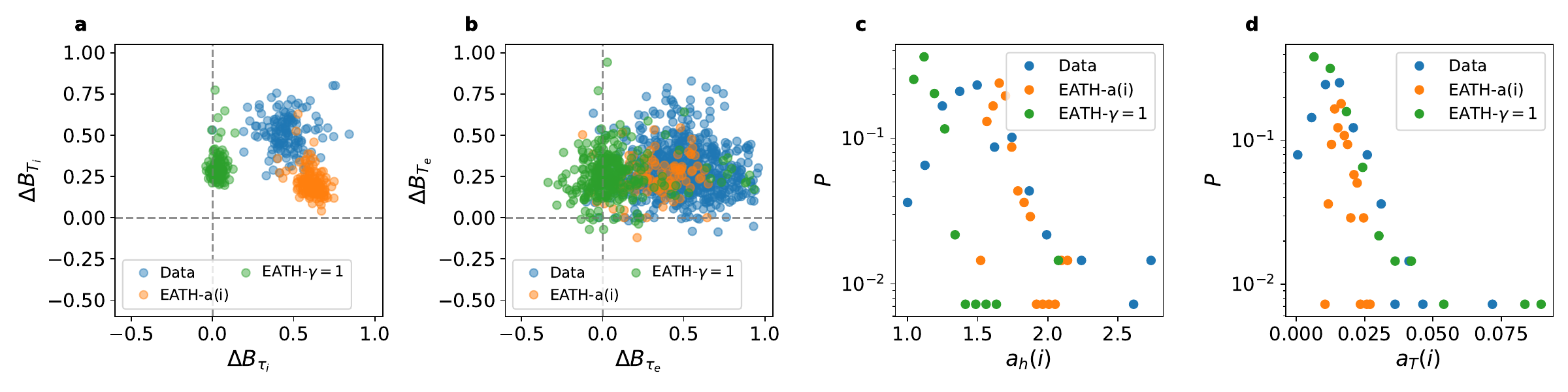}
\caption{\textbf{Nodes and hyperedges heterogeneity.} For each node and hyperedge we evaluate the burstiness $\Delta B_{\tau}$ and $\Delta B_{T}$ of their inter-event time and duration distributions. In panel \textbf{a} (\textbf{b}), each point corresponds to a node (hyperedge) and we show the correlations between $\Delta B_{\tau}$ and $\Delta B_{T}$. We consider only the nodes and hyperedges with at least $10$ activation events. In panels \textbf{c},\textbf{d} we show respectively the distribution of the instantaneous activity $P(a_h(i))$ and of the persistence activity $P(a_T(i))$, measured as described in the main text for the empirical and synthetic hypergraphs. In each case we consider the WS16 dataset and the corresponding synthetic hypergraphs: EATH-$a(i)$, EATH-$\gamma=1$.}
\label{fig:figure38}
\end{figure*}

\begin{figure*}[ht!]
\includegraphics[width=0.85\textwidth]{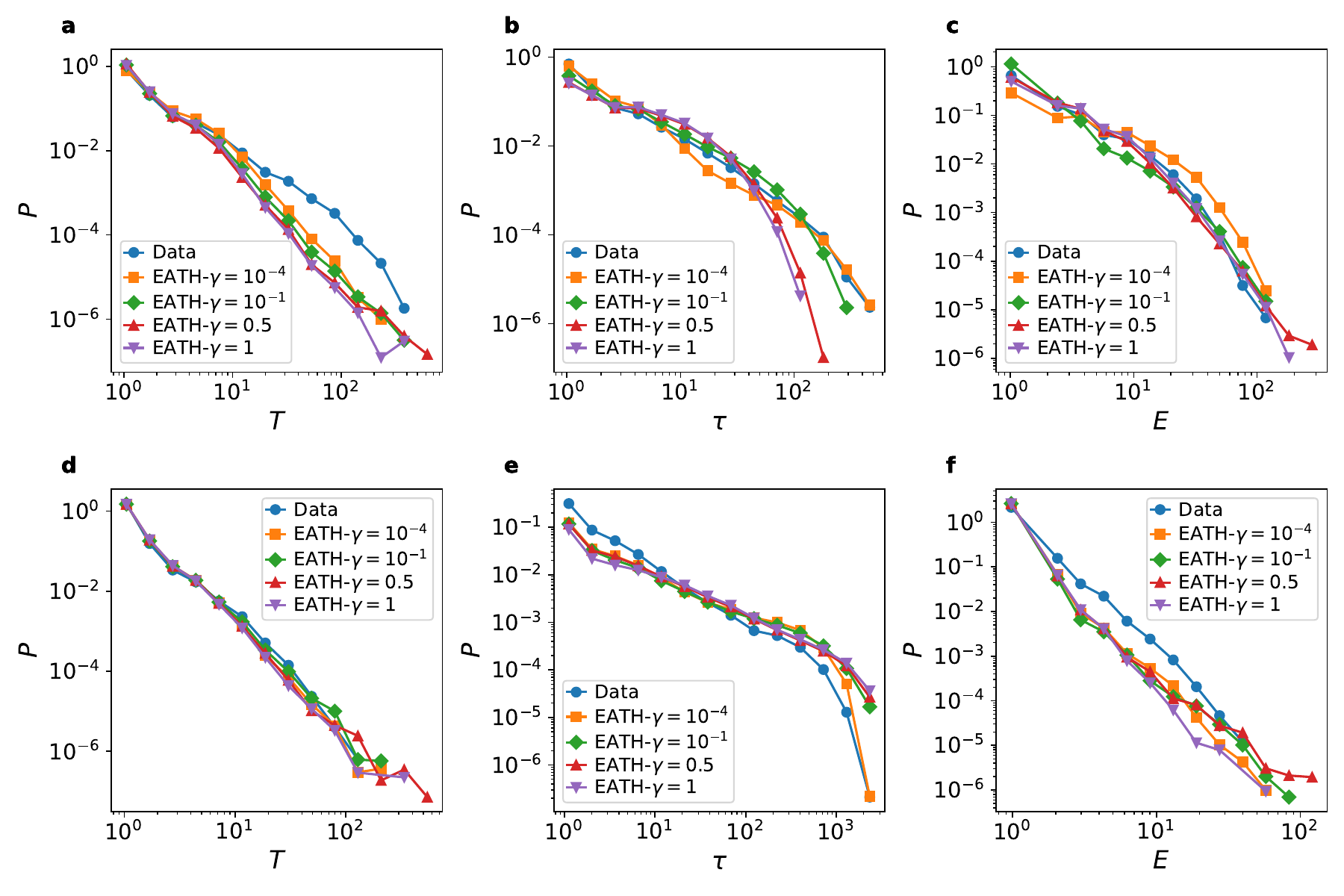}
\caption{\textbf{Temporal properties of the hypergraphs.} The first and second rows show respectively the temporal properties of nodes and hyperedges for the WS16 dataset and for different synthetic hypergraphs, EATH-$\gamma$ with $\gamma \in [10^{-4},10^{-1},0.5,1]$. \textbf{a},\textbf{d}: distribution of event durations $P(T)$;  \textbf{b},\textbf{e}: distribution of inter-event times $P(\tau)$;  \textbf{c},\textbf{f}: distribution of the number of events in a train of events $P(E)$, where a train is defined with $\Delta = 15 \delta t $.}
\label{fig:figure39}
\end{figure*}

\begin{figure*}[ht!]
\includegraphics[width=\textwidth]{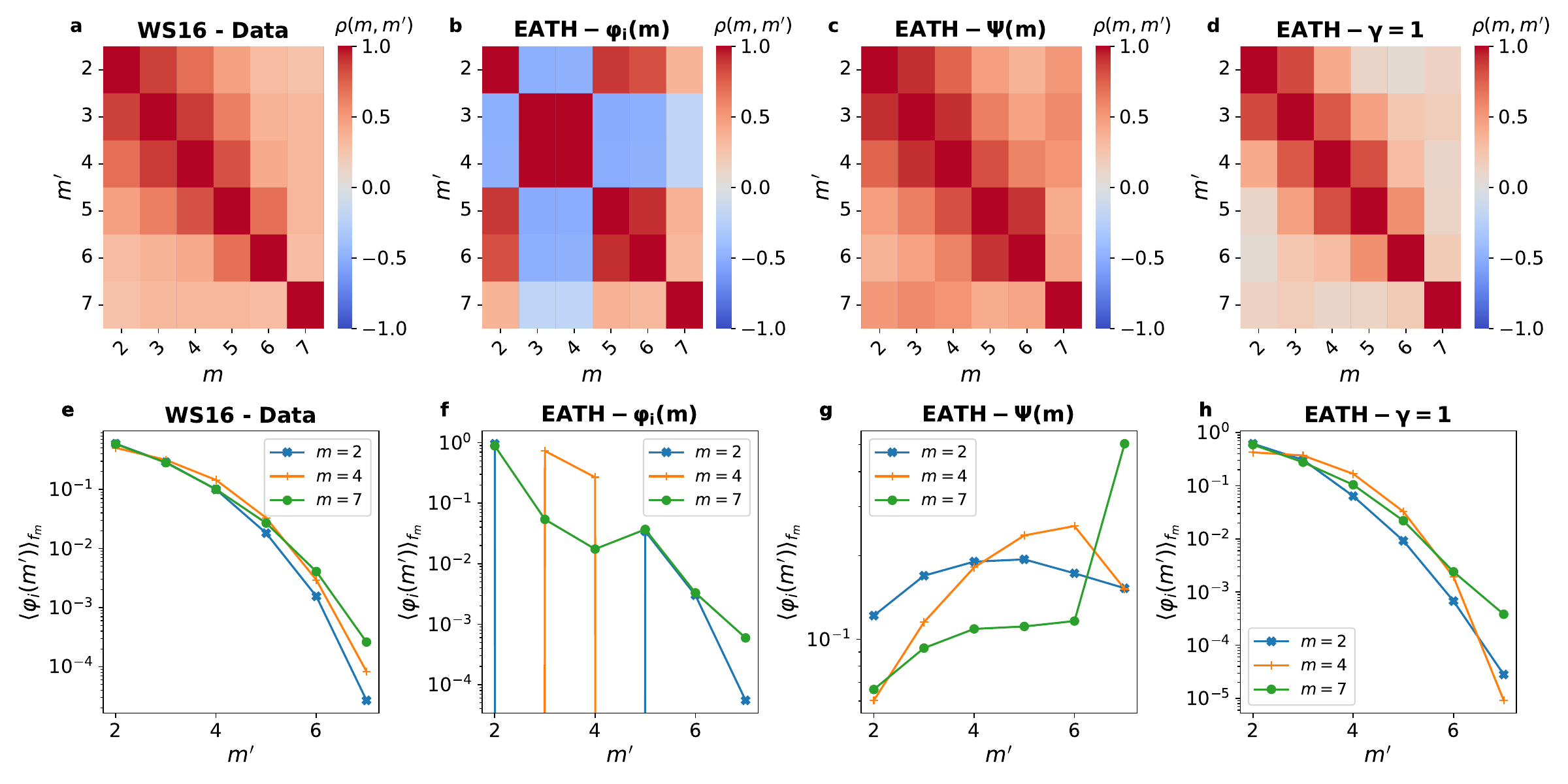}
\caption{\textbf{Participation of nodes at different interaction orders.} We consider the WS16 dataset and the corresponding synthetic hypergraphs: EATH-$\varphi_i(m)$, EATH-$\Psi(m)$, EATH-$\gamma=1$ (see title). For each of them, at each order $m$ we rank the nodes based on the time they spend interacting at size $m$ and we show the Pearson's correlation coefficient $\rho(m,m')$ between the rankings obtained at order $m$ and $m'$ (panels \textbf{a}-\textbf{d}). Moreover, we show the average order propensity $\langle \varphi_i(m') \rangle_{f_m}$ as a function of $m'$, averaged over the nodes occupying the top $f N$ positions of node rankings at order $m$, for different $m$ (panels \textbf{e}-\textbf{h}), fixing $f=0.1$.}
\label{fig:figure40}
\end{figure*}

\begin{figure*}[ht!]
\includegraphics[width=\textwidth]{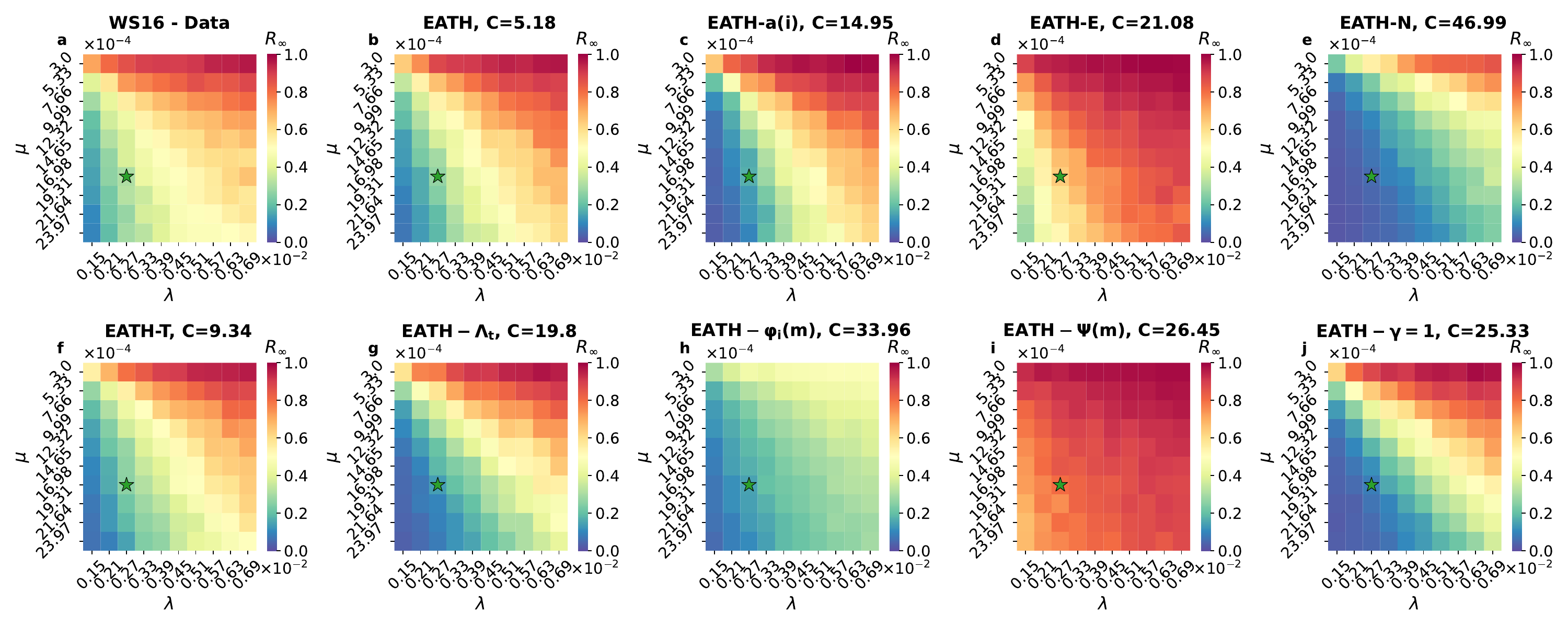}
\caption{\textbf{Higher-order SIR dynamics - I.} Each panel show the epidemic final-size $R_{\infty}$ as a function of the epidemiological parameters $(\lambda,\mu)$. The results are obtained by averaging over 400 simulations, fixing $\nu=4$, and considering the empirical WS16 hypergraph (Data) and all the synthetic hypergraphs generated (fully tuned to the dataset with memory - EATH - or with perturbed properties - EATH-*, see title). $C$ indicates the Canberra distance between the empirical and synthetic matrices. In all the panels the blue star indicate $(\lambda,\mu)=(0.27 \, 10^{-2},16.98 \, 10^{-4})$.}
\label{fig:figure41}
\end{figure*}

\begin{figure*}[ht!]
\includegraphics[width=\textwidth]{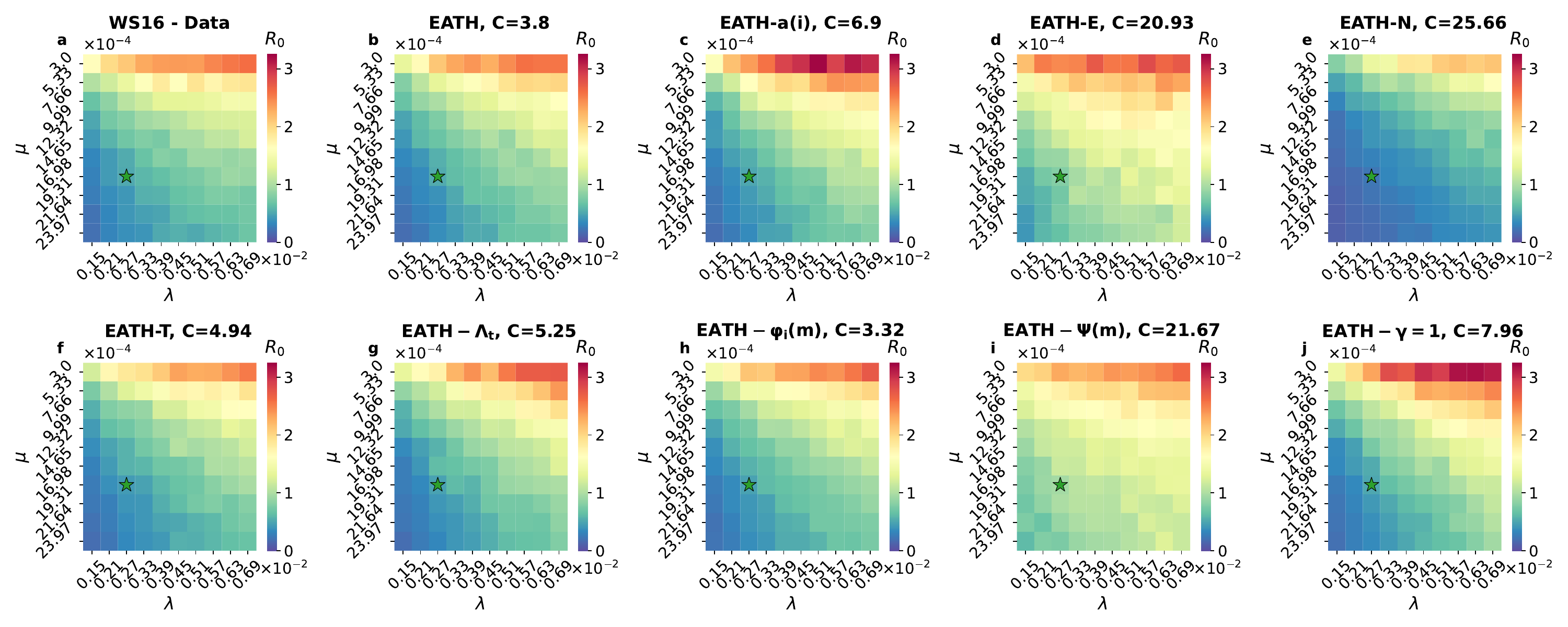}
\caption{\textbf{Higher-order SIR dynamics - II.} Each panel show the basic reproduction number $R_0$ as a function of the epidemiological parameters $(\lambda,\mu)$. The results are obtained by averaging over 400 simulations, fixing $\nu=4$, and considering the empirical WS16 hypergraph (Data) and all the synthetic hypergraphs generated (fully tuned to the dataset with memory - EATH - or with perturbed properties - EATH-*, see title). $C$ indicates the Canberra distance between the empirical and synthetic matrices. In all the panels the blue star indicate $(\lambda,\mu)=(0.27 \, 10^{-2},16.98 \, 10^{-4})$.}
\label{fig:figure42}
\end{figure*}

\begin{figure*}[ht!]
\includegraphics[width=\textwidth]{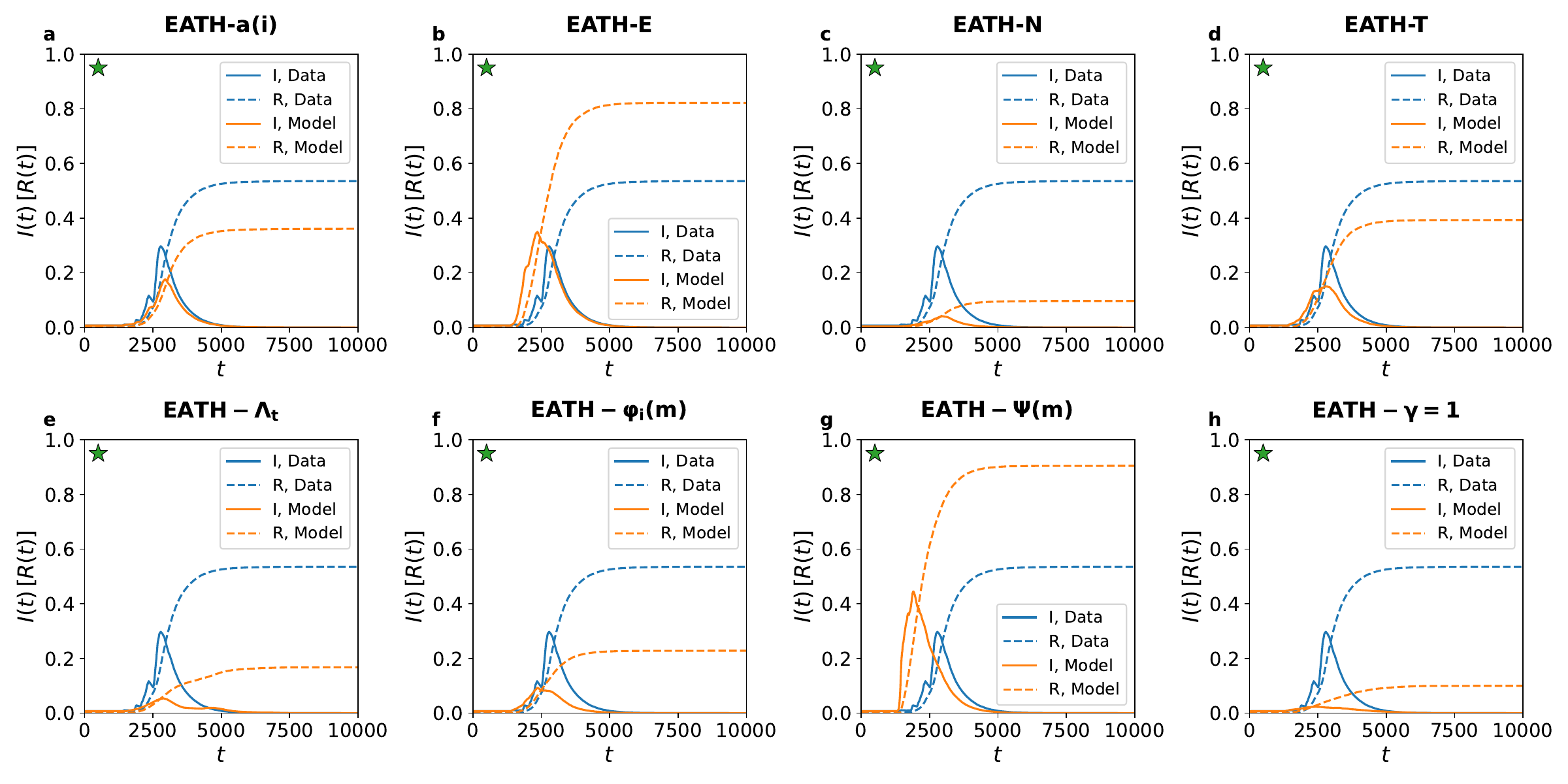}
\caption{\textbf{Higher-order SIR dynamics - III.} In each panel we show the results of numerical simulations, showing the fraction of infected $I(t)$ and recovered $R(t)$ nodes as a function of time, when the initial seed is infected at $t_0=0$ and averaging the curves over 200 realizations. In all the panels we fix $(\lambda,\mu)=(0.27 \, 10^{-2},16.98 \, 10^{-4})$, as indicated by the blue stars in Supplementary Figs. \ref{fig:figure41}, \ref{fig:figure42}, and we show the epidemic dynamics for the WS16 dataset and for synthetic hypergraphs with perturbed properties - EATH-* (see title).}
\label{fig:figure43}
\end{figure*}

Here we show the results of the generation of artificial temporal hypergraphs using the EATH model by setting synthetically some of the hypergraphs properties. We consider the following configurations (all the properties and parameters not explicitly indicated in each point are extracted from the corresponding dataset as described in the main text):
\begin{itemize}
    \item \textbf{EATH-a(i)}: we consider equal instantaneous activity $a_h(i)$ and persistence activity $a_T(i)$ for all the nodes, hence generating a homogeneous population;
    \item \textbf{EATH-E}: we consider a larger number of interactions, i.e. $\langle E_t \rangle = 2 \langle E_t \rangle_d$, where $\langle E_t \rangle_d$ is the interaction level in the dataset;
    \item \textbf{EATH-N}: we consider a larger population, i.e. $N=2N_d$, where $N_d$ is the number of nodes in the dataset. Each new node $j$ is an "image individual" that fully replicates the individual properties of another node $i$ of the empirical datasets selected randomly in the original population, i.e. with the same activities and order propensity.
    \item \textbf{EATH-T}: we consider a longer generation, i.e. $\mathcal{T}=2 \mathcal{T}_d$, where $\mathcal{T}_d$ is the dataset time-span, accounting for periodicity in the datasets, such as weekends and working hours;
    \item \textbf{EATH-}$\bm{\Lambda_t}$: we consider a different external modulation of the system activity, i.e. we consider $\Lambda_t=[\sin(3 \pi t/n)]^2$, where $n=\mathcal{T}/\delta t$;
    \item \textbf{EATH-}$\bm{\varphi_i(m)}$: we consider a synthetic version of the order propensity $\varphi_i(m)$. We divide randomly the nodes in two groups of equal size, $\mathcal{A}$ and $\mathcal{B}$, and we fix a block-shaped $\varphi_i(m)$ so that nodes in group $\mathcal{A}$ interact only at low and high orders, while nodes in group $\mathcal{B}$ interact only at intermediate orders.
    \item \textbf{EATH-}$\bm{\Psi(m)}$: we consider a hyperedge size distribution uniform over the interval $m \in [2,M]$;
    \item \textbf{EATH-}$\bm{\gamma}$: we consider different values of the $\gamma$ parameter, considering $\gamma \in [10^{-4},10^{-1},0.5,1]$.
\end{itemize}
For each of these configurations, we compare the generated properties with the corresponding empirical ones, showing how the topological (Supplementary Figs. \ref{fig:figure37}, \ref{fig:figure38}, \ref{fig:figure40}), temporal (Supplementary Figs. \ref{fig:figure35},\ref{fig:figure36},\ref{fig:figure38},\ref{fig:figure39}) and dynamical (Supplementary Figs. \ref{fig:figure41}-\ref{fig:figure43}) properties are modified. In each comparison we consider only the cases that are specifically constructed to modify the investigated properties, moreover we consider the WS16 dataset (see Methods) \cite{Genois2023}. In Supplementary Table \ref{tab:table4} we report the values of burstiness for the temporal distributions shown in Supplementary Fig. \ref{fig:figure39}, considering the hypergraphs generated with EATH-$\gamma$, for different $\gamma$ values.

\begin{table}[h!]
    \begin{tabular}{c|c|c|c|c}
    \hline
    \hline
         & \multicolumn{2}{c}{Nodes} \vline & \multicolumn{2}{c}{Hyperedges} \\ \hline
         & $\Delta B_{T}$ & $\Delta B_{\tau}$ & $\Delta B_{T}$ & $\Delta B_{\tau}$ \\ \hline
         Data & 0.52 &0.48 & 0.62 & 0.36 \\
         EATH-$\gamma=10^{-4}$ & 0.42 &0.57 & 0.61 & 0.11 \\
         EATH-$\gamma=10^{-1}$ & 0.60 & 0.29 & 0.70 & 0.15 \\
         EATH-$\gamma=0.5$ & 0.80 &0.10 & 0.88 & 0.21 \\
         EATH-$\gamma=1$ & 0.54 & 0.07 & 0.72 & 0.18 \\
    \hline
    \hline
    \end{tabular}
    \caption{\textbf{Burstiness values of temporal distributions - Artificial hypergraphs.} For each value of the parameter $\gamma$ of the EATH model, we report the burstiness of the overall duration distribution, $\Delta B_T$, and the burstiness of the overall inter-event times distribution, $\Delta B_{\tau}$, for both nodes and hyperedges. In all cases we consider the WS16 dataset.}
    \label{tab:table4}
\end{table}

\newpage

\newpage

\newpage

\newpage